\pdfoutput=1
\RequirePackage{ifpdf}
\ifpdf % We are running pdfTeX in pdf mode
\documentclass[pdftex]{sigma}
\else
\documentclass{sigma}
\fi

\usepackage{bbm}

\newcommand{\su}{\mathfrak{su}}

\def\unit{\mathbbm{1}}

\def\f{\frac}
\def\mone{^{-1}}
\def\tl{\widetilde}
\def\pp{\partial}

\def\eps{\epsilon}
\def\vareps{\varepsilon}
\def\thet{\vartheta}

\def\C{{\mathbbm C}}
\def\N{{\mathbbm N}}
\def\Z{{\mathbbm Z}}

\def\calA{{\mathcal A}}
\def\calC{{\mathcal C}}

\def\calG{{\mathcal G}}
\def\calH{{\mathcal H}}

\def\calO{{\mathcal O}}
\def\calT{{\mathcal T}}

\def\l({\left(}
\def\r){\right)}

\DeclareMathOperator{\cyl}{Cyl}
\DeclareMathOperator{\im}{im}
\DeclareMathOperator{\Ad}{Ad}

\DeclareMathOperator{\Hom}{Hom}

\DeclareMathOperator{\id}{id}

\DeclareMathOperator{\tor}{tor}

\DeclareMathOperator{\tr}{tr}
\DeclareMathOperator{\Tr}{Tr}

\DeclareMathOperator{\SU}{SU}

\DeclareMathOperator{\SO}{SO}

\DeclareMathOperator{\ISU}{ISU}

\numberwithin{equation}{section}

\begin{document}

\allowdisplaybreaks

\renewcommand{\thefootnote}{$\star$}

\renewcommand{\PaperNumber}{009}

\FirstPageHeading

\ShortArticleName{Lessons from Toy-Models for the Dynamics of Loop Quantum Gravity}

\ArticleName{Lessons from Toy-Models for the Dynamics \\ of Loop Quantum Gravity\footnote{This
paper is a contribution to the Special Issue ``Loop Quantum Gravity and Cosmology''. The full collection is available at \href{http://www.emis.de/journals/SIGMA/LQGC.html}{http://www.emis.de/journals/SIGMA/LQGC.html}}}

\Author{Valentin BONZOM~$^\dag$ and Alok LADDHA~$^\ddag$}
\AuthorNameForHeading{V.~Bonzom and A.~Laddha}

\Address{$^\dag$~Perimeter Institute for Theoretical Physics, 31 Caroline St. N,\\
\hphantom{$^\dag$}~ON N2L 2Y5, Waterloo, Canada}
\EmailD{\href{mailto:vbonzom@perimeterinstitute.ca}{vbonzom@perimeterinstitute.ca}}

\Address{$^\ddag$~Institute for Gravitation and the Cosmos,
Pennsylvania State University,\\
\hphantom{$^\ddag$}~University Park, PA 16802-6300, USA}
\EmailD{\href{mailto:alok@gravity.psu.edu}{alok@gravity.psu.edu}}

\ArticleDates{Received October 11, 2011, in f\/inal form February 24, 2012; Published online March 07, 2012}

\Abstract{We review some approaches to the Hamiltonian dynamics of (loop) quantum gravity, the main issues being the regularization of the Hamiltonian and the continuum limit. First, Thiemann's def\/inition of the quantum Hamiltonian is presented, and then more recent approaches. They are based on toy models which provide new insights into the dif\/f\/iculties and ambiguities faced in Thiemann's construction. The models we use are parametrized f\/ield theories, the topological BF model of which a special case is three-dimensional gravity which describes quantum f\/lat space, and Regge lattice gravity.}

\Keywords{Hamiltonian constraint; loop quantum gravity; parametrized f\/ield theories; topological BF theory; discrete gravity}

\Classification{83C45; 57R56; 83C27}

\renewcommand{\thefootnote}{\arabic{footnote}}
\setcounter{footnote}{0}

\section{Introduction}

The key feature of general relativity is the equivalence between gravity and geometry, so that geometry itself becomes a dynamical entity. Loop Quantum Gravity (LQG) \cite{alreview, nicolaireview, alejandroreview, ttbook} is a~quantization framework which particularly emphasizes those geometric content. It makes sense of quantum geometry of space, with quantum numbers of areas and volumes. And it is also more than that. It has been put to a rigorous mathematical status, which makes clear that it is a generic way to quantize background independent theories. This means that no background metric is available, and that general covariance must be satisf\/ied at all steps. A key theorem asserts the uniqueness of the quantization map when dif\/feomorphism invariance is required \cite{lost}. As a result, the basic excitations are supported by embedded graphs, known as spin networks.

After the completion of the kinematical framework, the task was to make sense of the Hamiltonian of general relativity. This implies two steps: $i)$ regularization within the loop quantization, and $ii)$ the continuum limit. In a remarkable series of papers, \cite{qsd1, qsd2, qsd3}, this task was carried out by Thiemann, and as a major success the quantization was shown to be anomaly-free, which means no classical degrees of freedom of the regularization are left over through quantization.

This operator in addition to its technical precision has a number of interesting properties like dif\/feomorphism covariance. It has also led to detailed quantization of mini-superspace models like Loop Quantum Cosmology (LQC). Although its use in the current formulation of dynamics in LQG is rather limited (spin foam models \cite{baez-sf-bf,carlo-newlook} and alternative approaches like the Master constraint in the canonical theory \cite{thiemann-master-constraint} have been introduced to circumvent the dif\/f\/iculties faced by Thiemann's construction), as far as canonical LQG is concerned it still retains a fundamental importance. There are already a large number of reviews on this subject. \emph{Our aim in this review is not merely to repeat what has been said above, but instead to offer a frank account of the state of the art in the field, focusing on the most difficult aspects, and some of the different ways in which researchers have tried to go beyond Thiemann's construction.}

Quickly after the release of Thiemann's proposal, several criticisms were made, see for instance~\cite{lm2, lm1, smolin-ultralocality}. As usual in quantum theory, the construction is plagued by numerous ambiguities, but those papers also call into question the physical content of the continuum limit induced by the quantum Hamiltonian. In a nutshell, the dif\/f\/iculty with Thiemann's Hamiltonian is that although the quantum algebra is mathematically anomaly-free, it was not possible to check explicitly that it implements the classical (Dirac) constraint algebra at the quantum level. In~\cite{smolin-ultralocality}, this issue was traced back to the fact that the quantum Hamiltonian acts in an ultra-local way, only at the vicinity of nodes of the spin network excitations, so that physical states lack long range correlations.

In spite of those criticisms, there has been very few modif\/ications of the proposal (see~\cite{alesci-rovelli-hamiltonian} for a recent improved regularization). However, as usual in physics, toy models are useful to get insights which should survive in the full theory, because they enable to disentangle the main issues. Here we will focus on
\begin{itemize}\itemsep=0pt
\item parametrized f\/ield theory in two dimensions,
\item topological BF f\/ield theory,
\item as a specif\/ic case of the latter, 2+1 gravity,
\item a canonical approach to discrete gravity (Regge calculus).
\end{itemize}
They aim at going deeper into the structure of the loop quantization, and explore new alternatives which are, sometimes surprisingly, fully amenable. Let us say a few words on those models (to avoid overf\/lowing of references at this stage, we put them in the corresponding sections).

Due to the inf\/inite number of ambiguities of the standard quantum Hamiltonian, there is a~vast room for improvement in the def\/inition of the operator. In the context of parametrized f\/ield theory, we will argue that the of\/f-shell closure of the constraint algebra is a key requirement which not only has the potential to reduce the number of ambiguities, but also that in simple models, this requirement leads to physically correct results. The main results appeared in \cite{amhamconstraint, alokme2, amdiffconstraint}.

Three-dimensional gravity is well-known as an exactly solvable topological f\/ield theory~\cite{witten-3d-gravity}, where transition amplitudes have been evaluated~\cite{witten-amplitude-3d}. However, one can try to forget as most as possible that it is topological, and treat it instead just as geometrodynamics with the loop quantization. The main question is then: What is the Wheeler--DeWitt equation, which would quantize the scalar Hamiltonian of the theory? Hence, this model addresses, in a successful way as we will show, the issue of the regularization in the loop quantization, and proves that it is possible to reproduce non-trivially the expected results of the topological formulation. The main references for the work we review are~\cite{semiclass-paper, 3d-wdw, witten-amplitude-3d}.

Three-dimensional gravity is a special case of the topological BF theory \cite{blau-thompson-bf, horowitz-bf} which can be def\/ined in arbitrary dimensions. Though gravity and BF theory are not equivalent in dimension four and higher, since they dif\/fer by the presence and the lack of local degrees of freedom, it turns out that they are still related \cite{symmetries-bf-gravity, freidel-puzio, freidel-speziale-bf-gravity, horowitz-bf}. One of these relations consists in formulating gravity as a BF model supplemented with constraints, and this is the main road to the spin foam quantization attempts to quantum gravity. If one wants to understand how spin foams may be able to circumvent the dif\/f\/iculties faced in the Hamiltonian quantization, one should therefore start with a good understanding of the topological model. However, until recently, its loop (and spin foam) quantization was not so much developed. It is mainly known in the three-dimensional case, thanks to \cite{freidel-louapre-pr1, noui-perez-ps3d}, and progress in higher dimensions have only appeared very recently~\cite{sorting-out, bf-anydim}. A pedagogical review is \cite{baez-sf-bf} but it is now ten years old. \emph{We consider the section on the BF model of the present review as an update to this review, proving it is now much better understood.}

Since general relativity describes the dynamics of the geometry, it is interesting to see how the quantization of the 2+1 scalar constraint describes f\/lat quantum geometry. This geometry is naturally discrete since it is supported by graphs, so that it makes contact with the canonical approach to discrete gravity, which looks for a Hamiltonian respecting the gauge symmetries, and hence the physical degrees of freedom of the full theory. Such a Hamiltonian can be reached through the so-called perfect discrete action, obtained after coarse-graining. The material of the review is borrowed from~\cite{bahr-broken-sym, bahr-improved-regge}.

The literature is quite rich and we have certainly left interesting works behind. For instance, we have not been able to cover most of the work done in certain mini-superspace models. Similarly, very little will be said on spin foam models, though parts of the work presented in the review were motivated by linking canonical LQG to spin foams, especially \cite{3d-wdw, noui-perez-ps3d} in 2+1 dimensions. In higher dimensions, the most recent works are \cite{alesci-noui, alesci-thiemann, recurrence-paper}.

As we will show, the technology used in loop quantum gravity, especially in the case of quantum f\/lat space, relies heavily on $\SU(2)$ re-coupling methods. While the basic tools are very well-known for a long time \cite{varshalovich, jucys}, research in loop quantum gravity has led over the last years to interesting side-products on advanced quantum angular momentum theory. They include exact results, such as new recursion formulae for Wigner $6j$-symbols and more complicated objects inspired by spin foam models \cite{yet-another, recurrence-paper}, and also new asymptotic expansion in the regime of large angular momenta. In particular, the expansion of the $6j$-symbol is now understood beyond the usual leading order \cite{6jnlo, pushing6j, 6jmaite}, asymptotics of arbitrary Wigner coef\/f\/icients with some large and small quantum numbers have been obtained in \cite{3njsmall} completing the exciting results of \cite{yu15j, yu12j, yu9j} based on an improved WKB approximation. The coherent states technology introduced in the context of quantum gravity in \cite{livine-coherentBF} has been very useful in deriving new asymptotics for $15j$-symbols and other objects from spin foams \cite{barrett-asym-summary, barrett-asym15j}.

Such developments in quantum angular momentum theory have not appeared only in the quantum gravity community \cite{3nj-marzuoli, qm6j}, and is now considered as a key topic in modern physics~\cite{spinnets-marzuoli}, beyond the traditional use in spectroscopy and atomic/molecular physics, for instance in quantum computing \cite{quantum-tet-marzuoli}, in the presence of topological order~\cite{levin-wen-condensation}. This is not so surprising since the reference model for topological order and fault-tolerant computing is the Kitaev model~\cite{kitaev-code}, which is actually a lattice BF model with the group~$\Z_2$. We think those progress on quantum angular momenta in the study of quantum f\/lat space is one of the major consequences of quantum gravity outside of the f\/ield and would deserve their own review.

Such models lying at the border between toy models for condensed matter and for quantum gravity have been recently proposed in order to investigate the second key issue, the continuum limit in background independent quantization schemes. Since the project has been started very recently, we will not comment further, but refer to \cite{bianca-finite, bianca-spinnets} instead.

A dif\/ferent road towards a generic framework for taking the continuum limit is under active development: tensor models and group f\/ield theories, which generalize matrix models to dimensions higher than two. A paragraph on those models is included but we refer to the reviews~\cite{freidel-gft, oriti-review} and more recently~\cite{gurau-review} for more details.

Since we review quite advanced works, on the dynamics itself, we include for completeness some basic material of the loop quantization and its kinematical aspects in  Section~\ref{sec:backmat}. Then, we give an introduction to Thiemann's construction in Section~\ref{sec:thiemann}. In the interest of pedagogy, we restrict to the Euclidean Hamiltonian constraint. This is because the full Lorentzian constraint is the sum of this Euclidean term with an additional piece, so that it is in any cases necessary to understand the Euclidean part. A small paragraph on the master constraint program which was launched by Thiemann after criticisms of its construction is included. Throughout this section, we deliberately focus on the unsatisfactory aspects of the construction, for two reasons: $i)$~its mathematical consistency is already emphasized in previous reviews, $ii)$~to stress the issues which we address later using toy models.

While our presentation of the standard quantum Hamiltonian may sound a bit pessimistic, this is to contrast with the present situation which seems better to us. In the following sections, we move on to present some exciting results which have been obtained in the very last years. In Section~\ref{sec:pft}, we consider a simple two-dimensional generally covariant f\/ield theory to extract some interesting lessons. In Section~\ref{sec:diffeo}, the dif\/feomorphism constraint itself is shown to admit a~quantization rather satisfactorily if one is ready to examine all possible choices we are af\/f\/licted with in LQG. In Section~\ref{sec:bf}, we present a dif\/ferent approach based on the topological BF theory. It contains some review material on the BF model, together with the most recent quantization of the scalar Hamiltonian. A discussion on how to get beyond the topological model is also included. Some conclusions are presented in Section~\ref{sec:con}.

\section{Background material} \label{sec:backmat}

This section presents some background material on loop quantum gravity. More details can be found in the reviews \cite{alreview, alejandroreview, ttbook}.

{\bf Yang--Mills phase space.} The (unreduced, i.e.\ before imposing the constraints which generate the dynamics) phase space of 3+1 general relativity in Ashtekar--Barbero variables is quite similar to the Yang--Mills phase space. On the canonical surface $\Sigma$, we have a $\SU(2)$ connection 1-form $A_a^i$, where $i=1,2,3$ is a $\su(2)$ index and $a$ is the form index, and the canonical momentum $E^a_i$, which satisfy the Poisson brackets
\begin{gather*}
\big\{ A_a^i(x), E^b_j(y)\big\} = \gamma\ \delta^b_a\,\delta^i_j\,\delta^{(3)}(x-y).
\end{gather*}
The momentum is the non-Abelian version of the electric f\/ield. In 3+1 general relativity, it is the triad of the intrinsic geometry on $\Sigma$, related to the 3-metric $q_{ab}$ by $q q^{ab}= \delta^{ij} E^a_i E^b_j$. Here $\gamma$ is a free parameter of the classical theory known as the Immirzi parameter. The Yang--Mills phase space is the same with $\gamma=1$. In three-dimensional Riemannian gravity, the phase space is also the same, with $\gamma=1$ (but the form index just takes two values when $\Sigma$ is two-dimensional).

{\bf Holonomies.} From a connection, locally seen as 1-form on $\Sigma$ with special rules for gauge transformations, one can build holonomies, whose geometric meaning is to enable parallel transport of vectors. The holonomy along a path $\ell:[0,1]\rightarrow \Sigma$ is $U_\ell(A) = \mathcal{P} \exp -\int_\ell A$, where $\mathcal{P}$ denotes the path ordering. The loop quantization also features \emph{generalized connections}, which are objects which assign holonomies to any paths.

{\bf Kinematical states of the loop quantization.} The loop quantization provides a natural kinematical Hilbert space to quantize the Yang--Mills phase space. The idea is to probe the connection only through a f\/inite number of variables, similarly to a Wilson line which probes the connection along a given path. We introduce the holonomies $U_e(A)$ of $A$ along the path $e$. Given a graph $\Gamma$, with $E$ edges and $V$ vertices, and a function $f$ over $(\SU(2))^E$, we form the \emph{cylindrical function} $\psi_{\Gamma,f} \in \cyl_\Gamma$ supported on $\Gamma$ as
\[
\psi_{\Gamma,f}(A)  =  f\bigl( U_{e_1}(A),\dots, U_{e_E}(A)\bigr).
\]
We are moreover interested in $\SU(2)$ gauge invariant states. Gauge transformations act on holonomies only on their endpoints. If $h$ is a map from $\Sigma$ to $\SU(2)$, then the holonomy transforms as $U_e(A^h) = h(t(e)) U_e(A) h(s(e))\mone$, with $t(e)$, $s(e)$ being respectively the source and target points of the path~$e$. When focusing on a single graph $\Gamma$, this reduces gauge transformations to an action of $\SU(2)^V$ on the set of cylindrical functions over~$\Gamma$. So from any function~$f$ over $\SU(2)^E$, one gets an invariant function by averaging over the $\SU(2)^V$ action.

The algebra of such functions has a natural $\SU(2)^V$-invariant inner product which comes from the Haar measure on $(\SU(2))^E$,  $d\mu_\Gamma = \prod_e dg_e$, giving
\[ %\label{ps graphe}
\langle \phi_{\Gamma,h} | \psi_{\Gamma,f}\rangle  =  \int \prod_{e=1}^E dg_e\, \bar{h}(g_1,\dots,g_E) f(g_1,\dots,g_E) .
\]
By completing this space for the corresponding norm, one obtains the Hilbert space $\mathcal{H}_\Gamma = L^2\bigl(\SU(2)^E/\SU(2)^V, d\mu_{\Gamma}\bigr)$.

The union of all functions which are cylindrical with respect to some graph is denoted $\cyl=\cup_\Gamma \cyl_\Gamma$. It comes equipped with a natural Hermitian inner product inherited from the measures~$d\mu_\Gamma$. The completion of $\cyl$ then leads to the kinematical Hilbert space of the loop quantization~$\mathcal H_{\rm kin}$. The latter can be seen as the set of square-integrable functions over the set $\bar{\mathcal{A}}$ of generalized connections, $\mathcal{H}_{\rm kin} = L^2(\bar{\mathcal{A}},d\mu_{\rm AL})$, where $\mu_{\rm AL}$ is known as the Ashtekar--Lewandowski measure. $\mathcal{H}_{\rm kin}$ can also be constructed from the set of Hilbert spaces $\mathcal{H}_\Gamma$ for all possible graphs $\Gamma$, using projective techniques. Importantly, the measure $\mu_{\rm AL}$ is \emph{invariant} under dif\/feomorphisms of $\Sigma$.

{\bf The spin network basis.} It is an orthogonal basis of the above $L^2\bigl(\SU(2)^E/\SU(2)^V, d\mu_{\Gamma}\bigr)$. The f\/irst idea is to expand any element of $L^2(\SU(2)^E,d\mu_\Gamma)$ onto the matrix elements of the irreducible representations of $\SU(2)^E$, labeled by $E$ spins $j_e\in\N/2$. Then, rotation invariance at vertices imposes to contract the matrix elements of edges meeting at a vertex with intertwiners~$\iota_v$. If $v$ is a node of $\Gamma$ with ingoing links $(e_{\rm in})$, and outgoing links $(e_{\rm out})$, then an intertwiner~$\iota_v$ is a map: $\otimes_{e_{\rm in}}\calH_{j_e} \rightarrow \otimes_{e_{\rm out}}\calH_{j_e}$, (where $\calH_j$ is the carrier space of the spin~$j$ representation) which commutes with the group action. Thus, to form the spin network basis on $\calH_\Gamma$, we need to assign spins to edges and a basis of intertwiners at each node. For a trivalent vertex with ingoing edges for example, there is a single intertwiner $\iota_{j_1 j_2 j_3}: \calH_{j_1}\otimes \calH_{j_2} \otimes \calH_{j_3} \mapsto \C$ up to normalization, its components in the standard magnetic number basis being the Wigner $3jm$-symbol $\langle j_1, m_1 ; j_2, m_2 ; j_3, m_3 | \iota_{j_1 j_2 j_3}| 0\rangle = \left(\begin{smallmatrix} j_1 &j_2 &j_3 \\m_1 &m_2 &m_3 \end{smallmatrix}\right)$. Hence, spin network states are the following $\SU(2)^V$-invariant functions
\[ %\label{def spinnet}
s_\Gamma^{\{j_e,\iota_v\}}(g_1,\dots,g_E)  =  \sum_{\{m_e, n_e\}} \prod_{e=1}^E \langle j_e, m_e\lvert g_e\rvert j_e, n_e\rangle  \prod_{v=1}^V \langle \otimes_{e\,{\rm in}} \, j_e,m_e \lvert \iota_v \rvert \otimes_{e\,{\rm out}}  j_e, n_e\rangle ,
\]
which form an orthogonal set. This enables a transform between gauge invariant functions on~$\Gamma$ and functions over the colorings $(j_e,\iota_v)$.

{\bf Phase space on a graph.} There is a natural classical phase space on $\Gamma$, inherited from the full phase space, whose quantization leads to $L^2\bigl(\SU(2)^E/\SU(2)^V, d\mu_{\Gamma}\bigr)$. Each link $e$ carries a~group element $g_e\in\SU(2)$, which can be seen as the holonomy of the connection. In this picture, each node carries a local frame, and $g_e$ performs parallel transport from the frame of the source node of $e$ to its target node. The momenta $E^a_i$ give rise to f\/luxes when integrated along surfaces of codimension~1 on $\Sigma$. If it is done inf\/initesimally close to a node, source of a~link~$e$, we get the f\/lux variable $E^i_e\in\su(2)$, def\/ined with respect to the frame over the source node. The Poisson bracket is the structure of the cotangent bundle over $\SU(2)^E$,
\[
\bigl\{E_e^i,E_e^j\bigr\} = \eps^{ij}_{\phantom{ij}k} E_e^k ,\qquad \bigl\{E_e^i, g_e\bigr\} = g_e \tau^i ,
\]
where the matrices $(\tau_i)_{i=1,2,3}$ are anti-hermitian matrices\footnote{They read: $\tau^i = -\f{i}{2}\sigma^i$, in terms of the Pauli matrices $(\sigma^i)$.} satisfying $[\tau^i,\tau^j] = \eps^{ij}_{\phantom{ij}k}\tau^k$. All other brackets vanish. One can obviously def\/ine a f\/lux variable in the frame of the target node of $e$, using the adjoint action of the holonomy on the initial f\/lux, $\widetilde{E}_e = \Ad(g_e)\,E_e$.

{\bf Geometric content.} For simplicity, we consider $\Gamma$ to be dual to a triangulation of $\Sigma$. Fluxes, which come from the triad, encode the \emph{intrinsic geometry of the triangulation}. The main reason is that the Gau\ss{} law, which generates $\SU(2)$ transformations at a node $v$, reads
\begin{gather} \label{closure}
\sum_{e\text{ outgoing}} E^i_e - \sum_{e\text{ ingoing}} \widetilde{E}_e^i = 0.
\end{gather}
Thinking of the node as dual to a tetrahedron, and the links dual its triangles, each of them carries a f\/lux and this allows to interpret the above relation as the closure of the tetrahedron, where f\/luxes are the (oriented) normals to the triangles (in the frame of the tetrahedron). Areas are given by their norms, and dihedral angles by their dot products. This way, the geometry of each tetrahedron is completely specif\/ied. However, dif\/ferent tetrahedra will generically assign dif\/ferent lengths to their common edges. Such geometries are described in \cite{phase-space-dittrich, simplicity-dittrich, twisted-freidel, twisted-carlo-simone} and are known as \emph{twisted} geometries.

\begin{figure}[t]\centering
\includegraphics[scale=0.65]{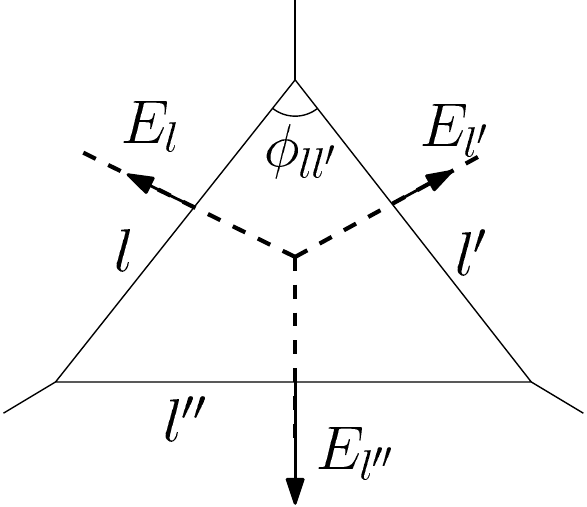}
\caption{The dual graph, in dashed lines, which supports the phase space, and the triangulation. The f\/luxes are the normals to the edges of the triangle, they encode the local embedding into f\/lat 3-space. Their dot products are proportional to the cosines of the angles $\phi_{ee'}$.} \label{fig:dualgraph}
\end{figure}

A similar interpretation holds in 2+1 gravity. Since f\/luxes are still three-dimensional vectors, one should rather think of nodes as dual to triangles locally embedded in f\/lat 3-space. The Gau\ss{} law on a node is the closure of the dual triangle, so that f\/luxes based on a node represent the normals to the dual edges in the frame of the dual triangle, as shown in Fig.~\ref{fig:dualgraph}. Norms and dot products of f\/luxes give access to the lengths and $2d$ angles within the triangles. The closure relation in particular asserts that
\begin{gather} \label{alkashi}
\cos\phi_{e_1e_2} = \frac{E_{e_1}^2 + E_{e_2}^2 - E_{e_3}^2}{2  \vert E_{e_1}\vert\,\vert E_{e_2}\vert},
\end{gather}
which is the standard formula to compute the angles of f\/lat triangles from their lengths, namely the law of cosines.

In contrast with the 3+1 case, the geometries encoded into f\/luxes in 2+1 dimensions assign lengths unambiguously to the whole triangulation and are thus Regge geometries.

It is also possible to introduce dihedral angles between the top-simplices of the triangulation, which give a notion of \emph{extrinsic geometry}, using f\/luxes and holonomies. The Hamiltonian of general relativity is precisely a relation between extrinsic and intrinsic geometries, and will be described later in the f\/lat case.

\section{Quantum Hamiltonian constraint: Thiemann's def\/inition} \label{sec:thiemann}

In this section we review the quantization of Hamiltonian constraint, as proposed by Thiemann in~\cite{qsd1, qsd2}. Instead of focusing on the details, we focus on the basic underlying ideas and what in our opinion are its major shortcomings.

In a remarkable series of papers, Thiemann proposed to quantize the Lorentzian Hamiltonian constraint of canonical gravity by using in a crucial way certain relations involving Poisson brackets at the classical level,
together with suitable choices of intermediate regularization schemes. The f\/inal Hamiltonian constraint turned out to be well def\/ined on ${\cal H}_{{\rm kin}}$. It showed glimpses of generating an anomaly free quantum Dirac algebra, as the commutator of two Hamiltonian constraints, when acting on any spin network state $T_{s}$ in ${\cal H}_{{\rm kin}}$, gave a null state when averaged over spatial dif\/feomorphisms.

However as we will see below, the def\/inition given by Thiemann is unsatisfactory for several reasons. This is not a criticism of the construction itself as it was the f\/irst ever rigorous def\/inition of the Hamiltonian constraint, but as it has been over f\/ifteen years since Thiemann's seminal work, it is perhaps good to take stock of the current state of the art in the f\/ield.

As we argue below, the most worrisome aspect of Hamiltonian constraint is that there are inf\/initely many ambiguities which feed into its def\/inition. Each such choice could give rise to (in the continuum limit) an inequivalent quantization of the Hamiltonian constraint. The preferred choices currently used in the literature are based on subjective criteria rather then appealing to certain physical requirements like local Lorentz invariance.

The classical, density one, Euclidean Hamiltonian constraint is given by
\begin{gather*}
H_E[N] = \int_{\Sigma} d^3x\, N(x) \epsilon_i^{\phantom{i}jk} \frac{F_{ab}^{i} E^{a}_{j} E^{b}_{k}}{\sqrt{q}}(x),
\end{gather*}
(we have set $G = 1$).
As we will see shortly, it is only for density weight one, that the continuum constraint operator can be well def\/ined on ${\cal H}_{{\rm kin}}$. However it is important to note that, it is a~priori not required that the domain for the quantum Hamiltonian should be in ${\cal H}_{{\rm kin}}$, and could very well be a subspace of $\cyl^{*}$ in which case other density weighted constraints could also admit a continuum limit.

One of the ingenious features of Thiemann's construction is the following. The density one constraint, although  non-polynomial in the basic f\/ields could be written as
\begin{gather*}
H_{E}[N]  =  \int_{\Sigma} N(x) \Tr (F(x) \wedge \{A(x), V(R)\}),
\end{gather*}
where $R$ is any open region containing the point $x$. $V(R)$ is the volume of this region, as a~function of the f\/ield $E$.

As the elementary variables in the theory are holonomies along (piecewise analytic) edges and f\/luxes along (semi-analytic) surfaces, we need to f\/irst approximate $H_{E}[N]$ by a sequence of functionals of holonomies and f\/luxes such that the limit of the sequence (in the suitable sense) is $H_{E}[N]$. This is accomplished as follows. Choose a cell decomposition
of the underlying manifold~$\Sigma$ (it can be simplicial or cubical for convenience).  For each 3-cell $\triangle$ in~$T$,  choose a vertex $v(\triangle)$, a family of edges ${\cal S}_{e} = \{s_{1}(\triangle),\dots,s_{M_{s}}(\triangle)\}$ and a family of loops ${\cal S}_{L} = \{\alpha_{1}(\triangle),\dots,\alpha_{M_{L}}(\triangle)\}$ and consider
\begin{gather*}%\label{eq:classicalham}
H_{T(\epsilon) E}[N] = \sum_{\triangle\in T(\epsilon)}\sum_{i=1}^{M_{s}}\sum_{J=1}^{M_{L}}{\cal C}^{iJ}\,\Tr( h_{\alpha_{J}(\triangle)}h_{s_{i}(\triangle)}\big\{h_{s_{i}(\triangle)}^{-1}, V(R_{v(\triangle)})\big\},
\end{gather*}
where $C^{iJ}$ are constants independent of $\epsilon$. Each choice of $( T(\epsilon), \{C^{iJ}\}, {\cal S}_{e}, {\cal S}_{L}\ )$ such that,
\begin{gather*}
\lim_{\epsilon\rightarrow 0}H_{T(\epsilon) E}[N]  =  H_{E}[N]
\end{gather*}
is an acceptable net of regulators. Classically one choice is as good as the next (in the limit of inf\/inite ref\/inement) but quantum mechanically it is not clear which of the choices are  indistinguishable in the continuum limit.

As $H_{T(\epsilon)}[N]$ is a function of holonomies and volume, it can be quantized on ${\cal H}_{{\rm kin}}$. Given a~spin network ${\bf s} = (\gamma, \vec{j}, \vec{c})$ let $T(\epsilon,\gamma)$ be a one parameter family of triangulations adapted to the graph $\gamma$. By adapted all we mean is that each vertex $v'$ in $V(\gamma)$ should be either in the interior of precisely one cell $\triangle \in T(\epsilon, \gamma)$ $\forall\, \epsilon$, or $v'$ be a vertex of $T(\epsilon, \gamma)$ such  that for all $\epsilon > 0$,  the number of 3-cells saturating $v'$ remain constant.  One can now prescribe the action of $\hat{H}_{T(\epsilon, \gamma)}[N]$ on $|{\bf s}\rangle$ as,
\begin{gather}\label{eq:hamaction}
\hat{H}_{T(\epsilon,\gamma)}[N]|{\bf s}\rangle = \sum_{\triangle\in T(\epsilon,\gamma)}\sum_{i=1}^{M_{s}}\sum_{J=1}^{M_{L}}{\cal C}^{iJ}\,\Tr\big( \hat{h}_{\alpha_{J}(\triangle)}\hat{h}_{s_{i}(\triangle)}\big[ \hat{h}_{s_{i}(\triangle)}^{-1}, \hat{V}(R_{v(\triangle)}) \big] \big)|{\bf s}\rangle.
\end{gather}
Let ${\cal H}_{\gamma}$ be the subspace of ${\cal H}_{{\rm kin}}$ spanned by spin networks based on graph $\gamma$. There are two def\/initions of volume operators available in the literature. One is known as AL (Ashtekar--Lewandowski) volume operator and the other is known as RS (Rovelli--Smolin) volume operator. We will restrict ourselves to the AL volume operator in this review\footnote{If one insists on def\/ining continuum limit of $\hat{H}_{T(\delta)}[N]$ on ${\cal H}_{{\rm kin}}$ then it is essential to use AL volume operator. However if one instead def\/ines continuum limit on some subspace of $\cyl^{*}$ then as shown in \cite{lm1}, RS volume operator is a perfectly viable choice.}.
The equation \eqref{eq:hamaction} describes the action of $\hat{H}_{T(\epsilon,\gamma)}[N]$ on ${\cal H}_{\gamma}$ for any $\gamma$. Whence we have a densely def\/ined operator $\hat{H}_{T(\epsilon,\gamma)}[N]$ on the (dense domain of) entire Hilbert space ${\cal H}_{{\rm kin}}$.

So far the parameter space regularizing the Hamiltonian constraint is uncountably inf\/inite dimensional. This is by any account quite huge. This itself is not an issue because it might happen that dif\/ferent regularization schemes lead to same continuum operator (or at least the the same physical spectrum of the theory).

One would also expect that putting physical requirements like dif\/feomorphism covariance will restrict the choice of triangulation. It turns out that the requirement of dif\/feomorphism covariance places the following restrictions on the allowed choices of triangulations. Given a pair of quintuples $(\gamma, v, \triangle\in T_{\epsilon},  {\cal S}_{e}, {\cal S}_{L})$ and $(\gamma', v', \triangle'\in T(\epsilon'), {\cal S}_{e}', {\cal S}_{L}')$ such that $(\gamma, v)$ is dif\/feomorphic to $(\gamma', v')$, we require the entire quintuples to be related by some dif\/feomorphism for any choice of $\epsilon$, $\epsilon'$.

Let us introduce $\hat{U}(\phi)$ as the unitary operator corresponding to a dif\/feomorphism $\phi$ which acts on spin network states like $\hat{U}(\phi)T_{{\bf s}} = T_{\phi^{-1}{\bf s}}$. It can be easily shown that the above requirements imply that $\forall\, \phi  \in {\rm Dif\/f}(\Sigma)$
\begin{gather}\label{eq:sep25-1}
\hat{U}(\phi)\hat{H}_{T(\gamma, \epsilon)}[N]\hat{U}(\phi^{-1})  =  \hat{U}(\phi')\hat{H}_{T(\phi^{-1}(\gamma), \epsilon)}[\phi^{*}N],
\end{gather}
where $\phi'$ is a dif\/feomorphism not necessarily equal to $\phi$. As we will see in the next section, this would imply that the continuum Hamiltonian constraint in dif\/feomorphism covariant.

However the covariance requirement, though non-trivial, does not reduce the parameter space of allowed regulators signif\/icantly\footnote{Although it should be emphasized that to the best of our knowledge the only explicit construction of a~covariant family of regulators available in the literature is given in~\cite{qsd1}, but that of course does not mean that it is the only choice, in fact as explained in~\cite{alreview}, there are inf\/initely many.}.

\emph{We emphasize that this is the only physical requirement one places on regularization.}

\subsection{Continuum limit}

There are various  operator topologies in which one can examine the convergence (or lack thereof) of $\hat{H}_{T(\epsilon,\gamma)}[N]$.  The f\/inest topology used in the literature so far is the so called URST topology\footnote{URST is an abbreviation for Uniform by Rovelli, Smolin and Thiemann.}, in which one def\/ines the continuum limit on ${\cal H}_{{\rm kin}}$ itself.  This is quite an important construction so let us review it a bit more.

One corollary of requiring the one parameter family of regulators ${\bf R}_{\epsilon, \gamma}$ to be covariant is that for any two $\epsilon, \epsilon' > 0$ and for any vertex $v \in V(\gamma)$ the corresponding family of loops ${\cal S}_{L, v}(\epsilon)$, ${\cal S}_{L, v}(\epsilon')$ are dif\/feomorphism covariant. This implies that
\begin{gather*}%\label{eq:urst}
{\bf (}\Psi|\hat{H}_{T(\epsilon)}[N]|{\bf s}\rangle = {\bf (}\Psi|\hat{H}_{T(\epsilon')}[N]|{\bf s}\rangle
\end{gather*}
$\forall \, \Psi \in {\cal H}_{\rm dif\/f}$. One can now def\/ine the continuum limit of the Hamiltonian constraint (on ${\cal H}_{{\rm kin}}$) the following way. $\hat{H}[N]$ is a continuum limit of $\hat{H}_{T(\epsilon)}[N]$ if, given any $\delta > 0$, there exists $\epsilon_{0}(\delta) > 0$ such that forall $\epsilon < \epsilon_{0}$
\begin{gather}\label{eq:ursttopology}
\vert{\bf (}\Psi| \hat{H}[N] |{\bf s}\rangle  -  {\bf (}\Psi |\hat{H}_{T(\epsilon, \gamma({\bf s}))} [N] |{\bf s}\rangle \vert  <  \delta.
\end{gather}
This is trivially true because any $\hat{H}_{(\epsilon_{1})}[N]\vert{\bf s}\rangle$ and $\hat{H}_{(\epsilon_{2})}[N]\vert{\bf s}\rangle$ are related by a dif\/feomorphism, so that ${\bf (}\Psi|\hat{H}_{T(\epsilon, \gamma({\bf s}))}[N]|{\bf s}\rangle$ is actually independent of $\epsilon$.

Whence in the above mentioned topology, $\hat{H}_{T(\epsilon_{0})}[N]$ is already a continuum operator for any~$\epsilon_{0}$. Two properties of this construction are worth noting.
\begin{enumerate}\itemsep=0pt
\item Since $\epsilon_{0}$ is independent of ($|{\bf s}\rangle$, $\Psi$ and $N$), the convergence is uniform.
\item If two operators when acting on any state in ${\cal H}_{{\rm kin}}$  give dif\/feomorphically related states, they are \emph{identical}.  This in conjunction with the equation~(\ref{eq:sep25-1}) implies that the continuum operator is dif\/feomorphism covariant!
\end{enumerate}
If we do not care about demanding uniform convergence (and it is not clear to us if there is any physical reason to care), then we can def\/ine the continuum limit on inf\/initely many subspaces of~$\cyl^{*}$. That is, if we def\/ine continuum limit via,
\begin{gather*}
[\hat{H}[N]'\Psi]|{\bf s}\rangle  =  \lim_{\epsilon\rightarrow 0}\Psi\big[\hat{H}_{T(\epsilon)}[N]|{\bf s}\rangle\big],
\end{gather*}
where the convergence is only point wise (the notion of continuity used is same as the one in~(\ref{eq:ursttopology}) except that $\epsilon_{0} = \epsilon_{0}(\delta, \Psi, {\bf s})$) then on a large class of states in $\cyl^{*}$ $\hat{H}[N]'$ could be well def\/ined. Let us illustrate this point using a following schematic example.
Let
\begin{gather*}
\Psi  =  \sum_{s'\in {\cal A}} f(s')\langle s'|,
\end{gather*}
where ${\cal A}$ is some set of spin networks (${\cal A}$ can be uncountable) and $f$ are functions of the spin networks def\/ined on a suitable domain. (e.g.~$f$ could be functions of vertices of~${\bf s}'$, or they could be functions of vertices and germs of edges incident on those vertices, or they could be functions of contractible loops contained in ${\bf s'}$ etc.). We refer to $f$ as vertex functions\footnote{See \cite{lm2,lm1} for more details.}.
Schematically
\begin{gather*}
\hat{H}_{T(\epsilon)}[N]|{\bf s}\rangle = \sum_{v\in V(\gamma({\bf s}))}\sum_{I=1}^{m}N(v)a_{v,I}|{\bf s}_{v I}^{\epsilon}\rangle,
\end{gather*}
where due to dif\/feomorphism covariance of regulators, the coef\/f\/icients $a_{v,I}$ are independent of $\epsilon$ and $|{\bf s}_{v I}^{\epsilon}\rangle = \hat{U}(\phi_{\epsilon}) |{\bf s}^{0}_{v I}\rangle$ for some reference spin network $\vert {\bf s}^{0}_{v I}\rangle$.

Now let the set {\cal A} be such that if it contains some state then it necessarily contains all its dif\/feomorphic images. Furthermore let us for the sake of illustration assume that there exists precisely one vertex $v_{0}$  for which $|{\bf s}_{v_{0} I}^{\epsilon}\rangle \in {\cal A}$. Whence
\begin{gather*}
{\bf (}\Psi|\hat{H}_{T(\epsilon)}[N]|{\bf s}\rangle =  {\bf (}\Psi|\sum_{v\in V(\gamma({\bf s}))}\sum_{I=1}^{m}N(v)a_{v,I}|{\bf s}_{v I}^{\epsilon}\rangle\\
\hphantom{{\bf (}\Psi|\hat{H}_{T(\epsilon)}[N]|{\bf s}\rangle}{}
= \sum_{v\in V(\gamma({\bf s}))}\sum_{I=1}^{m}N(v)a_{v,I}\sum_{{\bf s}'\in {\cal A}}f({\bf s}')\langle{\bf s}'|{\bf s}_{v I}^{\epsilon}\rangle
= N(v_{0})\sum_{I=1}^{m} a_{v_{0},I} f({\bf s}_{v_{0}, I}^{\epsilon}).
\end{gather*}
Thus
\begin{gather*}
\hat{H}[N]'\Psi|{\bf s}\rangle = N(v_{0})\sum_{I=1}^{m} a_{v_{0}, I}\lim_{\epsilon\rightarrow 0} f({\bf s}_{v_{0}, I}^{\epsilon}).
\end{gather*}
Such subspaces of $\cyl^{*}$ on which continuum limit of the Hamiltonian constraint is well def\/ined are known as the habitats. Our schematic computation illustrates that potentially there are inf\/initely many such habitats. One explicit construction of a habitat is given in the seminal paper~\cite{lm1}. Essentially it is characterized by the property that the function $f$ above is a (smooth) functions of vertices of the~${\bf s}$.  This habitat turns out to be a suf\/f\/iciently small extension of~${\cal H}_{\rm dif\/f}$. We will have many occasions to use this habitat in this review and henceforth refer to it as Lewandowski--Marolf (LM) habitat.

This f\/inishes the summary of the construction of candidate Hamiltonian constraints in LQG. Let us, for the benef\/it of the reader summarize some of the worrying aspects of the construction.

1.~The parameter space underlying allowed choices of regulators is inf\/inite dimensional. This would not be a problem if the continuum limit of the Hamiltonian constraint is independent of such choices, however this is far from being the case, e.g.\ in the URST topology, each pair of choices which is not dif\/feomorphically related to each other will give rise to a~dif\/ferent operator.

2.~It is a priori not clear what the domain of the Hamiltonian should be. It could either be a~dense subspace of kinematical Hilbert space, or a~habitat.

 3.~There are further ambiguities. In all the constructions available in the literature so far, Lapse is a multiplicative factor. However one could conceive type of regularizations where the lapse function is included in the def\/inition of the plaquette one uses to def\/ine curvature~\cite{amdiffconstraint}.

4.~There is a huge amount of choice available in def\/ining curvature operators. One such choice in particular has received some attention in the literature. This has to do with the representation underlying the curvature approximant in the Hamiltonian constraint~\mbox{\cite{alejandroreview,perezhamconstraint}}. The simplest choice for such a curvature approximant is
\begin{gather*}
F_{ab}^{i}(v) \approx \frac{1}{\textrm{Ar}(\alpha_{ab})}\big( \Tr(h_{\alpha_{ab}}\tau^{i}) \big),
\end{gather*}
where the trace is in the fundamental $(j=\frac{1}{2})$ representation. One could instead choose approximant to be
\begin{gather*}
F_{ab}^{i}(v) \approx \frac{1}{\textrm{Ar}(\alpha_{ab}) d_{j}}\big( \Tr_{j}(h_{\alpha_{ab}}\tau^{i}) \big),
\end{gather*}
where $\Tr_{j}$ is the trace in some representation of spin $j$, and $d_{j}$ is the appropriate normalization factor. In fact as in LQG one often employs ``state dependent'' regularization schemes, this ambiguity could be generalized further where e.g.
\begin{gather*}
\hat{F}_{ab}^{i}(v)\ \pi_{j_{e}}[h_{e}](A) = \frac{1}{\textrm{Ar}(\alpha_{ab}) d_{j_{e}}}\big( \Tr_{j_{e}}(h_{\alpha_{ab}}\tau^{i}) \big) \pi_{j_{e}}[h_{e}](A).
\end{gather*}
Here $ \pi_{j_{e}}[h_{e}](A)$ is the gauge-variant spin network state based on a single edge~$e$, and $v\in e$. whence the choice of representations used to def\/ine curvature approximants could themselves be state dependent (and in this sense dynamical).

 5.~Yet another ambiguity which seemed to have gone un-noticed in the literature until the arrival of improved dynamics in LQC is the following. Notice that the f\/lux operator when acting on a spin network only depends on the intersections of the edges of the spin networks with the underlying surface. That is, even if we scale the coordinate area of the surface arbitrarily, the action of f\/lux operator on a spin network does not change as long as the intersection numbers remains the same. This means the following.
Consider latticized classical curvature is given by
\begin{gather*}
F_{ab}^{i}(v) \approx \frac{1}{{\rm Ar}\triangle} \Tr\big(h_{\triangle_{ab}}\tau^{i}\big).
\end{gather*}
However one could equally well consider a dif\/ferent approximant of the type
\begin{gather*}
F_{ab}^{i}(v) \approx \frac{1}{{\rm Ar}\triangle} \big[ \Tr\big(h_{\triangle_{ab}}\tau^{i}\big) + \Tr(h_{\triangle_{ab}})E^{i}(S_{\epsilon}) \big],
\end{gather*}
where $\triangle_{ab}$ is a plaquette based at $v$ of co-ordinate area $\epsilon^{2}$, and $S_{\epsilon}$ is a surface also of co-ordinate area $\epsilon^{2}$. Classically the second term is higher order in $\epsilon$ than the leading order term, however quantum mechanically the dif\/ference between states obtained by action of two approximants on a given spin network state will be a f\/inite norm state. Whence this is a genuine ambiguity in the choice of curvature which can drastically af\/fect the def\/inition of the Hamiltonian constraint. In essence, it is this choice which was exploited in~\cite{aplimproved} to def\/ine a hamiltonian constraint in LQC which gave rise to better semi-classical properties of LQC.

Thus the situation may look rather worrying. We have inf\/initely many possible def\/initions of continuum Hamiltonian constraint and it is a priori not clear which if any of these choices are physically preferred (better semi-classical properties).

It should be mentioned that as explained in~\cite{ttbook}, it is conceivable that as the f\/inal spectrum of the theory consists of dif\/feomorphism invariant states, at least some of the choices which give rise to the kernel of the Hamiltonian constraint that are dif\/feomorphically related may not matter. Although this argument is intuitive and to the best of our knowledge not analyzed in detail, it is an important counter-point to the  grim view that we have presented above.

Finally a worrisome feature of $\hat{H}[N]$ not necessarily related to the inf\/initude of ambiguities is its \emph{ultra-local character} (see \cite{alejandroreview,smolin-ultralocality}  for details). Consider a spin network state $\vert {\bf s}\rangle$ and let $N$ have support in the neighborhood of a vertex $v_{0} \in V(\gamma({\bf s}))$.  Then $\hat{H}[N]$ (assuming it is def\/ined on~${\cal H}_{{\rm kin}}$) will give rise to a linear combination of states each of which is ``dif\/ferent'' from ${\bf s}$ only in the neighborhood of $v_{0}$ (which means only the edges, which are incident on $v_{0}$ will undergo any change under $\hat{H}[N]$ action.) This statement would remain true even if we worked with arbitrarily f\/ine graphs and hence even when we consider say (kinematical) semi-classical states based on a~single graph. Naively  we expect the Quantum constraint equation in this case to ``approximate'' classical constraint equation which are elliptic and hence not ultra-local. Thus the semi-classical limit of such Hamiltonian constraint might turn out to be incorrect. One way around this issue suggested in~\cite{smolin-ultralocality} is to tailor the choice of loops underlying the curvature approximants in such a way that it intersects many vertices in a given graph. Such a construction has been done independently in the context of BF theory for 2+1 gravity, detailed in Section~\ref{sec:bf}. Finally the criticism regarding ultra-locality has itself been criticized in~\cite{ttbook} on the grounds that non-locality might emerge when one considers physical states which live in~$\cyl^{*}$.

\subsection[Action of $\hat{H}\lbrace N\rbrace$ on ${\cal H}_{{\rm kin}}$]{Action of $\boldsymbol{\hat{H}[N]}$ on $\boldsymbol{{\cal H}_{{\rm kin}}}$}

In the URST topology we have a continuum operator on ${\cal H}_{{\rm kin}}$,
\begin{gather*}
\hat{H}[N]\vert{\bf s}\rangle = \hat{H}_{T(\epsilon_{0})}[N]\vert {\bf s}\rangle = \sum_{I=1}^{N({\bf s})}a_{I}\vert{\bf s_{I}}\rangle,
\end{gather*}
where $\epsilon_{0}$ is f\/ixed once and for all\footnote{There is a minor subtlety here. As the $\hat{H}_{T(\epsilon_{0})}[N]$  when acting on $\vert{\bf s}\rangle$ must be f\/iner then
the graph $\gamma({\bf s})$. $\epsilon_{0}$~implicitly depends on $\gamma({\bf s})$.}.
Each ${\bf s}_{I}$ has two more vertices and one additional edge than ${\bf s}$. The two new vertices that are created are planar, and are in the kernel of AL Volume operator. This
feature plays a key role in the ``anomaly free'' nature of the Hamiltonian constraint, \emph{when it is defined on ${\cal H}_{{\rm kin}}$}.

\subsection{The commutator and its unexpected triviality}

One of the interesting properties of the Quantum Hamiltonian constraint def\/ined by Thiemann was its ``on-shell closure''.  That is the commutator of two continuum Hamiltonian constraints vanished.  Although consistent, a closer look at the computation shows that the underlying mechanism for this commutativity is not too restrictive (as far as the regularization choices are concerned) and does not rely on any deep or non-trivial structural aspects of the Hamiltonian constraint.

This is because the commutativity only relies on two features of the regularization.
\begin{enumerate}\itemsep=0pt
\item The density one nature of the constraints ensure that the corresponding operators are ``ultra-local'' (whether we def\/ine them on ${\cal H}_{{\rm kin}}$ or on $\cyl^{*}$) and can not give rise to one forms $N\nabla_{a} M - M\nabla_{a} N$ in the commutators.
\item The dif\/feomorphism covariance condition placed on the regularization scheme  (which is the only physical requirement fed into the quantization so far) guarantees that the commutators at f\/inite triangulation is anti-symmetric in the lapses. This can be seen as follows. As before, we will merely sketch the proof. Details can be found in~\cite{lm1, ttbook}.
\end{enumerate}
Let us schematically write the Hamiltonian constraint operator at f\/inite triangulation as
\begin{gather*}
\hat{H}_{T(\epsilon)}[N] \vert{\bf s}\rangle = \sum_{v\in V(\gamma)}N(v)\hat{O}_{\epsilon}(v) \vert{\bf s}\rangle.
\end{gather*}
Here $\hat{O}_{\epsilon}(v)$ is a non-local operator which depends on a set of loops (based at $v$), set of edges and the Volume operator.

The essence of dif\/feomorphism covariance of the regularization scheme is the statement that for any  graph $\gamma$ and $\epsilon$, $\epsilon'$,  $\exists$ a dif\/feomorphism $\phi_{\epsilon,\epsilon'}(\gamma)$ which preserves $\gamma$ and
\begin{gather}\label{eq:cov}
\hat{O}_{\epsilon}(v)\vert{\bf s}\rangle = \hat{U}(\phi_{\epsilon, \epsilon', \gamma})\hat{O}_{\epsilon'}(v)\vert{\bf s}\rangle
\end{gather}
$\forall\, v \in V(\gamma)$.
Consider two lapse functions $N$, $M$ and as the Hamiltonian constraint operator is linear in lapse we can (without loss of generality) assume that given a graph $\gamma$ $N$, $M$ have support only in the neighborhood of vertices $v_{N}$, $v_{M}$ respectively.  Clearly the commutator between $\hat{H}_{T(\epsilon)}[N]$, $\hat{H}_{T(\epsilon')}[M]$ is zero for all $\epsilon$, $\epsilon'$ if $v_{N} = v_{M}$,\footnote{This statement is true if we use AL volume operator. However as shown in~\cite{lm1}, on the LM habitat, even if we use RS volume operator, even though this statement will no longer hold, in the continuum limit $[H[N],H[M]]$ will vanish if $v_{N}=v_{M}$.} whence the continuum limit of such a commutator (in URST  topology on ${\cal H}_{{\rm kin}}$ or some weaker topology on a suitable habitat) will be zero.

If $v_{N} \neq v_{M}$ then it can be easily shown that (essentially using the covariance of regulator as in \eqref{eq:cov}),
\begin{gather*}
[ \hat{H}_{T(\epsilon)}[N]\hat{H}_{T(\epsilon')}[M] - N\leftrightarrow M ]\,\vert{\bf s}\rangle =
N(v_{N})M(v_{M}) \big(\hat{U}(\phi(\epsilon, \epsilon')) -1\big) \hat{O}_{\epsilon}(v_{N})\hat{O}_{\epsilon'}(v_{M})\vert{\bf s}\rangle.
\end{gather*}
Clearly in the URST, this operator vanishes as
\begin{gather*}
( \Psi\vert (\hat{U}(\phi) - 1)\vert{\bf s}\rangle = 0
\end{gather*}
$\forall\, {\bf (}\Psi\vert\in\ {\cal H}_{\rm dif\/f}$, $\phi \in {\rm Dif\/f}(\Sigma)$, $\vert{\bf s}\rangle \in {\cal H}_{{\rm kin}}$.
The argument can be extended for the continuum Hamiltonian constraint def\/ined on Lewandowski--Marolf habitat (see~\cite{lm1} for details). Although plausible, we are not aware of any arguments showing that this result will continue to hold when the domain of the continuum Hamiltonian constraint is any suitable subspace of~$\cyl^{*}$.

If one works on the LM habitat, this commutativity is puzzling at f\/irst. This is because states in the LM habitat have the following properties.
\begin{enumerate}\itemsep=0pt
\item It carried a representation of the algebra of spatial dif\/feomorphisms.
\item By its very construction it contained the dif\/feomorphism invariant states.
\end{enumerate}
Whence not all the states in the LM habitat are dif\/feomorphism invariant, and we do not expect the commutator of two Hamiltonian constraints to vanish on the habitat. In~\cite{lm2} it was shown that this apparent inconsistency could be avoided by a specif\/ic quantization of r.h.s. (Essentially what was shown in~\cite{lm2} was that there exists a quantization of $\hat{q}^{ab}$ which gave an identically zero operator on the habitat.) This happens because of the following simple reasons. Consider the r.h.s.\ of the Dirac algebra (for density one Hamiltonian constraint)
\begin{gather*}
\textrm{r.h.s.} = \int_{\Sigma}d^{3}x\, q^{ab}\left(N\nabla_{a}M - M\nabla_{a} N\right)H_{(\textrm{dif\/f}) b}(x).
\end{gather*}
It turns out that one can quantize $q^{ab}H_{b}$ using identities introduced by Thiemann and the f\/inal operator at f\/inite triangulation has the following form,
\begin{gather*}
\widehat{\textrm{r.h.s.}_{T(\epsilon)}} = \sum_{\triangle\in T}\epsilon\left(N\nabla_{a} M - M\nabla_{a} N\right) , \hat{O}^{a}_{\triangle(\epsilon)}(v(\triangle)),
\end{gather*}
where $\hat{O}_{\triangle(\epsilon)}(v(\triangle))$ is a composite operator built out of holonomies and f\/luxes and hence it is well def\/ined on ${\cal H}_{{\rm kin}}$. Thus if the limit $\epsilon\rightarrow 0$ exists on some space (whether on ${\cal H}_{{\rm kin}}$ or on LM habitat or some other subspace of~$\cyl^{*}$) then it will trivially be zero.

Naively one would think that the chief culprit in the game is density one character of the constraint. That is, if we were to work with higher density constraints then at f\/inite triangulation the operators will be of the type
\begin{gather*}
\hat{\tilde{H}}_{T(\epsilon)}[N]|{\bf s}\rangle = \frac{1}{\epsilon^{a}}\sum_{v\in V(\gamma({\bf s}))}N(v)\hat{O}_{\epsilon}(v) |{\bf s}\rangle,
\end{gather*}
where we have placed tilde on $H$ to emphasize on higher density nature. $a$ is a positive integer which depends on the density of $H$.

Due to the presence of $\epsilon$ in the denominator, it is certainly conceivable that commutator of two such Hamiltonians could give rise to factors of the type $\frac{N(v)M(v + \epsilon) - M(v)N(v + \epsilon)}{\epsilon}$.  If the limit~$\epsilon\rightarrow 0$ exists on some habitat, then the continuum limit of the commutator would contain factors of the type $M\nabla_{a}N - N\nabla_{a}M$. Whence higher density constraints have a chance to yield a faithful closure. However as argued in~\cite{lm2}, this is not enough. Unless the new vertices created by the action of Hamiltonian are such that, they can themselves admit a non-trivial action of~$\widehat{\widetilde{H}}[N]$, the commutator would \emph{never} produce inf\/initesimal dif\/feomorphisms.

Thus the necessary conditions to obtain of\/f-shell closure are, higher density weight of the classical constraint, and a choice of regularization which produces ``non-trivial'' vertices in the sense explained above. So far, no quantum Hamiltonian constraint satisfying these conditions has been found.

One could question the importance of of\/f-shell closure in the formulation of LQG. After all, we have at our disposal the dif\/feomorphism invariant Hilbert space, and working within this arena, of\/f-shell closure condition is super ceded.  However the reasons for requiring of\/f-shell closure are two fold.
\begin{enumerate}\itemsep=0pt
\item As we argued above, the allowed choices of regulators which yield well def\/ined continuum Hamiltonian constraint on either ${\cal H}_{{\rm kin}}$ or some habitat are inf\/inite in number.
 It is important to reduce this ambiguity, not by appealing to subjective criteria, but to certain physical conditions. A key physical requirement would be local Lorentz invariance, which in canonical gravity is encoded in the Dirac algebra. From this point of view, demanding Dirac algebra to be faithfully represented in the quantum theory seems very appealing.
\item Moreover, the fact that commutator continues to vanish even on non dif\/feomorphism invariant states seems like an extremely unpleasant feature of the operator. The fact that one can choose to quantize r.h.s.\ in such a way that it also annihilates such states does not seem very comforting as it is tantamount to the inverse metric being zero operator.
\end{enumerate}
It is sometimes argued, \cite{ttbook}, that the LM-habitat is too small an extension of ${\cal D}^{*}_{\rm dif\/f}$ to see any non-triviality or anomaly (${\cal D}^{*}_{\rm dif\/f}$ is a dense subspace of $\cal H^*_{\rm dif\/f}$ which contains the dif\/feomorphism invariant distributions).  However this is not true, as has been explicitly demonstrated in simpler contexts like Parametrized Field Theories (PFT) \cite{amhamconstraint}.  We will summarize results obtained in this simple model below in Section~\ref{sec:pft}.

Whence as far as quantization of the Hamiltonian constraint in LQG is concerned, this is where things stand. One has a slue of regularization ambiguities each of which can potentially give rise to a dif\/ferent continuum operator. Non-trivial checks like the  of\/f-shell closure of the constraint algebra do not seem to go through and in simple models like Loop quantum cosmology and PFT, it has been shown that the regularization choices used in the full theory so far leads to either incorrect (in PFT case) or physically unsatisfactory (in LQC case) results.

One of the key lessons that LQC teaches us with regards to LQG is the need to write the curvature operator as a concomitant of holonomy and f\/lux operators. Even though the reason as to why this curvature operator is selected in LQC comes from detailed semi-classical analysis of physical observables, it is interesting to note that the curvature operator used in that case does not ``fall into'' class of regularizations considered for curvature in the full theory.

We can now pose a following question. Is there a subset of choice of regularization ambiguities (choice of triangulation, choice of curvature, choice of operator ordering etc.) and a habitat in~$\cyl^{*}$ such that quantization of (certain density weighted) Hamiltonian constraint \emph{and} certain quantization of the dif\/feomorphism constraint gives a faithful representation of the Dirac algebra on the habitat. If we managed to f\/ind such regularization scheme, we could further ask if the choice of curvature operator had any (at least conceptual) similarity with the curvature operator used in LQC.

\subsection{The master constraint program}

Before going into the toy models at the core of the review, let us say a few words on the master constraint program.

Criticisms on Thiemann's construction appeared quite quickly after publication of his work. But due to the dif\/f\/iculty of the challenge, few modif\/ications aiming at improving it have been proposed, and ef\/forts have instead concentrated on developing the path integral spin foam formalism. In \cite{thiemann-master-constraint}, Thiemann himself introduced the \emph{master constraint} program, to revive the interest into the canonical quantization. He traced back the dif\/f\/iculties in the well-known peculiarities of the Dirac constraint algebra. In the latter, the scalar constraint does not generate a~subalgebra, is not spatially dif\/feomorphism invariant, and the whole algebra closes with structure functions which are not polynomials in the metric components.

Hence, Thiemann introduced in \cite{thiemann-master-constraint} a re-writing of the scalar constraint as
\[
M = \int_\Sigma d^3x \, \frac{\vert H(x)\vert^2}{\sqrt{\det (q(x))}},
\]
The density-weight one Hamiltonian $H(x)$ is squared, rescaled to $\sqrt{\det (q(x))}$ to make it back density-wight one, and f\/inally integrated over the canonical surface $\Sigma$. That stands for the inf\/inite number of constraints $H(x)=0$. $M$ commutes with the generator of spatial dif\/feo\-mor\-phisms.

The price to pay for getting a single master equation is that a (Dirac) observable $O$ is now def\/ined by a non-linear equation,
\begin{gather} \label{observable-MC}
\{O,\{O,M\}\}_{\vert M=0} = 0.
\end{gather}

The program has been successfully tested in various situations, \cite{Dittrich:2004bn, Dittrich:2004bp, Dittrich:2004bq, Dittrich:2004br, Dittrich:2004bs} and an explicit comparison between the Dirac and path integral quantizations is presented in \cite{Han:2009aw}. The loop quantum gravity case is developed in \cite{Thiemann:2005zg}.

At the technical level, the master constraint program suf\/fers from ambiguities just like the previous attempt. At the conceptual level, its point is to avoid the requirement of reproducing the Dirac algebra. It is appealing since the dif\/feomorphism algebra indeed encodes gauge, non-physical degrees of freedom. However, the generators of the algebra also form the full Hamiltonian of general relativity, which means that solving the dynamics is equivalent to f\/inding the equivalence classes determined by dif\/ferent gauge orbits. It seems to us that this intuition becomes more complicated in the master constraint program, in particular since the equation for Dirac observables is not anymore invariance along the gauge orbits but~\eqref{observable-MC} instead. Obviously, because the master constraint program is a re-writing, the usual notions from gauge systems have some equivalent in this formulation, but it is not clear to us whether hiding the dif\/f\/iculty of working with the Dirac algebra is fully satisfactory.

\section {Parametrized f\/ield theory as an ideal toy model} \label{sec:pft}

The main question coming from our review of Thiemann's construction is whether there are some choices of regularization and a habitat such that the Dirac algebra can be faithfully represented. Needless to say, posing this question in LQG is going to be pretty hard, so as a warm-up we can pose it in some simplif\/ied model. The model has to be be a f\/ield theory so that it has spatial dif\/feomorphism covariance, and it has to be generally covariant, so that in the canonical theory one has Dirac algebra as it's constraint algebra.

Parametrized Field Theories (PFT) present us with just such a class of models. These are just free f\/ield theories on f\/lat spacetime  in $d\geq 2$ dimensions written in generally covariant guise by adding ``fake'' (pure gauge)  degrees of freedom.  For details we refer the reader to~\cite{hajicekpft, karelpft}.

Two dimensional parametrized massless f\/ield theories on a cylinder are the simplest toy models for canonical gravity as the constraint algebra just becomes the well known Witt algebra ($\textrm{Dif\/f}(S^{1})\oplus \textrm{Dif\/f}(S^{1})$), and hence is a true Lie algebra. The corresponding gauge group (which would not even be def\/ined as an ordinary Lie group \cite{weinstein}) in canonical gravity is ${\cal G}_{\rm PFT} = \textrm{Dif\/f}(S^{1})\times \textrm{Dif\/f}(S^{1})$. This is very nice as in LQG we know how to handle spatial dif\/feomorphisms very well.  Thus once we quantize the kinematical Phase space of the theory using background independent techniques that LQG provides for us, we can average over the entire gauge group of the theory ${\cal G}_{\rm PFT}$ using the Ref\/ined Algebraic Quantization method as outlined in~\cite{marolf}. So on one hand we have Hamiltonian constraint and dif\/feomorphism constraint just as in canonical gravity, and on the other hand we could complete the entire quantization program without worrying about the constraints and obtain physical Hilbert space of the theory. So we could pose the problem we posed above for LQG Hamiltonian constraint with a further consistency check. If we instead of averaging over ${\cal G}_{\rm PFT}$, quantized Hamiltonian and dif\/feomorphism constraints, is there a choice of their quantization such that the kernel of the constraints yield the physical Hilbert space obtained via group averaging techniques.  With this motivating monologue out of the way, we delve into more details.

\subsection{Brief review of polymer PFT}

Consider the following canonical f\/ield theory on a Lorentzian Cylinder $({\cal M} = S^{1}\times{\bf R}, \eta)$. Cauchy slices are oriented circles coordinatized by the angular coordinate $x\in[0,2\pi]$,
with the direction of angular increase agreeing with the orientation
of the circle.
The theory has the following sets of f\/ields. Two scalar f\/ields $X^{+}$, $X^{-}$ which are known as embedding f\/ields and their conjugate momenta, and a matter scalar f\/ield $f$ and it's conjugate momenta\footnote{We refrain from introducing the Lagrangian formulation of the theory from which the canonical structure is derived, and from where the f\/ields receive their titles (embedding f\/ields and matter f\/ields), as it is not relevant for our purpose. It suf\/f\/ices to note that we have a f\/ield theory with constraint algebra which is isomorphic to Dirac algebra in two dimensions and which has no true Hamiltonian.}.

The Structure of the model can be summarized as follows:

{\bf Canonically conjugate embedding variables}: $(X^{+}(x), \Pi_{+}(x))$, $(X^{-}(x), \Pi_{-}(x))$,
$X^{\pm}(2\pi)$ $= X^{\pm}(0) \pm 2\pi$.

{\bf Matter variables}: $Y^{\pm}(x):= \pi_f \pm f^{\prime}$, with the brackets $\{f(x), \pi_f (y)\}= \delta (x,y)$, $\{Y^+,Y^-\}=0$,
$\{Y^{\pm}(x), Y^{\pm}(y)\} = \pm (\partial_{x}\delta(x,y) - \partial_{y}\delta(y,x) )$.

{\bf Density weight 2 constraints}: ${{H_{\pm}}}(x)  =  [ \Pi_{\pm}(x)X^{\pm'}(x) \pm
\frac{1}{4}Y^{\pm}(x)^2 ]$.

The constraint algebra is isomorphic to the Lie Algebra of vector f\/ields on the
circle. This algebra is more commonly known as the Witt algebra.

Instead of working with the above constraints which form a true Lie algebra, we can also work with the ``traditional'' dif\/feomorphism and Hamiltonian constraints which form the Dirac algebra with structure functions.

{\bf Dif\/feomorphism constraint}: $C_{\rm dif\/f}$ generates spatial dif\/feomorphisms
\begin{gather*}
C_{\rm dif\/f}(x) = H_+ + H_- =\big[\Pi_{+}(x)X^{+}{}'(x) + \Pi_{-}(x)X^{-}{}'(x) + \pi_{f}(x)f'(x)\big].
%\label{diffconstraint}
\end{gather*}
{\bf Hamiltonian constraint}: $C_{{\rm ham}}$ generates evolution normal to the Cauchy slice,
\begin{gather}
C_{{\rm ham}}(x)  = \frac{1}{\sqrt{X^{+}{}'(x)X^{-}{}'(x)}}(H_+ -H_-) \nonumber\\
\phantom{C_{{\rm ham}}(x)}{} = \frac{1}{\sqrt{X^{+}{}'(x)X^{-}{}'(x)}}\big[\Pi_{+}(x)X^{+}{}'(x) - \Pi_{-}(x)X^{-}{}'(x) + \frac{1}{4}\big(\pi_{f}^{2}+f'^{2}\big)\big].
\label{hamconstraint}
\end{gather}
{\bf Constraint algebra}: The Poisson algebra generated by
 $C_{\rm dif\/f}$ and $C_{{\rm ham}}$ is the  Dirac algebra,
\begin{gather}
\{C_{\rm dif\/f}[\vec{N}], C_{\rm dif\/f}[\vec{M}]\} = C_{\rm dif\/f}[\vec{N},\vec{M}],\qquad
\{C_{\rm dif\/f}[\vec{N}], C_{{\rm ham}}[M]\} = C_{{\rm ham}}[{\it L}_{\vec{N}}M],\nonumber\\
\{C_{{\rm ham}}[N], C_{{\rm ham}}[M]\} = C_{\rm dif\/f}[\vec{\beta}(N,M)],\label{eq:structure}
\end{gather}
wherein $\vec{N}$, $\vec{M}$ are shift vectors, $N$, $M$ are lapse functions and
the structure function $\beta^{a}(N,M):= q^{ab}(N\nabla_{b}M - M\nabla_{b}N)$ in \eqref{eq:structure} is def\/ined by the induced spatial metric $q_{ab}$,
\begin{gather*}
q_{ab}dx^a dx^b = -X^{+\prime}X^{-\prime}(dx)^2.
%\label{qab}
\end{gather*}

Whence working with the density one constraints, we are led to a two dimensional avatar or constraint algebra of canonical gravity. On the other hand, working with  density tow constraints, one obtains an algebra without the structure functions. The Lie algebraic nature of algebra generated by density two constraints is special to two dimensions. This happens essentially because in two dimensions (with one dimensional Cauchy slices), the structure function changes from~$q^{ab}$  (for density one constraints) to~$q q^{ab}$ for density two constraints, but~$q q^{ab} = 1$. What we would like to check is if either of these two (or in principle any density constraints) admit an of\/f-shell closure in quantum theory.

\subsection{Quantum theory}

A {\em charge network} $s$ is a f\/inite collection, $\gamma (s)$, of colored, non-overlapping (except at vertices) edges,
$e$, which span the range of the angular coordinate $x$ (i.e.~$[0,2\pi]$), the colors being referred to as charges, and the collection of edges being referred to as a graph.
Charge network states are in correspondence with charge networks and
constitute an orthonormal basis similar to spin network states in LQG.

\subsubsection{Embedding sector}

\begin{itemize}\itemsep=0pt
\item
Charge network: $s^{\pm} = \{\gamma(s^{\pm}), (k_{e_{1}^\pm}^{\pm},\dots,k_{e_n^\pm}^{\pm})\}$
where $k_{e_{I}^{\pm}}^{\pm}$ are embedding charges whose range is specif\/ied by $k_{e_{I}^{\pm}}^{\pm}\in \frac{2\pi L}{\hbar A}{\bf Z}$, for all $I=1,\dots,n^\pm$. Here $A$ is a f\/ixed,  positive, integer-valued Barbero--Immirizi-like parameter.
It is useful to def\/ine the `minimum length increment', as $a:=\frac{2\pi L}{ A}$.

\item Elementary variables: $X^{\pm}(x)$,
$T_{s^{\pm}}[\Pi_{\pm}]:= \exp\big[{}-i\sum\limits_{e\in \gamma(s^{+})}k_{e^{\pm}}^{\pm}\int_{e^{\pm}}\Pi_{\pm}\big]$.

\item Representation: $T_{s_{1}^{\pm}}$ denotes an embedding charge network state. $\hat{X}^{\pm}(x),\hat{T}_{s^{\pm}}$
denote the operators corresponding to the classical quantities $X^{\pm}(x)$, $T_{s^{\pm}}[\Pi_{\pm}]$. Their action is given by
\begin{gather*}%\label{eq:33}
\hat{T}_{s^{\pm}}T_{s_{1}^{\pm}} = T_{s^{\pm}+s_{1}^{\pm}} \qquad \text{and}\qquad\hat{X}^{\pm}(x)T_{s^{\pm}} := \lambda_{x,s^{\pm}}T_{s^{\pm}},
\end{gather*}
where, for $\gamma (s^{\pm})$ with $n^{\pm}$ edges,
\begin{gather*} %\label{eq:33a}
\lambda_{x,s^{\pm}}:= \begin{cases}
                       \hbar\ k_{e_{I^\pm}^{\pm}}^{\pm} &\text{if $x\in$ Interior$(e_{I^{\pm}}^{\pm})$ for $1\leq I^{\pm}\leq n^{\pm}$,}\\
		      \frac{1}{2}\,\hbar\ \bigl(k_{e_{I^\pm}^{\pm}}^{\pm} + k_{e_{(I^\pm+1)}^\pm}^\pm\bigr) &\text{if $x\in e_{I^\pm}^{\pm}\cap e_{(I^{\pm}+1)}^{\pm}$ for $1\leq I^{\pm}\leq (n^{\pm}-1)$,}\\
		      \frac{1}{2} \hbar  \bigl(k_{e_{n^\pm}^\pm}^\pm \mp \frac{2\pi L}{\hbar} + k_{e_1^\pm}^\pm\bigr) &\text{if $x=0$,}\\
		      \frac{1}{2} \hbar  \bigl(k_{e_1^\pm}^\pm \pm \frac{2\pi L}{\hbar} + k_{e_{n^\pm}^\pm}^\pm\bigr) &\text{if $x=2\pi$.}
\end{cases}
\end{gather*}
\end{itemize}

\subsubsection{Matter sector}
\begin{itemize}\itemsep=0pt
\item Charge-network: %\begin{equation}
$s^{\pm} =  \{\gamma(s^{\pm}), (l_{e_{1}^{\pm}}^{\pm},\dots,l_{e_{n}^{\pm}}^{\pm})\}$, with the technical condition\footnote{The zero sum condition on the matter charges stems from technicalities related to the scalar f\/ield zero mode~\cite{alokme2}.} $\sum\limits_{I=1}^{n^{\pm}} l_{e_{I}^{\pm}}^{\pm}=0$ and $l_{e_I^\pm}^\pm\in \epsilon {\bf Z}$ for all $I=1,\dots,n^\pm$. Here $\epsilon$ is a f\/ixed real positive parameter with dimensions $(ML)^{-\frac{1}{2}}$. $\epsilon$ is also a Barbero--Immirizi-like parameter.

\item Elementary variables: $W_{s^{\pm}_{\lambda^{\pm}}}[Y^{\pm}] = \exp[i\sum_{e^{\pm}\in \gamma(s^{\pm})} l_{e^{\pm}}^{\pm}
\int_{e^{\pm}}Y^{\pm}]$.

\item Weyl algebra\footnote{The def\/inition of the Weyl algebra follows in the standard way from the Poisson brackets between $Y^{\pm}(x), Y^{\pm}(y)$ and an application of the Baker--Campbell--Hausdorf\/f lemma.} of operators, $\hat{W}(s^{\pm})\hat{W}(s'^{\pm}) =\exp[-i\frac{\hbar}{2}\alpha(s^{\pm}, s'^{\pm})] \hat{W}(s^{\pm} + s'^{\pm})$. Here the exponent in the phase-factor $\alpha(s^{\pm},s'^{\pm})$ is given by
\begin{gather*}
\alpha(s^{\pm},s'^{\pm}) := \sum_{e^{\pm}\in \gamma(s^{\pm})}
\sum_{e^{\prime \pm}\in \gamma(s^{\prime \pm})} l_{e^{\pm}}^{\pm} l_{e^{\prime \pm}}^{\pm}\ \alpha(e^{\pm},e^{\prime \pm}).
%\label{defalphass'}
\end{gather*}
Here $\alpha(e^{\pm}, e^{\prime \pm}) = (\kappa_{e'^{\pm}}(f(e^{\pm}))-\kappa_{e'^{\pm}}(b(e^{\pm})))-(\kappa_{e^{\pm}}(f(e^{\prime \pm}))-\kappa_{e^{\pm}}(b(e^{\prime \pm})))$. Here $f(e)$, $b(e)$ are the f\/inal and initial points of the edge $e$ respectively. $\kappa_{e}$ is def\/ined as
\begin{gather*}
\kappa_{e}(x) = \begin{cases} 1 &\text{if $x$ is in the interior of $e$,}\\
 \frac{1}{2} &\text{if $x$ is a boundary point of $e$.}
\end{cases}
\end{gather*}

\item Representation: $\hat{W}(s^{\pm})W(s'^{\pm}) = \exp(\frac{-i\hbar}{2}\alpha(s^{\pm}, s'^{\pm}))W(s^{\pm}+
s'^{\pm})$.
\end{itemize}

\subsubsection{Kinematic Hilbert space}
The kinematic Hilbert space ${\cal H}_{{\rm kin}}$ is the product of the plus and minus sectors, ${\cal H}^{\pm}_{{\rm kin}}$,
each of which is a product of the appropriate embedding and matter sectors.
${\cal H}^{\pm}_{{\rm kin}}$ is spanned by an orthornormal basis of   charge network states.

A charge network state in ${\cal H}^{\pm}_{{\rm kin}}$ is denoted by $|{\bf s}^{\pm}\rangle :=T_{s^{\pm}}\otimes W(s^{\prime\pm})$.

The label ${\bf s}^{\pm}$ is specif\/ied by ${\bf s}^{\pm}_{\lambda^{\pm}}:= \{\gamma ({\bf s}^\pm), (k^\pm_{e_1^\pm}, l^\pm_{e_1^\pm}),\dots,(k^\pm_{e_{n^\pm}^\pm},l^\pm_{e_{n^\pm}^\pm})\}$. Here we have used the equivalence of charge networks to set $\gamma {(\bf s}^{\pm}):=\gamma(s^{\pm})=\gamma(s^{\prime\pm})$ so that each edge of the charge network is labeled by an embedding charge and a matter charge.

We refer to the quantum theory as Polymer Parametrized f\/ield theory (PPFT).

\subsubsection{Unitary representation of gauge transformations}
Finite gauge transformations generated by the density 2 constraints act, essentially, as  two {\em independent} dif\/feomorphisms of the spatial manifold, one which acts only on the `$+$' f\/ields and one which acts only on the `$-$' f\/ields. Consequently, in analogy to spatial dif\/feomorphisms in LQG, their action on charge networks is to appropriately `drag' them around the circle\footnote{Due to the quasi periodic nature of $X^{\pm}$ it is more appropriate to think of these dif\/feomorphisms as being periodic dif\/feormorphisms of the real line. Consequently  the action of these gauge transformations in quantum theory also keeps track of `factors of $2\pi$' when embedding charge edges `go past $x=2\pi$'. We refrain from going into such technicalities in this review.}.

Then a unitary representation of the gauge group is given by
\begin{gather*}%\label{eq:56}
\hat{U}^{\pm}(\phi^{\pm})T_{s^{\pm}}  :=  T_{\phi^{\pm}(s^{\pm})}
\hat{U}^{\mp}(\phi^{\mp})T_{s^{\pm}}  :=  T_{s^{\pm}},\\
\hat{U}^{\pm}(\phi^{\pm})W(s^{\prime\pm})  :=  W((\phi^{\pm})(s^{\prime\pm}),\qquad
\hat{U}^{\mp}(\phi^{\mp})W(s^{\prime\pm})  :=  W(s^{\prime\pm}).
\end{gather*}
Denoting $T_{s^{\pm}}\otimes W(s^{\prime\pm})$ by $|{\bf s}^{\pm}\rangle$, the above equations can be written in a compact form as
\begin{gather*}
|{\bf s}^{\pm}_{\phi^{\pm}}\rangle := {\hat U}^{\pm}(\phi^{\pm})|{\bf s}^{\pm}\rangle .
%\label{bfsphi}
\end{gather*}

\subsubsection{Physical Hilbert space}
Physical states are obtained by group averaging the action of the f\/inite gauge transformations discussed in the
previous section. Henceforth we restrict attention to a physically relevant superselected sector of the physical
Hilbert space. This sector is obtained by group averaging a~superselected subspace, ${\cal D}_{ss}$ of
${\cal H}_{{\rm kin}}$, ${\cal D}_{ss}={\cal D}^+_{ss} \otimes {\cal D}^-_{ss}$.

${\cal D}^{\pm}_{ss}$ is def\/ined as follows. Fix a pair of graphs $\gamma^{\pm}$ with $A$ edges. Place the embedding charges $\vec{k}^{\pm}$ such that $k_{e_{I^{\pm}}^{\pm}}^{\pm}-k_{e_{I^\pm-1}^{\pm}}^{\pm}=\frac{2\pi}{A\hbar}$, for all $I^{\pm}$. Consider the set of all charge-network states $\{|{\bf s}^{\pm}\rangle =|\gamma^{\pm}, \vec{k}^{\pm}, (l_{e_{1}^{\pm}}^{\pm},\dots,l_{e_{A}^{\pm}}^{\pm})\rangle\}$, where
$l_{e_{I}^{\pm}}^{\pm}\in {\bf Z}\epsilon$ are allowed to take all possible values subject to the zero sum condition
$\sum_{I}l_{e_I^{\pm}}^{\pm}=0$. Let ${\cal D}_{ss}^{\pm}$ be f\/inite span of charge network states of the type
$\{|{\bf s}^{\pm}_{\lambda^{\pm} \phi^{\pm}}\rangle, \forall\, \phi^{\pm}\}$.

The action of the group averaging map $\eta^{\pm}$ on a charge network state in ${\cal D}^{\pm}_{ss}$ yields the
distribution
\begin{gather*}%\label{eq:120}
\eta^{\pm} (|{\bf s}^{\pm}\rangle) = \sum_{{\bf s}^{\pm}\in [{\bf s}^{\pm}]}
\langle  {\bf s}^{\prime\pm}| =  \sum_{\phi^{\pm}\in {\rm Dif\/f}_{[{\bf s}^{\pm}]}^{P}\mathbf{R}}\langle {\bf s}_{\phi^{\pm}}^{\pm}| .
\end{gather*}
Here $[{\bf s}^{\pm}]$ is the equivalence class def\/ined by $[{\bf s}^{\pm}] : =  \{ {\bf s}^{\prime\pm}\vert {\bf s}^{ \prime\pm}
=  {\bf s}_{\phi^{\pm}}^{\pm} \; {\rm for\ some\ }\phi^{\pm}\}$, and ${\rm Dif\/f}_{[\bf{s}^{\pm}]}^{P}\mathbf{R}$ is a set of gauge transformations such that for each  ${\bf s}^{\prime\pm}\in\ [{\bf s}^{\pm}]$ there is precisely one gauge transformation  in the set which maps ${\bf s}^{\pm}$ to ${\bf s}^{\prime\pm}$. The space of such  gauge invariant distributions comes equipped with
the inner product
\begin{gather*}
\langle \eta^{\pm}(|{\bf s}_{1}^{\pm}\rangle), \eta^{\pm}(|{\bf s}_{ 2}^{\pm}\rangle)\rangle _{\rm phys}=
\eta^{\pm}(|{\bf s}_{1}^{\pm}\rangle) [|{\bf s}_{2}^{\pm}\rangle ],
%\label{physip}
\end{gather*}
which can be used to complete $\eta^{\pm}({\cal D}_{ss}^{\pm})$ to the Hilbert space ${\cal H}_{\rm phys}^{ss\pm}$.
We shall restrict attention to ${\cal H}_{\rm phys}^{ss}:={\cal H}_{\rm phys}^{ss+}\otimes{\cal H}_{\rm phys}^{ss-}$.

\subsection[The construction of ${\cal H}_{\rm diff}$]{The construction of $\boldsymbol{{\cal H}_{\rm dif\/f}}$}

Spatial dif\/feormorphism invariant distributions are constructed by the action of the group averaging map,
$\eta_{\rm dif\/f}$ on the dense space of f\/inite linear combinations of charge network states as follows.
Let $[{\bf s}^+, {\bf s}^-]$ be the orbit of ${\bf s}^+$, ${\bf s}^-$ under all spatial dif\/feomorphisms
so that $[{\bf s}^+, {\bf s}^-]$ is the set of all distinct charge network labels obtained by the action of
spatial dif\/feomorphisms on ${\bf s}^+$, ${\bf s}^-$. Then
\begin{gather}
\eta_{\rm dif\/f} (\vert {\bf s}^+\rangle\otimes\vert {\bf s}^-\rangle )=
\eta_{[{\bf s}^+, {\bf s}^-]}\sum_{{\bf s}^{\prime +}, {\bf s}^{\prime -}\in [{\bf s}^+, {\bf s}^-]}
                              \langle {\bf s}^{\prime +}\vert\otimes\langle {\bf s}^{\prime -}\vert .
\label{diffavg}
\end{gather}
Here $\eta_{[{\bf s}^+, {\bf s}^-]}$ is a constant which depends only on the orbit of ${\bf s}^+$, ${\bf s}^-$.\footnote{The arbitrariness in the choice of this constant can be reduced by requiring that $\eta_{\rm dif\/f}$ commute with all dif\/feomorphism invariant observables.}

\subsection{The Hamiltonian constraint operator in PPFT}

This section is devoted to the construction of the Hamiltonian constraint as an operator on the space of
dif\/feomorphism invariant states. We follow the strategy used in LQG. Our aim is to f\/irst def\/ine a discrete approximant to
the Hamiltonian constraint on a triangulation of the spatial manifold, promote the expression to an operator on the
kinematic Hilbert space and then show that its dual action on dif\/feomorphism invariant distributions,
admits a well def\/ined continuum limit.

Quantization of Hamiltonian constraints suf\/fers from similar type of regularization dilemmas as the Hamiltonian constraint of LQG.  We would like to see if there exists any choice of regularization, such that
\begin{enumerate}\itemsep=0pt
 \item[(a)] The quantum operators carry a faithful representation of the Dirac algebra (or at least the Witt algebra).
\item[(b)] The kernel of the quantum constraints (or more precisely the kernel of the Hamiltonian constraint plus group averaging over spatial dif\/feomorphisms) contains (as a vector space)~${\cal H}_{\rm phys}$.
\end{enumerate}

From equation (\ref{hamconstraint}) the smeared Hamiltonian constraint with  lapse $N(x)$ is
\[
C_{{\rm ham}}[N]  =  \int N(x)\left[\Pi_{+}(x)X^{+}{}'(x)  -  \Pi_{-}(x)X^{-}{}'(x)  +  \frac{1}{4}\big(\pi_{f}^{2}+f'^{2}\big)\right]\frac{1}{\sqrt{X^{+}{}'(x)X^{-}{}'(x)}} .
\]
On a triangulation $T$, a discrete approximant to the above expression is given by
\begin{gather}
C_{{\rm ham}, T}[N]  =  \sum_{\triangle\in T}|\triangle|\ N(b(\triangle)) \frac{1}{\sqrt{X^{+}{}'X^{-}{}'}}(b(\triangle))\nonumber\\
\hphantom{C_{{\rm ham}, T}[N]  =}{}\times
\left[\Pi_{+}(b(\triangle))\left(\frac{X^{+}(m(\triangle)) - X^{+}(m(\triangle-1)
+ L\delta_{b(\triangle ),0})}{|\triangle|}\right) \right.\nonumber\\
\hphantom{C_{{\rm ham}, T}[N]  =}{}
- \Pi_{-}(b(\triangle))\left(\frac{X^{-}(m(\triangle))-X^{-}(m(\triangle-1)-L\delta_{b(\triangle ),0})}{|\triangle|}\right)  \nonumber\\
\left.
\hphantom{C_{{\rm ham}, T}[N]  =}{}
+\frac{1}{4}(Y^{+})^{2}(b(\triangle)) + \frac{1}{4}(Y^{-})^{2}(b(\triangle))\right],\label{eq:jun14-1}
\end{gather}
where  $b(\triangle)$ is the beginning vertex of simplex $\triangle$,
$|\triangle|$ is its length and $m (\triangle )$ its midpoint.
The symbol $\triangle-1$ denotes the simplex to the left of $\triangle$, and
it is understood that if $\triangle_{1}$ is the left-most simplex with $b(\triangle_{1}) = 0$ then
$m(\triangle_{1}-1)=m(\triangle_{N})$, with $\triangle_{N}$ being the right-most simplex such that $f(\triangle_{N})=2\pi$ where $f(\triangle )$ is the ending vertex of $\triangle $.

Since only the holonomies of $\Pi_{\pm}$, $Y^{\pm}$ are well def\/ined operators on ${\cal H}_{{\rm kin}}$, the local
f\/ields~$\Pi_{\pm}$, $(Y^{\pm})^2$ need to be approximated on $T$ by appropriate combinations of holonomies. The equation~(\ref{eq:jun14-1}) is a discrete approximant to the continuum expression and we may modify it by terms which
vanish in the continuum limit, $|\triangle |\rightarrow 0$. It is straightforward to see that the following expression
is one such modif\/ication,
\begin{gather}
C_{{\rm ham}, T}[N] = \sum_{\triangle\in T}\frac{-i\hbar N(b(\triangle))}{|\triangle |\sqrt{X^{+}{}'X^{-}{}'}} \left(
\left[1+\frac{|\triangle |^2}{-i\hbar}\Pi_{+}(b(\triangle))X^{+}{}'(b (\triangle )) \right] \right.\nonumber\\
%\hphantom{C_{{\rm ham}, T}[N] =}{}
{}\times \left[1+ \frac{|\triangle |^2}{i\hbar}\Pi_{-}(b(\triangle))(b (\triangle ))\right]
\left.\left[1 + \frac{|\triangle |^2}{-4i\hbar}(Y^{+})^{2}(b(\triangle))\right]
\left[1 + \frac{|\triangle |^2}{-4i\hbar}(Y^{-})^{2}(b(\triangle)) \right]\right).\label{prodstructure}
\end{gather}

In a nutshell our aim is to choose a quantization scheme for $C_{{\rm ham},T}[N]$ such that the quantum constraints are a faithful representation of the classical constraint algebra, and
the kernel of Hamiltonian constraint (after averaging over spatial dif\/feomorphisms) leads to~${\cal H}_{\rm phys}$.
One way to anticipate what kind of regularization choices one should make is to consider action of classical Hamiltonian vector f\/ield of the hamiltonian constraint on cylindrical functions. If we start with the density one constraint, this action is quite involved due to the presence of $\frac{1}{\sqrt{q}}$ factor. However as we will see below, in quantum theory the  spin-network states are eigenstates of inverse metric operator, and the non-triviality of quantum action basically comes from $\Pi_{+}X^{+} - \Pi_{-}X^{-} + (Y^{+})^{2} + (Y^{-})^{2}$. Whence let us focus on that part, which is really just the density two constraint.

It is a simple computation to show that the action of Hamiltonian vector f\/ield  of (density two) constraints on cylindrical functions can be written as
\begin{gather*}
{\cal L}_{\tilde{\tilde{H}}[N]}\ T_{s^{+}}[\Pi_{+}] = \lim_{\delta\rightarrow 0}\frac{1}{\delta}\big(T_{s^{+}}[(\phi_{N}^{\delta})^{*}  \Pi_{+}] - T_{s^{+}}[\Pi_{+}]\big),\\
{\cal L}_{\tilde{\tilde{H}}[N]} T_{s^{-}}[\Pi_{-}] = \lim_{\delta\rightarrow 0}\frac{1}{\delta}\big(T_{s^{-}}[(\phi_{N}^{-\delta})^{*}  \Pi_{-}] - T_{s^{+}}[\Pi_{-}]\big).
\end{gather*}

Whence we seek f\/inite triangulation holonomy approximants to the various local f\/ields of interest in such a way that
$C_{{\rm ham}, T}[N]$ is proportional to a combination of f\/inite gauge transformations minus the identity, with the f\/inite
gauge transformations being parametrized by $|\triangle|$ so that at $|\triangle|=0$ the gauge transformations are
just identity.

{\samepage We display exactly such  approximants to $\Pi_{\pm}$, $(Y^{\pm})^{2} $ in next section. The holonomies turn out to be state dependent\footnote{As we will see shortly, this dependence involves the eigenvalues of the
embedding operators ${\hat X^{\pm}}$. Since ${\hat X^{\pm}}$ are the analogs of the LQG densitized triad operators,
this feature is reminiscent of the $\bar{\mu}$ scheme employed in the improved dynamics in LQC \cite{aplimproved}.},
the approximants yield $\Pi_{\pm} (b( \triangle ))$ and $(Y^{\pm})^{2}(b(\triangle))$ to leading order in $|\triangle|$ and the resulting quantum constraint is a linear combination of  operators localised on vertices, each of which is proportional to a the dif\/ference of a f\/inite gauge transformation and identity.

}

\subsection{Suitable ``curvature approximants''}

We shall focus on the left moving $(+)$ variables. Given a state $T_{s^{+}}$, let $V_{E}(\gamma(s^{+}))$ be the set of all vertices $v$ such that $k^{+}_{e_{v}} - k^{+}_{e^{v}} \neq 0$ (where $e_{v}(e^{v})$ refer to edges which terminate (originate) at~$v$. Similarly let $V_{M}(\gamma(s^{+}))$ be the set of all vertices $v$ such that $l^{+}_{e_{v}} - l^{+}_{e^{v}} \neq 0$. Then as shown in~\cite{amhamconstraint}, the quantization of $\Pi_{+}(v)$ which def\/ines the Hamiltonian constraint whose kernel contains states in ${\cal H}_{\rm phys}$ is
\begin{gather}\label{newpi}
\hat{\Pi}_{+}^{\triangle}(v)  T_{s^{+}}  = \begin{cases} \dfrac{i}{\vert\triangle\vert(k_{e_{v}}^{+} -k_{e^{v}}^{+})} \big(\hat{h}^{(k_{e_{v}}^{+} - k_{e^{v}}^{+})} - 1\big)  T_{s^{+}} &\text{if $v \in V_{E}(\gamma(s^{+}))$,}\vspace{2mm}\\
\dfrac{i}{\vert\triangle\vert}(\hat{h}_{\triangle}^{+} - 1)  T_{s^{+}} &\text{if $v \notin  V_{E}(\gamma(s^{+}))$}.
\end{cases}
\end{gather}
Next, we show that the above choice (\ref{newpi}) directly leads to an operator action of the~`$+$' embedding part of the
$C_{{\rm ham}, T}$ which is a f\/inite dif\/feomorphism on the $+$ part of the embedding state. As we shall see,
this will f\/inally lead to a satisfactory def\/inition of the Hamiltonian constraint.

With the above def\/inition, the approximant to the $\Pi_+X^{+\prime}$ term is
\begin{gather*}
\hat{\Pi}_{+}\hat{X}^{+ '}(b (\triangle ))|_TT_{s^{+}}  :=  \frac{1}{|\triangle|}\hat{\Pi}_{+}^{\triangle}(b(\triangle))
\big(\hat{X}^{+}(m(\triangle))-\hat{X}^{+}(m(\triangle -1 ))\big)T_{s^{+}}\nonumber\\
\hphantom{\hat{\Pi}_{+}\hat{X}^{+ '}(b (\triangle ))|_TT_{s^{+}}}{}
 =  \begin{cases} \dfrac{-i\hbar}{|\triangle|^{2}}(\hat{h}_{\triangle}^{k^{+}_{e_{v}}-k^{+}_{e^{v}}}-1)T_{s^{+}} &\text{if $b (\triangle )\in V_E(\gamma^{+})$},\vspace{1mm}\\
0 &\text{if $b (\triangle )\notin\ V_E(\gamma^{+})$.}
\end{cases}
\end{gather*}

Whence for $x\in V_E(\gamma^{+})$,
\begin{gather*}
%\label{phideltaemb+}
\hat{\Pi}_{+}\hat{X}^{+ '}( b (\triangle ) )|_TT_{s^{+}} =\frac{-i\hbar}{|\triangle|^{2}}(\hat{h}_{\triangle}^{k^+_{e_{v}}-k^{+}_{e^{v}}}-1)T_{s^{+}} \\
\hphantom{\hat{\Pi}_{+}\hat{X}^{+ '}( b (\triangle ) )|_TT_{s^{+}}}{}
 = \frac{-i\hbar}{|\triangle|^{2}}(T_{s^{+}_{\phi_{\triangle}}} - T_{s^{+}})
= \frac{-i\hbar}{|\triangle|^{2}}({\hat U}^{+,E}( \phi_{\triangle}) -1) T_{s^{+}}.
\end{gather*}

Here, $\phi_{\triangle}$ is a dif\/feomorphism of the circle
(more precisely,  $\phi_{\triangle}$ is a periodic dif\/feomorphism of the real line) which is identity
in the neighborhood of all the vertices of~$T$ except~$b(\triangle )$,~$f(\triangle )$.\footnote{In addition, $\phi_{\triangle}$ also dif\/fers from identity at $x=2\pi$ (or $x=0$) if $b(\triangle )=0$ (or $f(\triangle )= 2\pi$); this is just a~consequence of the circular topology of space.} %\label{fnphidelta}
 Further, $\phi_{\triangle}$  maps $b (\triangle )$ to $f(\triangle )$  and its action on the charge network label $s^+$ is denoted by~$s^{+}_{\phi_{\triangle}}$. ${\hat U}^{+,E}(\phi_{\triangle})$ is the restriction of the unitary action of the f\/inite gauge transformation ${\hat U}^{+}(\phi^+=\phi_{\triangle} )$  to the left-moving embedding Hilbert space.

Similarly the quantization of $\widehat{(Y^{+})^{2})}$ that yields a satisfactory def\/inition of the Hamiltonian constraint is
\begin{gather*}
\widehat{(Y^{+})^{2}}(b (\triangle ))_TW_{s^{+}} =  \frac{-4i\hbar}{|\triangle |^{2}}\left[\exp\left(-i\frac{\hbar}{2}\big(l_{e_{I}^{+}}^{+}-l_{e_{I+1}^{+}}^{+}\big)^{2}\right)  \hat{h}_{\triangle}^{\bigl(l_{e_{I}^{+}}^{+}-l_{e_{I+1}^{+}}^{+}\bigr)}  -  1  \right]W_{s^+},
%\label{y2=h}
\end{gather*}
which can also be rewritten as
\begin{gather*}
\widehat{(Y^{+})^{2}}(b(\triangle ))_T =   \frac{-4i\hbar}{|\triangle |^{2}} \big({\hat U}^{+,M}({\phi_{\triangle}}) -1\big),
%\label{y2=u}
\end{gather*}
where $\phi_{\triangle}$ has been def\/ined above and ${\hat U}^{+,M}({\phi_{\triangle}})$ is the
restriction of the f\/inite gauge transformation operator  ${\hat U}^{+}(\phi^+={\phi_{\triangle}})$
to the matter Hilbert space. The analysis of the right-moving~$(-)$ mode proceeds in a similar way.

Substituting the above ``curvature approximants'' into \eqref{prodstructure} leads to the following def\/inition of $\hat{C}_{{\rm ham}, T}[N]$ on ${\cal H}_{{\rm kin}}$,
\begin{gather*}
{\hat C}_{{\rm ham}, T}[N]\vert {\bf s^+},{\bf s^-}\rangle =
\sum_{\substack{\triangle\in T\\ b(\triangle )\in V_E(s^+)\cup V_E(s^-)}}
 N(b(\triangle))\\
\times\Bigl[{\hat U}^{+,E}(\phi_{\triangle})\otimes{\hat U}^{-,E}\big(\phi^{-1}_{\triangle -1 }\big)\otimes{\hat U}^{+,M}(\phi_{\triangle})\otimes{\hat U}^{-,M}\big(\phi^{-1}_{\triangle -1 }\big) -1\Bigr]
\frac{-i\hbar}{|\triangle |\hat{\sqrt{X^{+}{}'X^{-}{}'}}} \vert {\bf s^+},{\bf s^-}\rangle.
\end{gather*}
Using the fact that the unitary operators in the above equation are just restricted actions of unitary operators associated with f\/inite gauge transformations, we obtain
\begin{gather*}
{\hat C}_{{\rm ham}, T}[N]\vert {\bf s^+},{\bf s^-}\rangle = \sum_{\substack{\triangle\in T\\ b(\triangle )\in V_E(s^+)\cup V_E(s^-)}}
N(b(\triangle)) \left(\frac{-i\hbar}{a^2}\right) \lambda (s^+, s^-, b(\triangle ))\\
\phantom{{\hat C}_{{\rm ham}, T}[N]\vert {\bf s^+},{\bf s^-}\rangle =}{}
\times \Bigl[{\hat U}^{+}(\phi_{\triangle})\otimes{\hat U}^{-}\big(\phi^{-1}_{\triangle -1 }\big)-1\Bigr]
\vert {\bf s^+},{\bf s^-}\rangle.
%\label{qhamt}
\end{gather*}
This action precisely mimics the action of the Hamiltonian vector f\/ield of $C_{{\rm ham}}[N]$. We could now consider the continuum limit of ${\hat C}_{{\rm ham}, T}[N]$ on ${\cal H}_{{\rm kin}}$ in the URST topology.  As shown in~\cite{amhamconstraint}, just like in LQG, this operator converges to a continuum operator ${\hat C}_{{\rm ham}}[N]$ on ${\cal H}_{{\rm kin}}$.

Specif\/ically, in the notation used above, we may choose the limit of the one-parameter family ${\hat C}_{{\rm ham}, T(\gamma, \delta)}[N]$ to be the operator ${\hat C}_{{\rm ham}}[N]$ where
\begin{gather*}%\label{urst}
{\hat C}_{{\rm ham}}[N]|{\bf s}^{+},{\bf s}^{-}\rangle  :=
{\hat C}_{{\rm ham}, T(\gamma, \delta_{0})}[N]|{\bf s}^{+},{\bf s}^{-}\rangle.
\end{gather*}
On one hand this is a very promising candidate for the Hamiltonian constraint of PFT. A~moment of thought reveals that given any ${\bf (}\Psi\vert$ in ${\cal H}_{\rm phys}$
\begin{gather*}
(\Psi\vert  {\hat C}_{{\rm ham}}[N])|{\bf s}^{+},{\bf s}^{-}\rangle  =  0,
\end{gather*}
for all $N$ and $\vert{\bf s}^{+},{\bf s}^{-}\rangle$. On the other hand, this Hamiltonian exhibits the same triviality properties that the LQG Hamiltonian does. Instead of going into technical details, we summarise the f\/indings of~\cite{amhamconstraint} in qualitative terms.

\subsection{Commutator of Hamiltonian constraints in PPFT}

(1) Precisely due to the reasons we elaborated upon in the earlier sections, the commutator of density one constraints ${\hat C}_{{\rm ham}}[N]$ vanishes on ${\cal H}_{{\rm kin}}$.

(2) One can def\/ine an analog of the LM habitat, on which both the r.h.s.\ and the l.h.s.\ of constraint algebra can be quantized. Due to the density one character of the constraints, the product of the two Hamiltonians is ultra-local in the lapses and hence vanishes under anti-symmetrization. It turns out that if we use the same curvature approximants that we used to def\/ine ${\hat C}_{{\rm ham}, T}[N]$ to def\/ine r.h.s., then two things happen. On one hand the operator at f\/inite triangulation $\widehat{C}_{{\rm dif\/f}, T}[\vec{q^{-1}[N,M]}]$ mimics the action of corresponding classical Hamiltonian vector f\/ield, and on the other hand $\widehat{C}_{{\rm dif\/f}, T}[\vec{q^{-1}[N,M]}]$
is proportional to $\vert\triangle\vert$, whence on LM habitat, its continuum limit is zero.

(3) As we had noted earlier, in their analysis of the LM habitat in LQG, the authors of~\cite{lm2} noted certain necessary conditions for the commutator to be non-trivial. One of them was the density ${\neq}$ 1 character of the constraint, and the other was, a def\/inition of curvature approximant such that the newly created vertices where themselves not in the ``kernel'' of the Hamiltonian. The choice of curvature approximants in PFT ensure the second condition is satisf\/ied. So one would hope that appropriately changing the density weight (which is dictated by how many powers of $\vert\triangle\vert$ we need in the denominator, so that the continuum limit contains correct number of derivatives)  would give us a non-trivial constraint algebra.

In two dimensions, in order to get the right powers of $\vert \triangle\vert$ in the denominator, one needs to work with density two constraints. But when we work with the density two constraints, several remarkable things happen.

$(3i)$ The continuum limit of $\hat{\tilde{C}}_{{\rm ham}, T}[N]$ (we use tilde to denote the density two character) is not well def\/ined on ${\cal H}_{{\rm kin}}$ or the LM habitat. However, the commutator of two such Hamiltonians {\emph is} well def\/ined on the LM habitat and has an anomaly!

$(3ii)$ There exists a dif\/ferent habitat, on which the density one constraints admit a continuum limit and the constraint algebra closes of\/f-shell without any anomaly! and all the states in ${\cal H}_{\rm phys}$ are contained in this new habitat.

Whence we see that at least in the case of our toy model, everything nicely f\/its into a coherent picture. The requirement that constraints close of\/f-shell is a very stringent requirement. It not only pins down the curvature approximants and the density weights, but also dictates that the right home for the constraints might be something quite dif\/ferent then ${\cal H}_{{\rm kin}}$ or even ${\cal H}_{\rm dif\/f}$. In the case of PPFT, it turns out that the kernel of such constraints rather remarkably contains the solution  obtained via completely dif\/ferent methods.

\section{Dif\/feomorphism constraint in LQG} \label{sec:diffeo}

If the lessons learnt above in the case of PPFT could be applied to LQG,  one could in principle obtain a quantization of the Dirac algebra. The key lessons one learnt in PPFT were
\begin{enumerate}\itemsep=0pt
\item Higher density objects could be well def\/ined in the theory if one worked on suitable habitats.
\item In the case that the constraints generate a true Lie algebra, the constraint operators at f\/inite triangulation (with triangulation parametrized by $\delta$) should be related to the action of f\/inite gauge transformation if the f\/inite gauge transformations were unitarily represented in the theory.
\end{enumerate}
It is unclear at this stage whether these lessons could be applied to the Hamiltonian constraint but as a warm-up one could apply these to the dif\/feomorphism constraint.

Recall that in the Dirac algebra, there is a Lie subalgebra generated by spatial dif\/feo\-mor\-phisms,
\begin{gather}\label{sep20-1}
\big\{H_{\rm dif\/f}[\vec{N}], H_{\rm dif\/f}[\vec{M}]\big\} = - H_{\rm dif\/f}[\vec{N}, \vec{M}].
\end{gather}
We can now ask if there exists a quantization of $H_{\rm dif\/f}[\vec{N}]$ on some subspace of $\cyl^{*}$ such that
\begin{enumerate}\itemsep=0pt
\item[$(a)$] The operator $\hat{H}_{\rm dif\/f}[\vec{N}]$ is a faithful representation of \eqref{sep20-1} on that subspace.
\item[$(b)$] The kernel of $\hat{H}_{\rm dif\/f}[\vec{N}]$ matches precisely (or at least contains) the vector space of dif\/feo\-mor\-phism invariant states obtained via Rigging map techniques.
\end{enumerate}

It is important to note that at the level of f\/inite triangulation, $\hat{H}_{T {\rm dif\/f}}[\vec{N}]$ is well def\/ined on~${\cal H}_{{\rm kin}}$ and also admits inf\/inite number of quantization choices (choice of triangulation, choice of loops to def\/ine operator for curvature, choice of surfaces to def\/ine operator for triads, choice of representation for the curvature operator, etc). Whence it is interesting to see if requiring the dif\/feomorphism constraint operator to satisfy~$(a)$ and~$(b)$  restricts these choices in anyway. Also as the r.h.s.\ of Poisson bracket of two Hamiltonian constraints has a dif\/feomorphism constraint, it might be essential to have an operator correspondent of $H_{\rm dif\/f}[\vec{N}]$.

As shown in \cite{amdiffconstraint},  there does exist a quantization of $H_{\rm dif\/f}[\vec{N}]$ which precisely generates inf\/initesimal dif\/feomorphisms on LM habitat, and hence satisf\/ies~\eqref{sep20-1}\footnote{Recall that on ${\cal H}_{{\rm kin}}$ only f\/inite dif\/feomorphisms are unitarily represented and the generator of such dif\/feomorphisms does not exist. However such generators could and indeed they do exist on certain distributional spaces.}.

The ideas underlying the quantization are sketched below. As always we set $G=\hbar =c=1$. We shall also set the Barbero--Immirizi parameter to unity. The dif\/feomorphism constraint $H_{\rm dif\/f}({\vec N})$ is
\begin{gather*}
H_{\rm dif\/f}({\vec N}) = \int_{\Sigma} {\cal L}_{\vec N}A_a^i {\tilde E}^a_i = V ({\vec N}) - {\cal G}(N^cA_c^i),
\end{gather*}
where $V({\vec N}) = \int_{\Sigma}N^aF_{ab}^i {\tilde E}^b_i$ and ${\cal G}(N^cA_c^i) = \int_{\Sigma}N^cA_c^i {\cal D}_a{\tilde E}^a_i$.

Let $T(\delta )$ be a one-parameter family of triangulations of $\Sigma$ with the continuum limit being \mbox{$\delta \rightarrow 0$} and let $H_{{\rm dif\/f}, T(\delta )}$, $V_{T(\delta )}$, ${\cal G}_{T(\delta )}$ be f\/inite triangulation approximants to the quantities  $H_{\rm dif\/f}$, $V$, ${\cal G} $ of the above equations. Thus $D_{T(\delta )}$, $V_{T(\delta )}$, ${\cal G}_{T(\delta )}$ are expressions which yield $H_{\rm dif\/f}$, $V$, ${\cal G}$ in the continuum limit.

For simplicity consider a (non- gauge invariant) spin network state consisting of a single edge~$e$ with spin label $j$ so that the state is just the $m$, $n$  component of an edge holonomy $h_{e\; m}^{(j)\;\;n}$ of the (generalized) connection along the edge~$e$ in the representation~$j$, the indices $m$, $n$ taking values in the set $1,\dots,2j+1$. In what follows we shall suppress some of these labels and denote the state simply by $h_e$.

\begin{figure}[t]\centering
\includegraphics[scale=0.7]{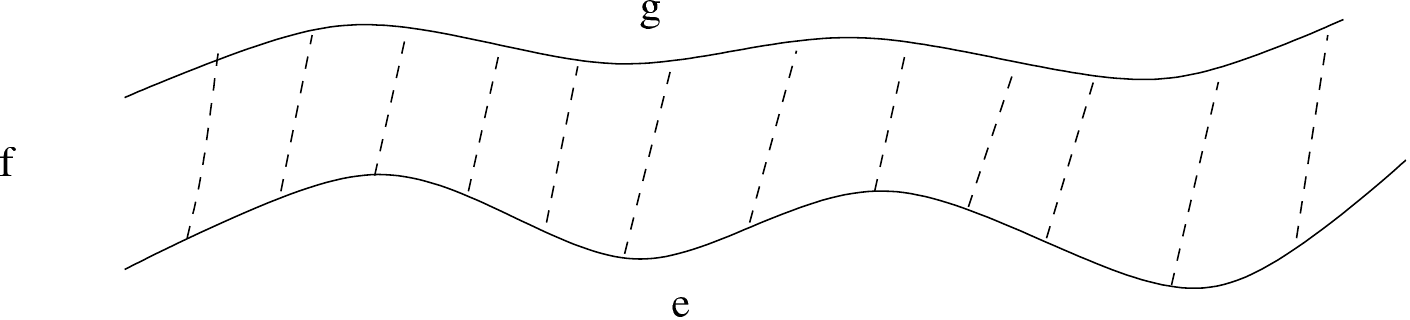}
\caption{\label{fig:1a}}
\end{figure}

As we want the operator at f\/inite triangulation to equal f\/inite dif\/feomorphism minus the identity, our f\/irst goal is
\begin{gather*}
\big(1+ i\delta {\hat H}_{{\rm dif\/f}, T(\delta )}\big) h_e = h_{\phi (\delta, {\vec N})\circ e},
%\label{desired}
\end{gather*}
where $\phi (\delta, {\vec N})\circ e$ is the image of $e$ by the dif\/feomorphism $\phi (\delta, {\vec N})$ which translates $e$ by an amount $\delta$ along the integral curves of the shift vector f\/ield $N^a$ (see Fig.~\ref{fig:1a}).

\begin{figure}[t]\centering
\includegraphics[scale=0.7]{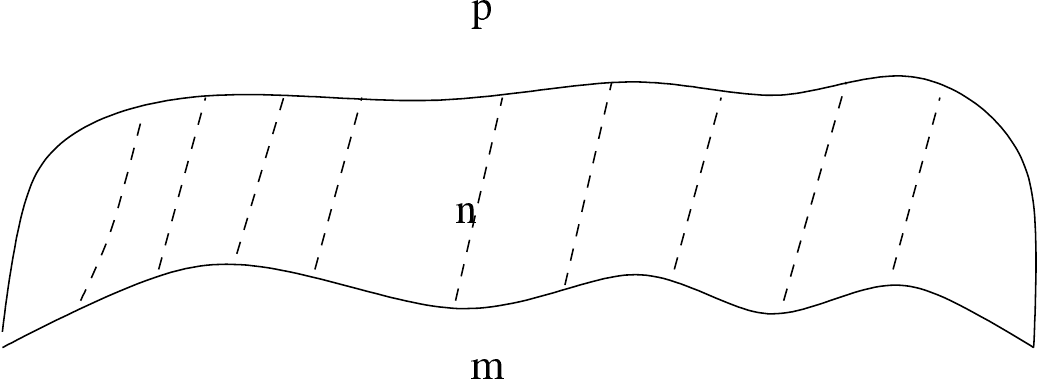}
\caption{\label{fig:1b}}
\end{figure}

\begin{figure}[t]\centering
\includegraphics[scale=0.7]{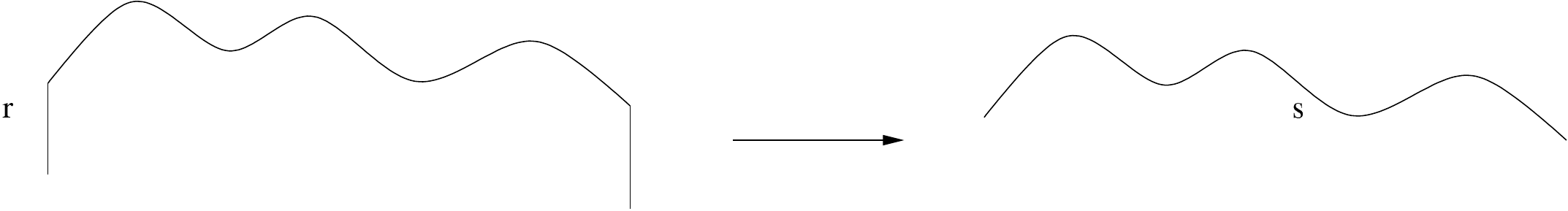}
\caption{\label{fig:1c}}
\end{figure}

As shown in \cite{amdiffconstraint} the desired result is obtained in the following steps.

\begin{enumerate}\itemsep=0pt
 \item[$(i)$] First we set
\begin{gather*}
(1+ i\delta \hat{H}_{{\rm dif\/f}, T(\delta )}):= (1- i\delta \hat{{\cal G}}_{T(\delta )})(1+ i\delta {\hat V}_{T(\delta )}).
\end{gather*}
\item[$(ii)$] Next, one shows that
\begin{gather}
(1+ i\delta {\hat V}_{T(\delta )})h_e = h_{{\bar e}({\vec N}, \delta )}.
\label{bare}
\end{gather}
Here ${\bar e}({\vec N}, \delta )$ has the same end points as $e$ (as it must by virtue of the gauge
invariance of $V$) and is obtained by joining the end points of
$\phi (\delta, {\vec N})\circ e$ to those of $e$ by a pair of segments which are aligned with integral
curves of $N^a$ as shown in Fig.~\ref{fig:1b}.
\item[$(iii)$] Finally,  one shows that the Gauss Law term, $(1- i\delta {\hat{\cal G}}_{T(\delta )})$ removes these two extra
segments (see Fig.~\ref{fig:1c}).
\end{enumerate}

The major part of the analysis concerns the derivation of the identity (\ref{bare}) in step (ii) above. ${ V}_{T(\delta )}$ is written as a sum over contributions $V_{\triangle}$ where $\triangle$ denotes a 3-cell of the triangulation {\em dual} to $T (\delta )$, and $V_{\triangle}$ is a f\/inite triangulation approximant to the integral
$\int_{\triangle}N^aF_{ab}^i {\tilde E}^b_i$.

{\bf Choice of operator ordering.} Order the triad operator to the right in ${\hat V}_{\triangle}$
so that only those 3-cells contribute which intersect $e$.

{\bf Choice of triangulation.} The triangulation $T(\delta )$ is adapted to the edge $e$ so that its restriction to $e$ def\/ines a triangulation of $e$. Thus, there is a triangulation of $e$ by 1-cells and vertices of $T(\delta )$ so that each of these vertices $v_I$, $I=1,\dots, N$  is located at the center of some 3-cell $\triangle = \triangle_I$.

\begin{figure}[t]\centering
\includegraphics[scale=0.6]{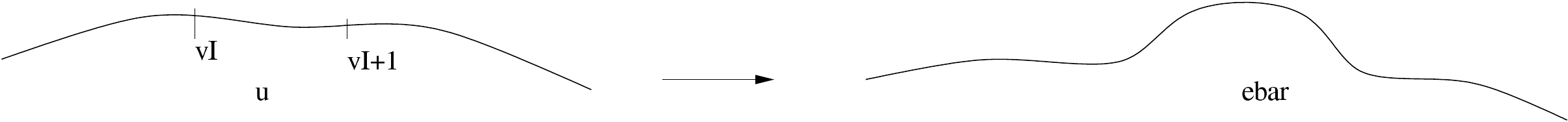}
\caption{\label{fig:1d}}
\end{figure}

{\bf Choice of curvature approximant.}  We def\/ine a f\/inite triangulation
approximant to $F_{ab}^i$ in a such a way that the following identity holds:
\begin{gather*}
\big(1+ i\delta {\hat V}_{\triangle_{I}}\big)h_e = h_{{\bar e}(\triangle_I)}
\end{gather*}
$\forall\, I \in \{1,\dots,N-1\}$. Here ${\bar e}(\triangle_I)$ is  obtained by moving the segment of $e$ between $v_I$ and $v_{I+1}$ along the integral curves of~$N^a$ by an amount~$\delta$ and joining this segment to the rest of~$e$ at the points $v_I$, $v_{I+1}$ by a pair of segments which run along the integral curves of~$N^a$  as shown in Fig.~\ref{fig:1d}.

\begin{figure}[t]\centering
\includegraphics[scale=0.6]{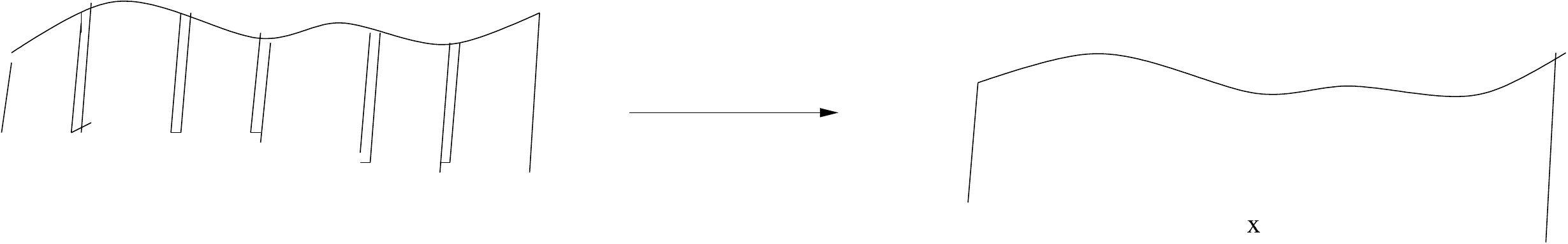}
\caption{The double lines indicate retraced paths.}\label{fig:1e}
\end{figure}

This can be accomplished by using the full quantization ambiguity that is available to us when def\/ining curvature operator in LQG. More precisely not only does one have to tune the spin (in which the trace of holonomy is evaluated) to the spin of the underlying state, but one also has to add the ``higher-order operators'' obtained by involving the f\/luxes. For more details we refer the reader to~\cite{amdiffconstraint}.

{\bf Sum to product reformulation}: Finally the  contributions from all the ${\hat V}_{\triangle_I}$ yield the edge
${\bar e}({\vec N}, \delta )$ of Fig.~\ref{fig:1b}. Recall that ${ V}_{T(\delta )}$ is obtained by summing over all the cell
contributions $V_{\triangle}$. However, summing over the action of all the ${\hat V}_{\triangle_I}$ on $h_e$
only yields a sum over states of the type~$h_{{\bar e}(\triangle_I)}$. In order to obtain the desired result, $h_{{\bar e}({\vec N}, \delta )}$, the sum over $\triangle$ is f\/irst converted to a~{\em product} over $\triangle$, i.e.\ to leading order in $\delta$, we show that
\begin{gather*}
1+ i\delta\sum_{\triangle} V_{\triangle} \sim \prod_{\triangle}(1+ i\delta V_{\triangle}).
\end{gather*}
Hence, replacing  the sum over the corresponding operators by the product provides an equally legitimate
def\/inition of ${\hat V}_{T(\delta )}$. The replacement then leads, to the following identity
\begin{gather*}
\prod_{\triangle}(1+ i\delta {\hat V}_{\triangle})h_e
= \prod_{I=1}^{N-1} (1+ i\delta V_{\triangle_I})h_e .
\end{gather*}
The def\/inition of  ${\hat V}_{\triangle_I}$ is such that each factor
in the product acts independently, the $I$th factor acting only on the part of $e$ between $v_I$ and $v_{I+1}$.
We are then able to show that the result~(\ref{bare}) follows essentially through the mechanism which is
illustrated schematically in Fig.~\ref{fig:1e}.

Thus we see that by making some very specif\/ic choices of triangulation, curvature  and ope\-ra\-tor ordering, one obtains
\begin{gather*}
\hat{H}_{{\rm dif\/f}, T(\delta)}[\vec{N}]  \vert  \gamma, \vec{j}, \vec{c}\rangle = \frac{1}{\delta}\big(\hat{U}(\phi^{\vec{N}}_{\delta}) - 1\big) \vert \gamma, \vec{j}, \vec{c}\rangle.
\end{gather*}
It can be easily shown that $\hat{H}_{{\rm dif\/f}, T(\delta)}[\vec{N}]$ admits a continuum limit on LM habitat where it precisely generates Lie action by~$\vec{N}$.

\section{New progress and insights from the BF topological model} \label{sec:bf}

We now present a dif\/ferent family of toy-models, namely the topological BF f\/ield theories in arbitrary dimensions. It is a f\/ield theory which has the same unreduced phase space as Yang--Mills or general relativity with Ashtekar--Barbero variables. But as a topological theory it is an exactly solvable theory and it has been considered as the main testing ground for the loop quantum gravity program since its early days, and even more after the birth of the spin foam program. Just like in the previous sections, the idea is to quantize {\it \`a la loop} the unreduced phase space, and then impose the topological symmetries at the quantum level. One expects to reproduce in this simplif\/ied situation a quantum algebra of constraints, and the resulting quantization should be equivalent to the one which has been performed using standard path integral methods.

This section f\/irst reviews the BF theory itself, in its standard quantization, then the most recent works in the loop framework, emphasizing the geometric aspects which are expected to show up similarly in four-dimensional quantum gravity. A nice review from the loop and spin foam perspective is~\cite{baez-sf-bf}. The present section should be considered as a new review on the state of the art for the loop approach to this topological model, ten years after.

We focus on the symmetry group $\SU(2)$, but arbitrary spacetime dimension $n$. The BF f\/ield theory was introduced by Blau and Thompson in \cite{blau-thompson-bf} and by Horowitz in~\cite{horowitz-bf}. It also appears in a number of interesting f\/ield theories. When $n=2$ \cite{witten-2dym} and $n=4$ \cite{bfym}, it appears as the zero coupling limit of Yang--Mills theory, an approach known as BF Yang--Mills (and as such it was observed early in lattice gauge theory). In three dimensions, it describes pure Riemannian gravity (with degenerate metrics), and remarkably can be recast in a Chern--Simons formulation~\cite{witten-3d-gravity}. In four dimensions, it is also possible to establish a link with gravity (which is crucial to spin foam models for quantum gravity) through the BF Plebanski formalism~\cite{bf-pleb}, or the BF MacDowell--Mansouri formulation~\cite{bf-mdm}.

It is also an interesting model exhibiting a relationship between f\/ield theory and topological invariants. The partition function produces the Ray--Singer torsion~\cite{blau-thompson-torsion}, and expectation values of observables are studied in~\cite{cattaneo-3d-4dBF}.

\subsection{Transition amplitudes}

\subsubsection{Field equations and f\/lat connections}

Let $P$ be a principal $\SU(2)$ bundle, over a $n$-dimensional manifold $M$. We consider a connection~$A$ on~$P$, locally seen as a $\su(2)$-valued 1-form over $M$ with specif\/ic transformation rules. We also introduce a $(n-2)$-form taking values in the adjoint bundle, i.e.~$B$ is $\su(2)$-valued and transform under the adjoint action of $\SU(2)$. When $d=3$, $B$ physically corresponds to the (Riemannian) triad, usually denoted~$e$, and the spacetime metric is obtained from $g_{\mu\nu}=\delta_{ij} e_\mu^i e_\nu^j$.

To write the BF action, we consider the non-degenerated bilinear invariant form `$\tr$' on the algebra, and
\[ %\label{BFaction}
S_{\mathrm{BF}}\bigl(B,A\bigr) = \int_M \tr\bigl(B\wedge F_A\bigr),
\]
with $F_A=dA+\f{1}{2}[A,A]$ being the curvature of $A$. The group $\calG$ of usual gauge transformations consists in $G$-valued f\/ields, which generate the following transformations,
\[ %\label{gauge transfo}
A \mapsto \Ad_g A + g\, dg\mone, \qquad B \mapsto \Ad_g B.
\]
In matrix notation the adjoint action reads $\Ad_g X = gX g\mone$. Since $S_{\rm BF}$ is the integral of a~dif\/ferential form, it is naturally invariant under dif\/feomorphisms of $M$.

The equations of motion are particularly simple,
\begin{gather*}
d_A\, B = 0, \qquad
F_A = 0, %\label{flatness}
\end{gather*}
where $d_A = d + [A,\cdot]$ is the covariant derivative. It turns out that all solutions are locally identical up to gauge transformations. Notice that the second equations comes from the fact that $B$ is a Lagrange multiplier for the curvature, so that the connection has to be f\/lat. We know then that all f\/lat connections are locally gauge equivalent. We also get that~$B$ is locally exact, $B=d_A\psi$. Now it is time to state that the action has an extra invariance under
\[ %\label{translation sym bf}
A \mapsto A, \qquad B \mapsto B + d_A\,\eta,
\]
for any $(n-3)$-form $\eta$ over $M$ taking values in $\su(2)$. That additional gauge symmetry is often called translation symmetry. Denoting $\calT$ the group of such translations, the full group of gauge symmetries is the semi-direct product $\calG\ltimes \calT$. It is then possible to transform any solution $B$ into the trivial solution. The physical consequence is that the theory is deprived of local degrees of freedom.

Those equations admit a useful cohomological interpretation. Squaring the covariant derivative gives $d_A^2\omega = [F_A,\omega]$ which vanishes on-shell. So $d_A$ def\/ines cohomology classes. The set of non-gauge-equivalent solutions to the classical equations of motion consists in non-gauge-equivalent f\/lat connections with elements of the $(n-2)$th cohomology group, the latter being isomorphic to the second cohomology group.

It is interesting to note that general relativity and the BF model share some essential properties which enable to make geometry dynamical. They are both formulated using dif\/ferential forms, without any spacetime metric, and are dif\/feomorphism invariant. Hence, it should be clearly understood that what makes the BF model special is the translation symmetry, so that dif\/feomorphisms are actually encoded into the full set $\calG\ltimes\calT$. Suppose~$\xi$ is a vector f\/ield on~$M$, then using the inner product $i_\xi$, we can form the function $\phi_\xi=i_\xi A$ and the $(n-3)$-form $\eta_\xi= i_\xi B$. The translation symmetry and dif\/feomorphisms are related in the following way
\begin{gather*}
  \delta_\xi A  = \mathcal{L}_\xi A = d_A\,\phi_\xi + i_\xi F,\qquad
\delta_\xi B  = \mathcal{L}_\xi B = d_A\,\eta_\xi + [\phi_\xi,B] + i_\xi\,d_A B,
\end{gather*}
where $\mathcal{L}_\xi$ is the Lie derivative along $\xi$. When written in this way, inf\/initesimal dif\/feomorphisms appear as gauge transformations up to terms proportional to the variations of the action \mbox{$\delta S/\delta B = F$}, so that the two symmetries are \emph{de facto} equivalent. Obviously in general relativity, it becomes impossible to rewrite dif\/feomorphisms as gauge symmetries plus terms which vanish on-shell.

In the context of the BRST (or Batalin--Vilkovisky) formalism, the symmetry of the BF model has been interpreted as a vector supersymmetry \cite{symmetries-bf-bv, maggiore-sorella-perturbative-4dbf}, which turns out to be crucial to solve the local cohomology of the BRST operator, and in the end to prove the perturbative renormalizability \cite{renormalisability-bf}! An important aspect in that algebraic approach is that the BRST algebra does not close~\emph{off-shell}. The BRST operator is nilpotent only on-shell when $n\geq4$, due to the fact that the translation symmetry is on-shell reducible. Indeed, the transformation of $B$ is the same for $\eta$ and $\eta+d_A\,\psi$. As a consequence, care is due when trying to quantize it in dimension four and higher.

\subsubsection{The Hamiltonian analysis and f\/lat connections again} \label{sec:bf-hamiltonian}

The Hamiltonian analysis is straightforward. One f\/inds that one canonical variable is the pull-back of the connection to the canonical surface $\Sigma$, denoted~$A_a^i$. The other canonical variable is the analog to the electric f\/ield, $E^a_i$, transforming in the adjoint representation. Those variables satisfy the canonical bracket
\[
\bigl\{ E^a_i(x), A_b^j(y)\bigr\} = \delta^a_b \delta^j_i \delta^{(3)}(x-y).
\]
The phase space is actually identif\/ied as the cotangent space $T^* \calA$ to the space of connections~$\calA$, exactly like in Yang--Mills theory.

However, the Hamiltonian is dif\/ferent since it is a combination of constraints imposed with Lagrange multipliers. The constraints are simply the restrictions of the equations of motion to~$\Sigma$, which do not involve any time derivatives. We integrate the local constraints as follows
\begin{gather*}
\calC_G(\Lambda) = \int dx\, \tr\bigl(\Lambda(x) D_a E^a(x)\bigr),\qquad \calC_F(N) = \int dx\, \eps^{abc}  \tr\bigl(N_c(x)\,F_{ab}(x)\bigr),
\end{gather*}
where $D$ is the covariant derivative with respect to $A_a^i$. Through the Poisson brackets, those constraints generate the whole set of gauge transformations on the canonical variables. Hence they satisfy the expected algebra,
\begin{gather*}
 \bigl\{ \calC_G(\Lambda), \calC_G(\Lambda')\bigr\} = \calC_G([\Lambda,\Lambda']),\qquad
 \bigl\{ \calC_G(\Lambda), \calC_F(N)\bigr\} = \calC_F([\Lambda,N]),\\
\bigl\{ \calC_F(N), \calC_F(N')\bigr\} = 0,
\end{gather*}
which is recognized as the Poincar\'e algebra $\ISU(2)$. The algebra is f\/irst class and the local number of degrees of freedom is easily evaluated to zero, recovering the result of the co\-va\-riant analysis. $\calC_G$ is the Gau\ss{} law which generates local $\SU(2)$ rotations, while $\calC_F$ generates translations as expected since curvature is zero.

To completely solve the theory classically, one can f\/irst reduce the phase space by the Gau\ss{} law $\calC_G$, just requiring invariance under the standard $\SU(2)$ gauge transformations. That leads to the reduced Yang--Mills phase space $T^*(\calA/\calG)$. Reducing with respect to the constraint $\calC_F$ implies that we consider only the set $\calA_0$ of f\/lat connections. It is also necessary to identify any two momenta which dif\/fer by a transformation generated by $\calC_F$, i.e.\ $E^a_i\sim E^a_i + \eps^{abc}D_b\eta_c$. The relevant set of momenta is thus the second cohomology group of~$D$ (which squares to zero on~$\calA_0$). In conclusion, the reduced phase space is the set of f\/lat connections modulo the $\calG$-action together with the second cohomology group over each f\/lat connection. This is the set $T^*(\calA_0/\calG)$, as described in \cite{horowitz-bf,witten-amplitude-3d}.

There exists a description of the set of f\/lat connections which shows that the reduced phase space is often f\/inite dimensional and which proves useful in practice. Holonomies of a f\/lat connection only depend on the homotopy types of its paths. Hence a f\/lat connection def\/ines a~morphism from the fundamental group $\pi_1(\Sigma)$ into the structure group~$\SU(2)$. Under a gauge transformation, such a morphism gets conjugated by the value of the transformation at the reference point of $\pi_1(\Sigma)$. We thus get
\[
\calA_0/\calG \subseteq \Hom\bigl(\pi_1(\Sigma),G\bigr)/G,
\]
where $G$ acts by conjugation. The set $\Hom(\pi_1(\Sigma),G)/G$ is known as the moduli space of f\/lat $\SU(2)$-bundles. When $\pi_1(\Sigma)$ has a f\/inite presentation, then it is a real algebraic variety. It also usually has singularities which prevent it from being a smooth manifold. Later, we will ignore those singularities and consider the case where $\calA_0/\calG$ really coincides with the moduli space of f\/lat bundles.

This space has been mainly studied when $\Sigma$ is a two-dimensional closed surface, where a~symplectic structure is available. That has been quite ef\/f\/icient to solve Yang--Mills theory in~$2d$~\cite{witten-2dym}, Chern--Simons theory in the canonical formalism~\cite{witten-3d-gravity}. More recently \cite{barrett-pr, cell-homology, twisted, frohman-torsion}.

\subsubsection{Quantization}

{\bf The path integral.} Formally, without gauge f\/ixing, it reads
\[
``Z(M) = \int DA DB  \exp\Bigl( i\int_M \tr\bigl(B\wedge F_A\bigr)\Bigr) = \int DA  \prod_{x\in M} \delta\bigl( F_A(x)\bigr)\text{''},
\]
from which we learn two things.
\begin{itemize}\itemsep=0pt
 \item The functional integral is exactly peaked on f\/lat connections over $M$.
 \item Gauge f\/ixing is crucial to get a topological invariant (if any). Indeed, if we naively continue the above calculation, we f\/ind (after some regularization) the determinant of $\delta F_A/\delta A$ which is the the covariant derivative $d_A$ acting on 1-forms and which is not a topological invariant. The intuitive reason is that $d_A$ here only probes the topology going from 1-forms to 2-forms. The f\/inal topological invariant should instead probe all dimensional forms (or cells) over $M$.
\end{itemize}

The path integral can be evaluated using the usual Faddeev--Popov gauge-f\/ixing in three dimensions, and the Schwarz' resolvent in higher dimensions (to deal with the reducibility), see~\cite{blau-thompson-torsion, blau-thompson-bf}.

Let us sketch the three-dimensional situation. We assume for simplicity that the moduli space of f\/lat connections is f\/inite and that for any f\/lat connection $A_{(\alpha)}$ the cohomology of the operator $d_{A_{(\alpha)}}$ is trivial. We pick up a background metric $g$ and select a representative $A_{(\alpha)}$ of each equivalence class of f\/lat connections up to gauge which satisf\/ies $d_{A_{(\alpha)}} * A_{(\alpha)}=0$, where $*$ is the Hodge operator. Setting $A=A_{(\alpha)}+a$ in the neighborhood of $A_{(\alpha)}$, we get
\begin{gather*}
Z(M) = \sum_\alpha \mu(A_{(\alpha)}),
\end{gather*}
with
\begin{gather*}
\mu(A_{(\alpha)})=\int Da DB D\omega D\bar \omega D\chi D\bar \chi Du Dv\\
{} \times \exp\biggl(i\int \tr\,\bigl( B\wedge d_{A_{(\alpha)}} a + u\wedge d_{A_{(\alpha)}}* a
+  v\wedge d_{A_{(\alpha)}}* B + \bar \omega\wedge \Delta_{A_{(\alpha)}} \omega + \bar \chi\wedge \Delta_{A_{(\alpha)}} \chi
\bigr)\biggr),
\end{gather*}
where $\Delta_{A_{(\alpha)}} = d_{A_{(\alpha)}}* d_{A_{(\alpha)}}$. $u$, $v$ are Lagrange multipliers imposing gauge conditions on~$a$,~$B$, and $\omega$, $\chi$ and $\bar \omega$, $\bar \chi$ are the corresponding ghosts and anti-ghosts. We refer to~\cite{blau-thompson-torsion, blau-thompson-bf} for details. The main point is that $\mu(A_{(\alpha)})$ can be evaluated to be the Ray--Singer analytic torsion. It is a~topological invariant, which does not depend on the choice of the background metric.

{\bf The wave-functions.} From the canonical analysis on $M=\Sigma\times[0,1]$, it is clear that they are square-integrable functions over the set of f\/lat connections modulo gauge transformations~\cite{horowitz-bf}.

{\bf The measure.} The inner product is obtained by considering the path integral on $M=\Sigma\times[0,1]$. The measure is given by the Ray--Singer torsion on $M$ which in this case descents on~$\Sigma$, so that the measure is well def\/ined on the canonical surface.

{\bf A physical IR divergence.} The partition function is well-def\/ined for topologies where the cohomology $H^*(d_A)$ is trivial. At least in three dimensions, there is a nice physical interpretation, noticed by Witten \cite{witten-amplitude-3d}. There the partition function diverges after gauge f\/ixing if the moduli space of f\/lat connections is not a f\/inite set. Notice that the equation for the tangent vectors to that space is $\delta F = d_A \delta A = 0$. This is exactly the equation of motion for the triad \mbox{$d_A e=0$}. Hence, the partition function is f\/inite on those topologies where there are no non-trivial (equivalence classes of) solutions for the triad. The intuitive picture of the quantum mechanical consequence is that there are only some quantum f\/luctuations around the trivial solution. When restoring the units, it means that lengths are f\/luctuating around the Planck scale, explaining the f\/initeness of $Z(M)$. However, if there are some tangent directions to the moduli space of f\/lat connections, there are some non-trivial (equivalence classes of) triads satisfying $d_A e=0$. The integrals along those directions are non-compact, hence the divergence. Physically, that corresponds to an IR divergence, since the main contributions to the path integral come from large triads, hence spacetimes much larger than the Planck scale. The result is that a macroscopic physics can exist only on topologies with non-trivial cohomology and a divergent partition function.

\subsubsection{Loop quantization}

Since the loop quantization relies on cylindrical functions supported on graphs and since the measure on the set of f\/lat connections is the torsion, it would be convenient to evaluate the torsion using cell decomposition. That notion exists and is known as the Franz--Reidemeister torsion. This actually shows that the BF model admits an \emph{exact} quantum formulation in a~discrete setting. Some details can be found in~\cite{bf-anydim}.

{\bf Discrete f\/lat connections and covariant derivative.} The main idea is to f\/ind an equivalent on cells to the covariant derivative. Let us describe its parts which act on $0$-cells and $1$-cells. Further details can be found in \cite{barrett-pr, cell-homology, twisted}, in~\cite{frohman-torsion} in the context of knot invariants, and also in the Chern--Simons literature (where the measure is rather the square-root of the torsion) such as \cite{chern-simons-jeffrey, chern-simons-rozanski}.

Let $\Delta$ be a cell decomposition of $\Sigma$, and $\Delta_i$ the set of $i$-cells. Since we focus on holonomies, we can forget that they come from a connection, and consider instead the notion of discrete connection which is the assignment of $\SU(2)$ elements to edges of $\Delta$, denoted $A=(g_e)_{e\in\Delta_1}$.

Gauge transformations induce a $\SU(2)^{\vert \Delta_0\vert}$-action. A discrete gauge transformation $h$ is therefore a set of group elements $(h_v)_{v\in\Delta_0}$ acting at the vertices of $\Delta$ according to
\[
h \cdot  A = \bigl(h_{t(e)} g_e h^{-1}_{s(e)}\bigr)_{e\in\Delta_1}.
\]
To evaluate the ef\/fect of a small gauge transformation, we consider the dif\/ferential at $h=\unit$ of the map $\gamma_A$ which sends $h$ to $h\cdot A$,
\[
d\gamma_{A|\unit}  :\ \su(2)^{\vert \Delta_0\vert}  \rightarrow T_A \SU(2)^{\vert \Delta_1\vert}, \qquad v  \mapsto  \bigl(R_{g_e*}\bigl(\id-\Ad_{g_e}\bigr)v\bigr)_{e\in\Delta_1},
\]
where $\Ad$ stands for the adjoint representation of the group on its algebra, and $R_{g*}$ for the right translation. In matrix notation: $\Ad_g v = gvg^{-1}$ and $R_{g*}v = vg$. The kernel of $d\gamma_{A|\unit}$ is the algebra of the isotropy group $\zeta(A)$ of the connection $A$, while its image corresponds to the directions along which $A$ is changed by the group action. If $\calO_A$ denotes the orbit through $A$, then we know that
\[
\calO_A \simeq \SU(2)/\zeta(A),\qquad\text{and}\qquad T_A \calO_A = \im d\gamma_{A|\unit}.
\]

The curvature of a connection is encoded in the holonomies along the boundary of $2$-cells. It is represented by the map
\begin{gather*} %\label{curvature map}
H  :\ \SU(2)^{\vert \Delta_1\vert}\rightarrow \SU(2)^{\vert \Delta_2\vert},\qquad A\mapsto
\biggl( H_f(A)=\prod_{e\in\partial f} g_e^{\epsilon_{fe}} \biggr)_{f\in\Delta_2},
\end{gather*}
where $\epsilon_{fe}$ is the relative orientation of the face $f$ on the edge $e$, and $H_f(A)$ is the `holonomy' of the connection $A$ around the face~$f$. This provides a notion of f\/latness on $\Delta$: the connection is \emph{flat} if
\[ %\label{discreteflatness}
H(A) = \unit.
\]
For a discrete f\/lat connection $\phi$, the tangent space to the moduli space of f\/lat $\SU(2)$-bundles $\hom(\pi_1(\Sigma),\SU(2))$ is
\[
T_\phi \bigl(\hom(\pi_1(\Sigma),\SU(2))/\SU(2)\bigr) = \ker dH_\phi/ \im d\gamma_{\phi|\unit}.
\]
Moreover, gauge invariance implies that $dH_\phi \circ d\gamma_{\phi|\unit} = 0$. That equation means that cohomology classes can be def\/ined. The analogy with the continuum is that $d\gamma_{\phi|\unit}$ is the covariant derivative on $\su(2)$-valued $0$-form, and $dH_\phi$ the covariant derivative on $\su(2)$-valued $1$-form. Hence, the tangent space $T_\phi \hom(\pi_1(\Sigma),\SU(2))/\SU(2)$ is the f\/irst cohomology group. That provides with an algebraic description of the local properties of the moduli space of f\/lat connections.

{\bf Reidemeister torsion.} The above description gives a (co)chain complex which includes $0$-, $1$- and $2$-cells. If $\Delta$ is a cell decomposition of dimension $d$ greater than two, the complex can be extended for any discrete f\/lat connection $\phi$,
\begin{gather} \label{cochain complex}
0\longleftarrow C^d_\phi(\Delta) \xleftarrow{\ \delta^{d-1}_\phi(\Delta)\ } \dotsm \xleftarrow{\qquad}
C^2_\phi(\Delta) \xleftarrow{\ \delta^1_\phi(\Delta)\ }  C_\phi^1(\Delta) \xleftarrow{\ \delta^0_\phi(\Delta)\ } C_\phi^0(\Delta) \longleftarrow 0,
\end{gather}
where each cochain group basically contains assignments of a $\su(2)$ element to each cells, $C_\phi^i(\Delta)\simeq \su(2)^{\vert \Delta_i\vert}$. We have also relabel the previous $0,1$-cell covariant derivatives, $\delta^0_\phi = d\gamma_{\phi|\unit}$ and $\delta^1_\phi = dH_\phi$.

Assuming the cohomology of the above complex is trivial, one can def\/ine the Reidemister torsion as follows
\[
\tor_\phi(\Sigma) =  \prod_{j=0}^d \bigl( \det \bigl(\delta_\phi^{j\dagger} \delta^j_\phi  \bigr)\bigr)^{(-1)^j/2}.
\]
Here $\dagger$ denotes the adjoint with respect to the natural $\su(2)$ inner product. Obviously the operators $\delta_\phi^{j\dagger}\,\delta^j_\phi$ have some kernels so one must take care of removing them by restricting the operators to the orthocomplements of their kernels. Then the torsion only depends on the equivalence class $[\phi]$ of the f\/lat connection modulo gauge and is a topological invariant in the sense that it is independent\footnote{The full statement is that torsion is invariant under simple homotopy transformations of $\Delta$. The notion of simple homotopy is somehow intermediate between homeomorphisms and homotopy.} of the choice of cell decomposition $\Delta$.

{\bf Transition amplitudes.} That combinatorial torsion provides a natural analog to the Ray--Singer torsion which is well adapted to cylindrical functions. The latter depend on a f\/inite number of holonomies, just like those used to def\/ine the torsion. Moreover, provided we consider gauge invariant wave-functions, their value on a f\/lat discrete connection $\phi$ will actually only depend on the equivalence class $[\phi]$.

Hence, a natural candidate for BF transition amplitudes is built by gathering the torsion to get the measure and cylindrical functions evaluated on equivalence classes of f\/lat connections as quantum states. Consider two graphs $\Gamma_{1,2}$ supporting two cylindrical functions $\psi_{\Gamma_{1,2},f_{1,2}}$. Choose a cell decomposition of $\Delta$ which contains $\Gamma_{1,2}$. Then the natural extension of the formal path integral is
\[
\langle \psi_{\Gamma_1,f_1}\vert \psi_{\Gamma_2,f_2}\rangle = \sum_{[\phi]} \bar f_1([ \phi ]) \tor_{[\phi]}(\Sigma)  f_2([ \phi ]).
\]
Notice that the dependence of $f_{1,2}([\phi])$ on their graphs is only through their homotopy classes. It implies for instance that dif\/ferent knotting classes of a graph embedding cannot be distinguished by that invariant.

Such a formula def\/ines the physical inner product on the reduced Hilbert space, or equivalently the matrix elements of the projector onto those states which satisfy the f\/latness condition at the quantum level.

Obviously that proposal should be made more precise by building the full path integral on spacetime directly from a cell decomposition, identifying the gauge symmetries in that setting and gauge-f\/ixing them appropriately. This has been done at the covariant level when spacetime is three-dimensional (with boundaries): the full set of gauge symmetries is described in~\cite{freidel-louapre-pr1} while the relationship to torsion can be found in~\cite{barrett-pr}. At the canonical level, i.e.\ on a two-dimensional surface~$\Sigma$, the transition amplitudes have been evaluated in \cite{noui-perez-ps3d} by def\/ining the projector onto physical states which satisfy f\/latness. In that situation, there is no translation gauge symmetry to take care of, and the torsion is trivial, as well-known in $2d$ Yang--Mills theory \cite{witten-2dym}.

{\bf Coupling to extended matter.} As a topological f\/ield theory, the BF model is known to feature exotic statistics \cite{baez-fractional-bf, bergeron-fractional-bf}, which depending on the dimension involve particles, strings and/or branes. The dynamics and quantization of BF theory coupled to such extended matter are presented in \cite{baez-perez-strings-bf,fairbairn-perez-strings-bf}.

\subsection[The Wheeler-DeWitt equation for quantum flat space]{The Wheeler--DeWitt equation for quantum f\/lat space}

We showed in the previous section that the BF f\/ield theory admits an \emph{exact} discretization, which is well-adapted to cylindrical functions, and hence to the loop quantization. If $\Delta$ is a~cell decomposition of the canonical surface $\Sigma$, the projector on physical states, satisfying the f\/latness constraint $\calC_f$, reads formally
\[
P_\Delta = \prod_{f\in\Delta_2} \delta( H_f ).
\]
Let us say a few words on that projector. When written in this way, it only sees the 2-skeleton of~$\Delta$, so that one can try to study it on a arbitrary 2-complex, which may or may not be the 2-skeleton of a higher dimensional cell complex. Technically, divergences arise because they are too many delta functions in the above expression. If only a 2-complex is given, the complex~\eqref{cochain complex} is truncated to cells of dimension lower than two. Since gauge symmetries come from the action of dimensions three and higher onto the faces, divergences cannot be understood for the presence of gauge symmetries. However, the structure of the divergences are still encoded into the cohomology twisted by f\/lat connections (truncated at dimension~2), and the rate of divergence can be extracted as exactly the dimension of the cohomology group $H^2_\phi(\Delta)$ (see~\cite{cell-homology, twisted} and~\cite{sorting-out} for a summary and a comparison with other approaches to those divergences). This is an important result since the divergences are now well understood even when there cannot be interpreted as due to gauge symmetries. That may be helpful for the f\/lat sector of full quantum gravity, since the gauge symmetries of BF are generically broken in the full theory.

We now restrict to situations where this projector is well-def\/ined. That means either the gauge-f\/ixing is trivial, like when $\Delta$ is a decomposition of a closed orientable two-dimensional surface~$\Sigma_g$ of genus~$g$ greater than two, or it is almost trivial in the sense that it is suf\/f\/icient to remove a given number of faces which carry redundant delta functions. That is the case of the spherical topology, and such a gauge-f\/ixing is described in~\cite{freidel-louapre-pr1}.

The projector $P_\Delta$ on $\Sigma_g$ has been constructed in~\cite{noui-perez-ps3d}. Its matrix elements provide a def\/inition of the physical inner product. The projector can be unfolded in spacetime, using for instance the spin network basis. That mimics a time evolution for spin networks which actually corresponds to a spin foam model, namely the Ponzano--Regge model.

\subsubsection{The new Hamiltonian for 2+1 gravity}

Though the result of \cite{noui-perez-ps3d} was a major progress, it was still unclear whether a Wheeler--DeWitt equation for 2+1 gravity in the spin network basis can be written. Since the BF model describes in (2+1)-dimensions pure gravity (with degenerate metrics), the canonical formalism can be performed dif\/ferently than in  Section~\ref{sec:bf-hamiltonian}. It can be done like in general relativity, by introducing the shift and lapse. They are respectively Lagrange multipliers for the vector constraint which generates dif\/feomorphisms of the canonical surface and for the scalar constraint~$H$ which generates time reparametrizations,
\[
V_a = E^b_i F_{ab}^i,\qquad H = \epsilon_{ij}^{\phantom{ij}k} F^k_{ab}  E^a_i E^b_j.
\]
So we can trade the constraint $\calC_F$ for those two. Physical states have to satisfy the vector constraint $\widehat{V}_a \vert \psi\rangle =0$ and the Wheeler--DeWitt equation
\[
\widehat{H} \vert\psi\rangle = 0,
\]
for some regularization of $V_a$ and $H$. There are a few things we know about the Hamiltonian,
\begin{itemize}\itemsep=0pt
\item from the previous section, we know we can work on a single cell decomposition of spacetime, because $H$ should not change the graphs of cylindrical functions,
\item but $H$ should change spins in the spin network basis, so that it probably should generate dif\/ference equations.
\end{itemize}

Note that on a two-dimensional surface $\Sigma$ the curvature tensor $F_{ab}^i$ is just a vector $\vec{F}_{12}$. Since $B^i_\mu$ is the co-triad, the vector $\vec{n} = \vec{B}_1\times \vec{B}_2$ is the normal to the canonical surface. Then $H=\vec{n}\cdot \vec{F}_{12}/\vert \vec{n}\vert$. That is the origin of most dif\/f\/iculties to regularize $H$ on cylindrical functions, as we explain.

Assume that $\Delta$ is a complex dual to a triangulation. The natural variables are the Wilson lines along edges of $\Delta$ and their canonical momenta $E^i_e$ which are f\/luxes through those edges. The geometric picture is that the f\/luxes represent the edges of each triangle of the triangulation embedded into~${\mathbb R}^3$. The curvature at a node of the triangulation is regularized by the Wilson loop around the face dual to the node. However, there no clear def\/inition of the normal $\vec{n}$ to the node.

The way-out which has been proposed in~\cite{3d-wdw} is the following. If the node of the triangulation is for instance 3-valent, it is shared by three triangles, and one can def\/ine three regularized quantities looking like $H$ using the normal to each triangle. Since the three normals are generically linearly independent, that gives three constraints which are enough to impose the Wilson loop to be trivial, $H_f=\unit$. Hence that proposal enforces f\/latness around the node, and thus implicitly takes into account both the scalar and vector constraints.

\begin{figure}[t]\centering
\includegraphics[scale=0.5]{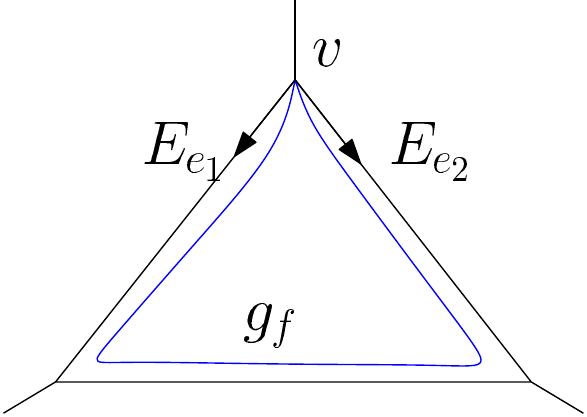}
\caption{One gets the new Hamiltonian by contracting the Wilson loop $g_f$ around the face with the f\/luxes meeting at the starting node of~$g_f$.}\label{fig:newH}
\end{figure}

Let $f$ be a face and $v$ a node of $\Delta$ on the boundary of~$f$ where $e_1$, $e_2$ meet. The curvature tensor is replaced with
\[
\epsilon^{ij}_{\phantom{ij}k} F_{ab}^k \  \longrightarrow\  \delta^{ij} - \bigl(R(g_f)\bigr)^{ij},
\]
where $g_f$ is the Wilson loop around~$f$ based at~$v$, and~$R$ the vector representation. The products of f\/ields $E_i^a E^b_j$ becomes the products of f\/luxes through~$e_1$ and~$e_2$, and we def\/ine (see Fig.~\ref{fig:newH})
\[ %\label{lqg hamiltonian}
H_{v,f} = E_{e_1}^i \bigl( \delta_{ij} - R(g_f)_{ij}\bigr) E_{e_2}^j .
\]
For a face with $k$ vertices, there are $k$ such constraints $(H_{v_n,f})_{n=1,\dots,k}$ around~$f$. Since they probe the matrix elements of $R(g_f)$ which has three independent degrees of freedom, we expect only three such constraints to be independent. That was proved in~\cite{3d-wdw}, and the key point is that the dif\/ferentials of these constraints around the solution $g_f=\unit$ is a system of rank~3. That result was used to prove that the constraint algebra is f\/irst class, hence it generates a gauge symmetry, on the 2-sphere triangulated by the boundary of a tetrahedron (the generic case being unclear so far).

\subsubsection{Euclidean f\/lat geometry}

%%%%%%%%%%%%%%%%%%%%%%%%%%%%%
\begin{figure}[t]\centering

\includegraphics[scale=0.4]{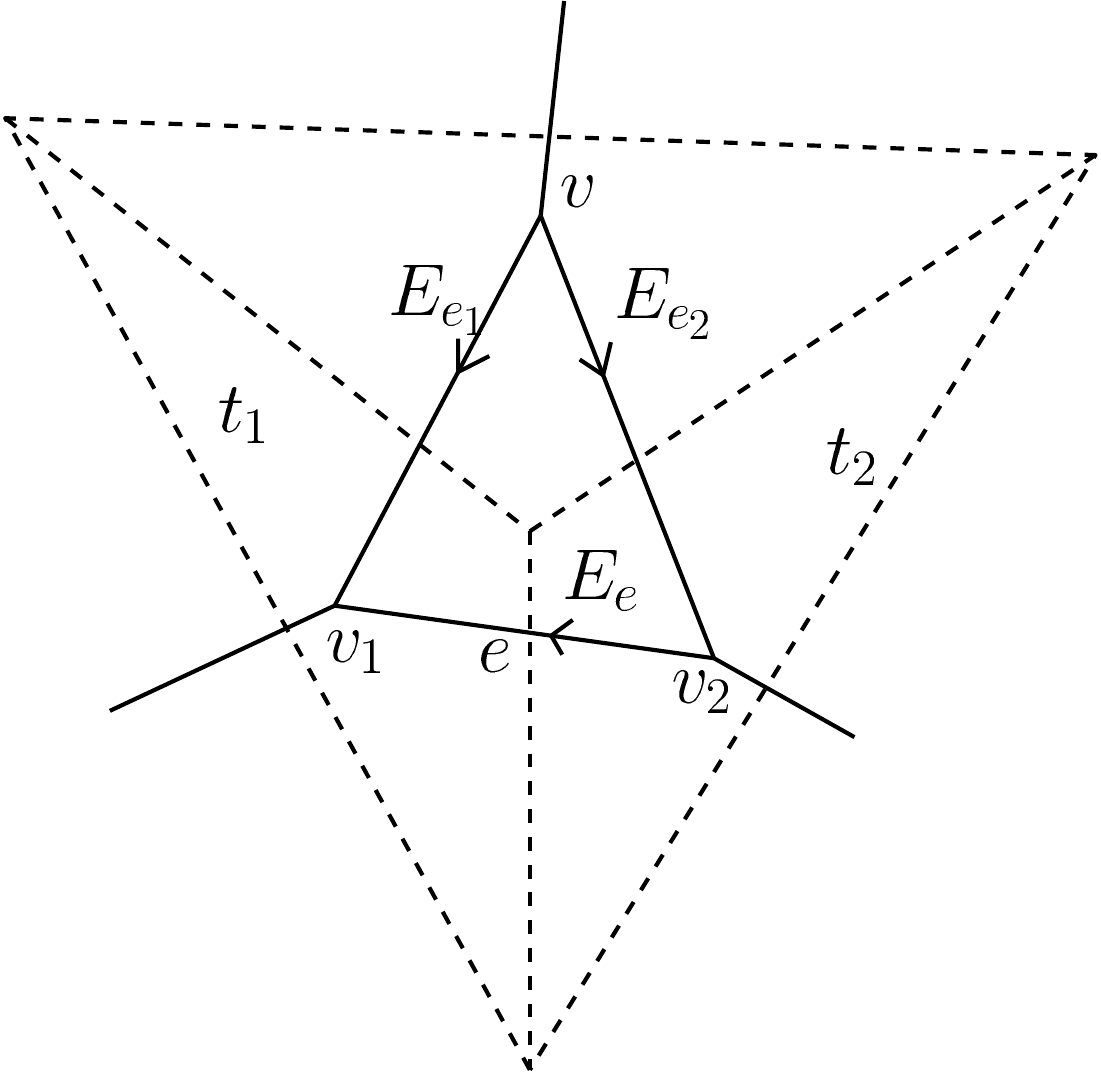}\qquad
\includegraphics[scale=0.55]{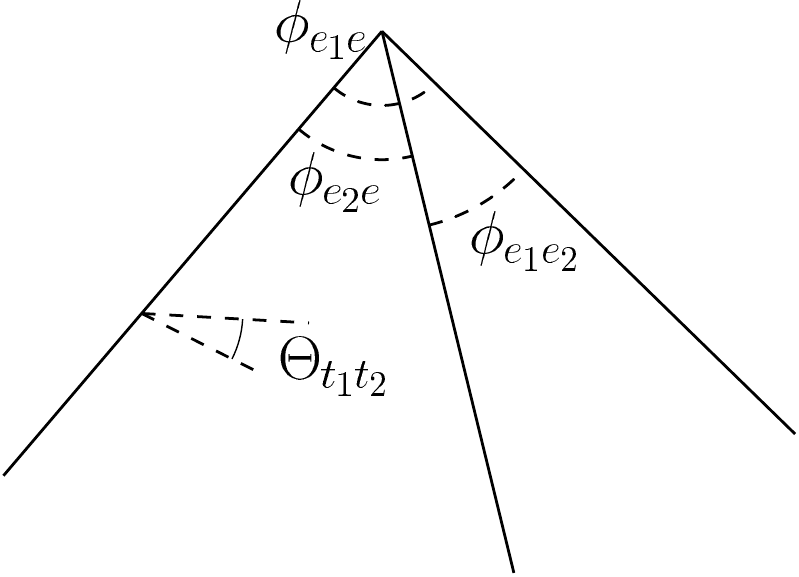}

\caption{On the left, the dashed lines represent three triangles meeting at a node, and the dual graph, in normal lines, depicts a face with three vertices. On the right, the geometry of the three triangles is pictured. Flat geometry is characterized by the fact that the dihedral angle $\Theta_{t_1 t_2}$ between the triangles~$t_1$,~$t_2$ is evaluated from the three angles $\phi_{ee'}$ at the apex, through the Euclidean formula \eqref{regge flat}.} \label{fig:3-cycle}
\end{figure}

The kinematical Hilbert space gives a quantization of the intrinsic geometry. Quantizing the Gau\ss{} law \eqref{closure} (i.e.\ imposing local $\SU(2)$ invariance) is indeed quantizing the law of cosines~\eqref{alkashi} (the Euclidean formula which evaluates angles of triangles in terms of their lengths). We now discuss the extrinsic curvature which is actually computed from the f\/latness constraint.

Physically, what we expect from the classical f\/latness constraint $F_{ab}^i=0$ is that $\Sigma$ with its geometry can be locally embedded into f\/lat 3-space. For the triangulation dual to $\Delta$, it means that the $(3d)$ dihedral angles $\Theta_{t_1 t_2}$ between two adjacent triangles $t_1$, $t_2$ should be given as a~standard function of the dihedral $2d$ angles $(\phi_{ee'})$. A typical situation is depicted at Fig.~\ref{fig:3-cycle}, for a~trivalent node of the triangulation. The triangles $t_1$, $t_2$ are dual to the vertices $v_1$, $v_2$ of~$\Delta$, which are linked by the edge~$e$. The classical angle between the two triangles~$t_1$, $t_2$ can be computed from the three $2d$ angles $\phi_{ee'}$ around the node which is dual to the face~$f$, using the formula
\begin{gather} \label{regge flat}
\cos \Theta_{t_1 t_2} (E,\tl{E}) = \f{\cos\phi_{e_1 e_2} - \cos\phi_{e_1 e} \cos\phi_{e_2 e}}{\sin\phi_{e_1 e} \sin\phi_{e_2 e}} .
\end{gather}
We have written it as a function of the f\/lux variables since all angles $\phi_{ee'}$ can be evaluated from them, without writing down holonomies\footnote{In addition, when the closure relation \eqref{closure} holds, the angles $\phi_{ee'}$ are determined as functions of the lengths $(\ell_e = \vert E_e\vert)$, and hence so are the angles $\Theta_{t_1t_2}$.}.

There is another natural notion of dihedral angles which makes use of f\/lux variables and holonomies. Since the f\/luxes represent the directions of the edges of the triangulation, the normal~$N_t$ to a triangle is given by the cross product of two of them. Following Fig.~\ref{fig:3-cycle}, $N_{t_1}^i = \epsilon^i_{\phantom{i}jk}  \tl{X}_e^j  \tl{X}_{e_1}^k$, and $N_{t_2}^i = \epsilon^i_{\phantom{i}jk} \tl{X}_{e_2}^j X_e^k$. It suggests to look at the dot product of these normals as being the cosine of the dihedral angle between $t_1$ and $t_2$. But for this, it is necessary to transport them in a common frame, say $N_{t_2}$ along the edge $e$ to the vertex $v_1$. So we def\/ine the angle $\theta_{t_1t_2}$ as
\[
\cos \theta_{t_1 t_2} (E, \tl{E}, g) = -\frac{N_{t_1}\cdot R(g_e) N_{t_2}}{\vert N_{t_1}\vert \vert N_{t_2}\vert}.
\]
This is a function of the holonomy $g_e$ since one has to compare the local embedding of~$t_1$ to that of~$t_2$. The result can be partially evaluated in terms of the $2d$ dihedral angles, because it is a~dot product of two vector products in ${\mathbb R}^3$,
\[
\cos \theta_{t_1 t_2} (E, \tl{E}, g) = \f{\tl{E}_1\cdot R(g_e)\tl{E}_2  - \cos\phi_{e_1 e} \cos\phi_{e_2 e}}{\sin\phi_{e_1 e}  \sin\phi_{e_2 e}}.
\]
Obviously, one would like to this notion of dihedral angle to coincide with the standard formula~\eqref{regge flat}. This is what the Hamiltonian does~\cite{3d-wdw}
\begin{gather} \label{geom H}
H_{v,f} = \sin\phi_{e_1 e}  \sin\phi_{e_2 e}  \bigl( \cos \Theta_{t_1 t_2}  -  \cos \theta_{t_1 t_2}\bigr).
\end{gather}
Hence, dihedral angles are indeed computed from the $2d$ angles with the usual formula of f\/lat Euclidean geometry.

\subsubsection[The Wheeler-DeWitt equation and the pentagon identity]{The Wheeler--DeWitt equation and the pentagon identity}

Let $\Delta$ be the dual graph to a triangulation of the 2-sphere by the boundary of a tetrahedron. It has four vertices (dual to the triangles), six links (dual to the six edges), and four triangular faces (dual to the nodes). A spin network function has six group elements as arguments, and six spins labeling the edges. Since the vertices are three-valent, the intertwiners are completely determined by the three spins meeting at each node, up to normalization. A common choice is to write the spin network function like
\begin{gather*}
s^{\{j_e\}}_{\rm tet}(g_1,\dots,g_6) = \sum_{\substack{ a_1,\dots,a_6 \\b_1,\dots,b_6}} \begin{pmatrix} j_1 &j_2 &j_3\\ b_1 &-a_2 &-a_3\end{pmatrix} \begin{pmatrix} j_1 &j_5 &j_6\\ -a_1 &b_5 &-a_6\end{pmatrix} \begin{pmatrix} j_3 &j_4 &j_5\\b_3 &-a_4 &-a_5\end{pmatrix} \\
\phantom{s^{\{j_e\}}_{\rm tet}(g_1,\dots,g_6) =}{}
\times \begin{pmatrix} j_2 &j_6 &j_4\\ b_2 &b_6 &b_4\end{pmatrix}   \left[\prod_{e=1}^6 (-1)^{j_e-a_e} \langle j_e, a_e \vert g_e \vert j_e,b_e\rangle\right].%\label{defspinnet}
\end{gather*}
The range of summation is $-j_e\leq a_e,b_e\leq j_e$ for the magnetic numbers on each link. The quantities into brackets are the components of the intertwiners in the usual spin and magnetic number basis, and are known as Wigner $3jm$-symbols \cite{varshalovich}, while $\langle j,a \vert g\vert j,b\rangle$ denotes the matrix elements of the Wigner $D$-matrix with spin $j$. Any state $\psi$ can be seen as a wave function and decomposed in that basis,
\[
\psi(g_1,\dots,g_6) = \sum_{j_1,\dots,j_6} \left[\prod_{e=1}^6 (2j_e+1)\right] \psi(j_1,\dots,j_6) s^{\{j_e\}}_{\rm tet}(g_1,\dots,g_6).
\]

\begin{figure}[t]\centering
\includegraphics[scale=0.45]{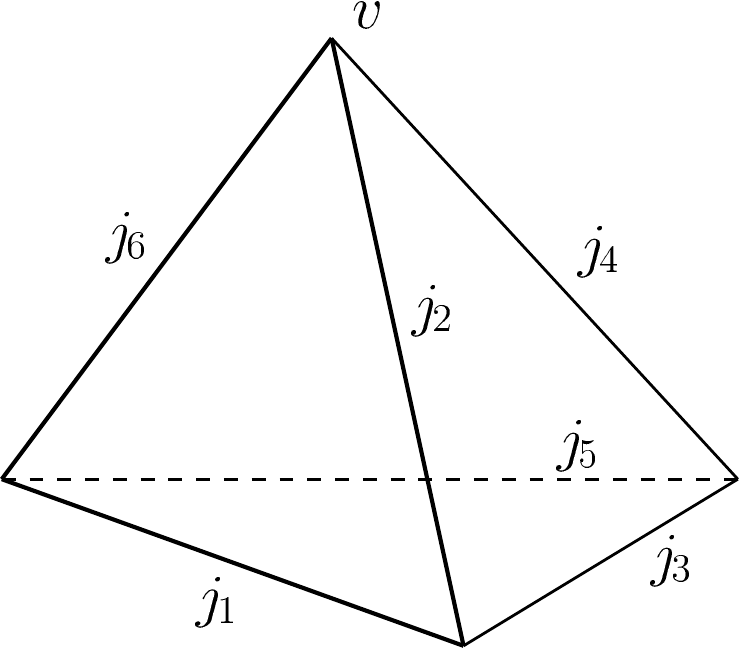}
\caption{The tetrahedral graph on the 2-sphere.} \label{fig:tet}
\end{figure}

We now consider the node $v$ where the edges 2 and 6 meet in the tetrahedral graph of Fig.~\ref{fig:tet} and the face $f$ with 1, 2, 6 on its boundary. The associated $H_{26} = E_2\cdot E_6 - E_2\cdot R(g_f) E_6$ has been quantized in \cite{3d-wdw}. $E_2\cdot E_6$ is diagonal on spin network functions, but $E_2\cdot R(g_f) E_6$ produces some shifts on the spin $j_1$ due to the presence of matrix elements of $g_1$ in $R(g_f)$. The latter acts by multiplication on spin networks, so that shifts are due to re-coupling between the spin $j_1$ and the spin $1$ carried by the vector representation $R$. In the end the Wheeler--DeWitt equation $H_{12}\vert \psi\rangle =0$ is in the spin network representation
\begin{gather} \label{rec6j}
A_{+1}(j_1) \psi(j_1+1,\dots,j_6) +
A_{0}(j_1) \psi(j_1,\dots,j_6) +
A_{-1}(j_1) \psi(j_1-1,\dots,j_6) = 0.
\end{gather}
The coef\/f\/icients are $A_+(j_1)= j_1 E(j_1+1)$ and $A_-(j_1)= (j_1-1)E(j_1)$, with:
\begin{gather*}  %\label{Apm1}
E(j_1) = \Bigr[\bigl((j_2+j_3+1)^2-j_1^2\bigr)  \bigl(j_1^2-(j_2-j_3)^2\bigr)  \bigl((j_5+j_6+1)^2-j_1^2\bigr)  \bigl(j_1^2-(j_5-j_6)^2\bigr)\Bigr]^{\f12},
\end{gather*}
and \label{A_0}
\begin{gather*}
A_0(j) = \bigl(2j_1+1\bigr)\Bigl\{2\bigl[j_2(j_2+1)j_5(j_5+1)+j_6(j_6+1)j_3(j_3+1)-j_1(j_1+1)j_4(j_4+1)\bigr] \\
\phantom{A_0(j) =}{} - \bigl[j_2(j_2+1)+j_3(j_3+1)-j_1(j_1+1)\bigr]\bigl[j_5(j_5+1)+j_6(j_6+1)-j_1(j_1+1)\bigr]\Bigr\}.
\end{gather*}
Hence the Wheeler--DeWitt equation is a second order dif\/ference equation, which can be solved from a single initial condition\footnote{It is an interesting feature that the solution is determined by one initial condition and instead of two. Looking at the lowest possible value of the spin $j_1$, that is: $j_1^{\min} = \max(\vert j_2-j_3\vert, \vert j_5-j_6\vert)$, it turns out that the lowering coef\/f\/icient $A_{-1}(j_1^{\min})$ evaluated on this spin is zero. Thus, the recurrence can be implemented starting from the initial value on this lowest spin. This property is important with respect to the asymptotic limit. There, it is known \cite{roberts, schulten-gordon2, varshalovich} that the $6j$-symbol oscillates like the cosine of the Regge action $S_{\rm R}(j_e) = \sum_e (j_e+\f12)\theta_e$, for a~tetrahedron with lengths $(j_e+\f12)$,
\[
\begin{Bmatrix} j_1 &j_2 &j_3\\j_4 &j_5 &j_6\end{Bmatrix} \approx \f1{\sqrt{12\pi V(j_e)}}\,\f12\,\Bigl(e^{i(S_{\rm R}(j_e)+\f\pi4)} + e^{-i(S_{\rm R}(j_e)+\f\pi4)}\Bigr) .
\]
Having a second order equation, it may have been expected to be solved in the asymptotics by both the positive and negative exponentials of the action. But the model is fully independent from the orientation: there is only one required initial condition, because of the vanishing of $A_{-1}(j_1^{\min})$, leading in the large spin regime to the cosine of the action.}.

This equation is sometimes taken as the def\/inition of the Wigner $6j$-symbol, $\psi(j_1,\dots,j_6) = \left\{ \begin{smallmatrix} j_1 &j_2 &j_3\\j_4 &j_5 &j_6\end{smallmatrix}\right\}$. It is a basic re-coupling coef\/f\/icient of quantum angular momenta theory. If one wants to f\/ind a basis of the tensor product $j_1\otimes j_2\otimes j_4\otimes j_5$, one can choose a spin $j_3$ which lies in both tensor products $j_1\otimes j_2$ and $j_4\otimes j_5$. Another basis is obtained by choosing a spin $j_6$ shared by $j_1\otimes j_5$ and $j_2\otimes j_4$. The $6j$-symbol gives the corresponding change of basis.

The dif\/ference equation is thus the standard recursion relation on the $6j$-symbol. The latter comes from the Biedenharn--Elliott (pentagon) identity~\cite{varshalovich}, which is the key to re-coupling $\SU(2)$ angular momenta. The above result is actually the achievement of some ef\/forts to prove a~conjecture relating the Biedenharn--Elliott identity to the Wheeler--DeWitt equation (see~\cite{barrett-crane-wdw}). The main idea behind the conjecture was that there exists a state-sum model, the Ponzano--Regge model~\cite{PR}, which is known to give the partition function of 2+1 gravity on a 3-manifold \cite{freidel-louapre-pr1, noui-perez-ps3d, ooguri-3d}. It is built on labeling edges of a triangulation with spins, assigning a $6j$-symbol to each tetrahedron and summing over all labelings. When f\/inite, it enjoys an invariance under some changes of the triangulation. Those changes are generated by 3-2 Pachner moves. The elementary move is the following. Take three tetrahedra glued along a common edge, each pair sharing a triangle, and remove those three triangles together with the common edge, and instead glue a new triangle dual to the removed edge. One ends up with only two tetrahedra glued along a triangle. This move preserves the topology, and the corresponding invariance of the Ponzano--Regge model is just due to the Biedenharn--Elliott identity. The latter thus appears as the algebraic translation of the Pachner move. As the 3-2 Pachner move is part of the topological invariance\footnote{The full topological invariance needs the 4-1 Pachner move, but unfortunately it is divergent, because the model as itself does not gauge f\/ix the non-compact gauge symmetry at the root of the topological invariance. Gauge-f\/ixing in the Ponzano--Regge model is described in~\cite{freidel-louapre-pr1} and f\/initeness in~\cite{barrett-pr}.}, a similar relation was expected in the Hamiltonian formulation.

The relationship to f\/lat connections on the tetrahedral graph is most conveniently seen by going back to wave functions of the holonomies,
\begin{gather*}
  \psi(g_1,\dots,g_6)  = \sum_{j_1,\dots,j_6} \left[\prod_{e=1}^6 (2j_e+1)\right]  \begin{Bmatrix} j_1 &j_2 &j_3 \\j_4 &j_5 &j_6\end{Bmatrix}\,s^{\{j_e\}}_{\rm tet}(g_1,\dots,g_6)\\
 \phantom{\psi(g_1,\dots,g_6)}{}
 = \delta\big(g_4\mone g_5 g_6\big) \delta\big(g_2\mone g_1\mone g_6\big) \delta\big(g_2\mone g_3 g_4\big).
\end{gather*}
The above state is exactly the projector onto f\/lat connections for the chosen graph and topo\-lo\-gy\footnote{If one gauge-f\/ixes, using the $\SU(2)$ action at each node, the group elements $g_1$, $g_2$, $g_3$ to the unit, the delta functions impose $g_4=g_5=g_6=\unit$, which is the unique f\/lat connection up to gauge on the sphere}.

As a remark, notice that the naive state would have a delta function on each of the four faces of the graph. However, that would be ill-def\/ined since one delta function would be trivially satisf\/ied thanks to the other. So it is usually removed by hand. Here, the state is well-def\/ined as a distribution and that is an outcome of our approach.

\subsubsection{Extra remarks}

{\bf Symmetry in the Boulatov model for $\boldsymbol{3d}$ gravity.} The latter is a group f\/ield theory, i.e.\ a non-local f\/ield theory on a group, whose Feynman expansion generates a sum over simplicial complexes supporting spin foam amplitudes, see~\cite{diffeo3dgft}, and more remarks on Section~\ref{sec:tensors}. Interestingly, a global symmetry of the model has been found in~\cite{diffeo3dgft} and there argued to be the translation of the dif\/feomorphism symmetry of $3d$ gravity in group f\/ield theory. The symmetry involves an action of the Drinfeld double of $\SU(2)$, and there is indeed a quite well established approach to quantize $3d$ gravity by going from the classical group $\operatorname{ISU}(2)$ (which takes into account both the standard gauge symmetry and the translation of the co-triad), to the quantum group $\mathcal{D}\SU(2)$ at the quantum level. We do not comment further as we are not expert, but refer to~\cite{diffeo3dgft} instead.

A direct connection with the quantization of the scalar constraint we have just presented is obtained from the way the global symmetry of \cite{diffeo3dgft} is expressed in the spin basis. It is a~recursion relation on the $6j$-symbol, which can be proved from~\eqref{rec6j}. It also derives directly from an operator acting on wave-functions, with the corresponding quantum constraint,
\[
\bigl[ \tr_j(g_f) - (2j+1)\bigr] \psi =0,
\]
where $g_f$ is the Wilson loop around a plaquette. In the spin basis, for a plaquette with $n$ links labeled $e=1,\dots,n$, it gives
\begin{gather}
\sum_{j_1,\dots,j_n}  (-1)^{j+\sum\limits_{e=1}^n j_e+k_e+l_e}\left[\prod_{e=1}^n d_{j_e}\right]
\begin{Bmatrix} k_2& j_2 & j \\ j_1& k_1 & l_1\end{Bmatrix}
\begin{Bmatrix} k_3& j_3 & j \\ j_2& k_2 & l_2\end{Bmatrix} \dotsm
\begin{Bmatrix}k_1& j_1 & j \\ j_n& k_n & l_n\end{Bmatrix} \psi(j_e)\nonumber\\
 \qquad{} = (2j+1)  \psi(k_e).\label{recholgen}
\end{gather}
That equation is obviously solved by the evaluation of the spin network on the identity.

There is an appealing geometric interpretation of that equation, as generating a quantum tent move evolution. We use the Ponzano--Regge interpretation~\cite{freidel-louapre-pr1} of the $6j$-symbol: it is a weight associated a tetrahedron whose edge lengths are given by $(j_e+1/2)$. A Ponzano--Regge amplitude is then obtained by taking products of $6j$-symbols corresponding to a gluing of tetrahedra, and summing over the spins of the internal dual edges, keeping those on the boundary f\/ixed.

That interpretation shows that the equation~\eqref{recholgen} is actually computing the Ponzano--Regge amplitude for a piece of a $3d$ triangulation built as follows. The plaquette is dual to a vertex~$s$ of a $2d$ triangulation. A new triangulation, identical to the f\/irst, is obtained by evolving~$s$ into another vertex~$s^*$. The piece of $3d$ triangulation is contained between the two triangulations, and is made from~$n$ tetrahedra all sharing the edge $(ss^*)$.

This evolution process in background independent approaches to lattice gravity is presented in \cite{bahr-broken-sym} at the classical level, and further details are given in  Section~\ref{sec:class-tent}. In the equation~\eqref{recholgen}, the evolved state $\psi(k_e)$ is obtained by summing over all admissible values of the spins $(j_e)$ on the initial surface, and the constraint just imposes that the physical state must be invariant under tent moves weighted by their Ponzano--Regge amplitudes.

Notice that since we are dealing with a f\/lat spacetime, the quantum length of $(ss^*)$, i.e.\ $j+1/2$, is a gauge degree of freedom, which simply corresponds to the lapse. It can be chosen arbitrarily, and naively summing over all values of this length would result in a divergence\footnote{For a triangular plaquette, this is nothing but the well-known divergence of the 1-4 Pachner move already observed by Ponzano and Regge. The quantum tent move presented here is a gauge-f\/ixed version.}.

{\bf Thiemann's Hamiltonian in $\boldsymbol{3d}$ gravity.} The regularization and quantization of the scalar constraint has been performed by Thiemann~\cite{thiemann-3d} following the prescriptions he used in 3+1 gravity. The construction is mathematically well-def\/ined, as the continuum limit exists and there are no anomalies. The aim was, like in the new proposal we have presented, to use 2+1 gravity as a testing ground, to be able to compare with the already known quantization. It turns out that the solutions to Thiemann's quantum constraints contain the usual solutions of~\cite{witten-amplitude-3d}, but also other, unusual solutions.

{\bf Non-zero cosmological constant.} The major remaining challenge for the loop quantization of pure three-dimensional gravity is to include the cosmological constant $\Lambda\neq 0$. The partition function and expectation values of observables are known in the case $\Lambda >0$ using the Chern--Simons formulation \cite{witten-3d-gravity} which makes it clear it is still topological. In particular the partition function is given by the Turaev--Viro model, which looks similar to the Ponzano--Regge model, expect that all objects from $\SU(2)$ representation theory are changed to their counterpart for the \emph{quantum group} $U_q(\SU(2))$, with $q=e^{i\sqrt{\Lambda}}$. Hence, it seems that from the loop quantization view, one has to trade of\/f Clebsch--Gordan coef\/f\/icients and $6j$-symbols (and maybe spin networks themselves) for their $q$-deformed versions. It is still unclear why and how that should be taken care of precisely by the constraints in the loop quantization. Some recent progress have been obtained in~\cite{perez-pranzetti1}, where the Kauf\/fman bracket is derived by combining standard $\SU(2)$ spin networks with the Chern--Simons (non-commutative) connection. Interestingly, this comes out just from kinematical considerations. While this strengthens the robustness of the loop quantization itself, it also leads to some puzzle. To extend the Kauf\/fman bracket, which is def\/ined for the spin~1/2, one has to tensor several spins~1/2. However, it must be followed by some symmetrization process, and the usual choice in LQG is the standard symmetrizer of~$\SU(2)$. But to get the correct extension of the Kauf\/fman bracket, one would have to use the $q$-symmetrizer instead. There is no reason at the kinematical level to do so, and it is further unclear why the dynamics would change the symmetrizer.

In the continuum, turning on $\Lambda$ implies a deformation of the constraint algebra where the Gau\ss{} law is unchanged but the curvature is obviously non-zero anymore,
\[
F_{ab}^i + \Lambda \epsilon^{i}_{\phantom{i}jk} e_a^j e_b^k=0,
\]
where $e_a^i$ is the pullback of the triad to the canonical surface. The algebra is f\/irst class, it is a~Lie algebra corresponding to $\SO(4)$, where the Gau\ss{} law generates rotations and the constraint on the curvature generates the Euclidean boosts.

The most natural regularizations of the constraints in the Hilbert space of the loop quantization produce some anomalies in the algebra \cite{perez-pranzetti1, perez-pranzetti2}. It seems to indicate that since the algebra itself is changed, the usual regularizations (used for $\Lambda=0$) must be amended to reproduce a~f\/irst class algebra.

A hint may come from representation theory again. When $\Lambda=0$, we have two types of constraints, the Gau\ss{} law and the f\/latness, whose geometric contents are respectively the law of cosines (the formula which evaluates angles of triangles from lengths) \eqref{alkashi} and a formula to evaluate dihedral angles between triangles from the angles within triangles \eqref{geom H}. At the quantum level, those constraints generate dif\/ference equations solved by classical $\SU(2)$ re-coupling objects. It is quite clear for the closure relation: if $e_1$, $e_2$, $e_3$ meet at a node, the constraint $(\widehat{E}_1 + \widehat{E}_2+\widehat{E}_3)\vert \psi \rangle=0$ produces recursion relations satisf\/ied by Clebsch--Gordan coef\/f\/icients (or $3mj$-symbols). Moreover, we have described above a regularization of the scalar constraint which leads to a dif\/ference equation on the $6j$-symbol.

Geometrically, the Gau\ss{} law regulated as the closure relation for the triangle \eqref{closure} only makes sense for a f\/lat triangle. If geometry is homogeneously positively curved, we would expect the regulated Gau\ss{} law to describe spherical triangles instead. In terms of lengths and angles that should be the spherical law of cosines,
\[
\cos\phi_{e_1 e_2} = \frac{\cos\ell_{e_3} - \cos\ell_{e_1} \cos\ell_{e_2}}{\sin\ell_{e_1} \sin\ell_{e_2}},
\]
where $\ell_e$ is the spherical length of the edge~$e$. From that view, the second constraint, that relating dihedral angles to angles of triangles, is actually unchanged, \eqref{regge flat},
\[
\cos \Theta_{t_1 t_2} = \f{\cos\phi_{e_1 e_2} - \cos\phi_{e_1 e} \cos\phi_{e_2 e}}{\sin\phi_{e_1 e}  \sin\phi_{e_2 e}}.
\]
To make those ideas more precise, in particular to quantize, it is necessary to know the expression of the above geometric relations in terms of holonomies and f\/luxes, or to have the symplectic structure which involves spherical lengths.

\begin{conjecture*} Based on the results we have reviewed in the case $\Lambda=0$, one can conjecture the following. Recursions satisfied by $q$-deformed Clebsch--Gordan coefficients and the $q$-deformed $6j$-symbol are quantizations of the above formulae for spherical triangles and tetrahedra.
\end{conjecture*}

Finding a regulated Gau\ss{} law whose geometric content is curved triangles is certainly the way to get an anomaly-free regularization. This is actually what comes from the covariant analysis of discretized gravity by Dittrich and collaborators which we describe in Section~\ref{sec:beyondbf}. In a nutshell, the discrete action with $\Lambda\neq0$ has no gauge symmetry when f\/lat simplices are used. To restore the exact symmetry at the discrete level while turning on $\Lambda>0$, it is possible to use a discretization based on homogeneously curved simplices. Geometrically, that is very natural and quite obvious: that means that translating the nodes of the lattice is a symmetry if it is done by moving on the curved manifold.

{\bf Four-dimensional extension.} A direct extension of the new Hamiltonian is presented in~\cite{semiclass-paper}. Its form is the same, built on holonomies and f\/luxes. It still acts on cycles independently, and thus necessarily related to the topological model which imposes f\/latness of the connection. On the boundary of a 4-simplex, triangulating the 3-sphere, the Wheeler--DeWitt equation is a~dif\/ference equation solved by Wigner $15j$-symbols.

More generally it is easily seen that the Hamiltonian we have discussed produces dif\/ference equations which are solved by Wigner symbols. It is actually not surprising, but it is instead a conf\/irmation that the formalism is under control. Indeed, Wigner symbols are evaluation of spin networks on the trivial connection, and as such they naturally arise in the quantization of the BF model.

{\bf Large spin asymptotics and loop quantum gravity geometries.} The large spin limit of the recursion~\eqref{rec6j} makes the geometric picture quite clear,
\begin{gather} \label{semiclass eq}
\bigl[\Delta_{e} +2  (1-\cos \Theta_{e} )\bigr] \psi(j_1,\dots,j_6) = 0,
\end{gather}
where $\Delta_e$ is the discrete Laplacian on the edge $e$, and $\Theta_e$ the dihedral angle at $e$, computed from the lengths. Such an equation can be straightforwardly derived from a simple dynamics in Regge calculus. The canonical variables are the edge lengths $l_e$ and momenta $\theta_e$, and the constraint reads
\[ %\label{semiclass constraint}
\cos \theta_{e}  -  \cos \Theta_{e}(l)  = 0,
\]
which means that the momenta are fully determined as dihedral angles between triangles for \emph{flat} space.

A similar constraint was proposed for f\/lat space in 3+1 dimensions in \cite{phase-space-dittrich}, with lengths replaced with triangle areas, and it was argued that they form a f\/irst class system. As a~convergence of ideas and approaches, it turns out that the BF model provides a realization at the quantum level of such a constraint for f\/lat space. Indeed, the large spin behavior of the dif\/ference equation coming from $H_{v,f}$ on the boundary of the 4-simplex, using coherent state methods, takes exactly the form \eqref{semiclass eq}, where spins label triangles and geometrically stand for quantum areas instead of quantum lengths~\cite{semiclass-paper}.

However, we would like to stress that the Hamiltonian on holonomies and f\/luxes goes beyond Regge calculus in the 3+1 dimensional case, in a sense we now explain. In three dimensions, all f\/lux conf\/igurations which satisfy the closure equation are Regge geometries, determined by length variables. We simply say that loop quantum gravity geometries are Regge. But the situation is dif\/ferent in four dimensions, because a tetrahedron is not determined by its areas. The result of the full analysis \cite{phase-space-dittrich} is that $\SU(2)$ loop quantum gravity geometries are discontinuous: each tetrahedron unambiguously describes a f\/lat Regge geometry, but dif\/ferent tetrahedra will generically describe their common triangles with dif\/ferent lengths. Such geometries were later called \emph{twisted geometries}~\cite{twisted-freidel}.

Instead of going further on the kinematical aspects (see \cite{phase-space-dittrich, twisted-carlo-simone}), we simply want to emphasize that, as far as f\/lat space is concerned, the Hamiltonian is well-def\/ined on the \emph{whole} phase space of loop quantum gravity, while the large spin behavior \eqref{semiclass eq} only describes the geometric sector, where geometries are really of the Regge type. For non-Regge conf\/igurations, the exact dif\/ference equation takes a dif\/ferent asymptotic form, which is consistent with asymptotics of dif\/ferent symbols in that regime~\cite{barrett-asym-summary}. In the geometric sector, \eqref{semiclass eq} can be taken as a criterion for a~quantum geometry model to be approximated with some quantum Regge calculus in the large spin limit.

{\bf More on recursion relations.} The study of $H_{v,f}$ and other recursion relations on Wigner symbols has suggested a nice feature of quantum BF theory, that geometric properties of quantum f\/lat space are encoded into recursion relations coming from group representation theory. An example, discussed in \cite{ semiclass-paper, yet-another, recurrence-paper} is the closure of the simplex. In the asymptotics, the idea is to combine \eqref{semiclass eq} with the classical closure which reads $\det(\cos \Theta_e(l))=0$, i.e. the vanishing of the Gram matrix. Remarkably, the $10j$-symbol at the core of the four-dimensional Barrett--Crane model is known to satisfy such an equation, but in an exact way~\cite{recurrence-paper}. This leads to expect that in the geometric sector of full quantum gravity, geometric properties are still probed through dif\/ference equations which arise from representation theory.

\subsubsection{What we learn}
In addition to its own interest as a new way of quantizing 2+1 gravity using the scalar Hamiltonian, this analysis gives the following insights.
\begin{itemize}\itemsep=0pt
\item The spin 1 is the natural spin to regularize the curvature. It makes it possible to solve the Wheeler--DeWitt equation for all values of the state on integers. Hence, that solves the model for $\SO(3)$. By contrast, working with a spin 2 regularization produces a dif\/ference equation with shifts $\pm 1$, $\pm2$, so that more initial conditions are needed to get the state on all integers~\cite{3d-wdw}. Solving the model for $\SU(2)$ instead of $\SO(3)$ however requires a spin 1/2 regularization of the curvature, which is accessible in the spinor formalism~\cite{spinor-lqg}.
\item The splitting between spatial dif\/feomorphisms and the scalar component f\/inds an unexpected realization in this exact model. Indeed, the usual tension is that the curvature at a node of a triangulation is regularized by Wilson loop around it, while the scalar Hamiltonian is def\/ined by contracting the curvature with the normal to the surface, $H\sim \vec{n}\cdot \vec{F}$. The way-out which turns out to be successful is to take a dif\/ferent normal for each triangle meeting at a node, three of them being linearly independent. From that view the above construction is more something which mimics the scalar constraint on spin networks.
\item The asymptotic equation \eqref{semiclass eq} is a key feature of the quantum simplex in the large quantum number regime, and can used as a criterion for a model to reproduce Regge calculus there.
\item Classical geometric properties of f\/lat space can be implemented at the quantum level as dif\/ference equations coming from group representation theory. This strengthens the core of spin foam models where spacetime dynamics is expressed with objects from group representations.
\item The loop quantization supports a non-trivial Hamiltonian, not only in the asymptotics but also for twisted geometries in four dimensions.
\item No divergences are present, because the generator of the gauge symmetry is quantized and not the projector. That improves the spherical case with respect to \cite{noui-perez-ps3d}, and it should be a good framework to study perturbations with respect to the topological order.
\end{itemize}

\subsection{Beyond the topological model} \label{sec:beyondbf}

\subsubsection{Broken symmetries} \label{sec:class-tent}

The above BF model is topological and enjoys an exact lattice realization which allows the evaluation of transition amplitudes between spin network states using a single cell decomposition of spacetime (containing the graphs). But generically in a theory which does have local degrees of freedom, a lattice formulation is a truncation which misses some degrees of freedom and hence breaks the symmetries. In a series a papers Bahr and Dittrich and collaborators~\cite{bahr-broken-sym, bahr-improved-regge, bahr-linearized-gauge, bahr-improved-quantum}, and Dittrich and H\"o{}hn~\cite{hohn-canonical-regge}, studied the discretization of reparametrization invariant systems, especially the Regge approximation to general relativity. Main issues are how to restore broken symmetries, and how to def\/ine a canonical discrete formulation which reproduces the dynamics of the discrete action. Here we sum up some ideas of~\cite{bahr-broken-sym}.

Starting from a triangulation of the initial canonical hypersurface, a local evolution with discrete time steps can be def\/ined in terms of \emph{tent moves}. Let $\Delta_{(d-1)}$ be the initial triangulation, and $s$ a node. A tent move on $s$ creates a new node $s^*$, and new links which join $s^*$ to all the nodes $s$ was connected to. That def\/ines a new canonical triangulation $\Delta^*_{(d-1)}$, on which the number of variables is unchanged. Equivalently, the move glues a piece $T$ of a $d$-dimensional triangulation between $\Delta_{(d-1)}$ and $\Delta_{(d-1)}^*$, and the link $(ss^*)$ is called the tent pole.

The independent variables on $T$ are the lengths. To get momenta and equations of motion on $\Delta_{(d-1)}$ and $\Delta_{(d-1)}^*$, one uses the action on $T$ as a generating function. When $d=3$, it reads
\[
S_{T} = -\sum_{e\subset \mathrm{int}T} l_e  \vareps_e - \sum_{e\subset \pp T} l_e  \psi_e^T + \Lambda \sum_v V_v.
\]
The def\/icit angles around the edges $e$ in the bulk of $T$ and on the boundary are
\[
\vareps_e = \biggl(2\pi-\sum_{v\supset e}\thet_{ev}\biggr), \qquad\text{and}\qquad \psi_e^T = \biggl(\pi - \sum_{v\supset e} \thet_{ev}\biggr),
\]
where $\thet_{ev}$ is the dihedral angle hinged at $e$ in the tetrahedron $v$. $V_v$ is the volume of $v$, and $\Lambda$ the cosmological constant. Denote $s_i$ the nodes connected to $s$, with lengths $l_i$ and $l^*$ the tent pole length. The momenta are then
\[
p_i = -\f{\pp S_T}{\pp l_i},\qquad
p_i^* =  \f{\pp S_T}{\pp l_i^{*}},\qquad
p = -\f{\pp S_T}{\pp l^*}.
\]
From the equations of motion, we get $p=0$. Moreover, performing a second tent move, it can be seen that there is a single well-def\/ined momentum per edge.

Symmetries on a solution can be observed from the existence of zero eigenvalues of the Hessian matrix evaluated on that solution. In the case of vanishing cosmological constant, a symmetry is found in~\cite{bahr-broken-sym} around the f\/lat solution. It corresponds to moving a internal vertex in f\/lat 3-space (after several tent moves are performed), and those degrees of freedom are interpreted as the lapse and the shift. That translates into constraints on the canonical data on $\Delta_{(d-1)}$ only,
\begin{gather} \label{tent-constraint}
p_i = -\pi + \thet_{s s_i}(l_i, l_{ij}),
\end{gather}
in agreement with the fact that $3d$ gravity supports a discretization with exact symmetries. Those constraints state that the momenta are identif\/ied with the dihedral angles determined by the lengths. This is analogous to the constraints~\eqref{geom H} considered in the previous section. We thus see a convergence of idea coming from dif\/ferent approaches to the same issue.

However, when turning on the cosmological constant, it was observed that there are no symmetry around homogeneously curved solutions. Therefore, there are no constraints anymore, but instead pseudo-constraints. That means that the equations of motion from $T$ translates in the canonical setting into equations which depend on the data living on two consecutive time slices, and not a single one like in~\eqref{tent-constraint}.

\subsubsection{Restoring symmetries and the continuum limit}

In that example (homogeneous curvature), it is nevertheless quite clear how one can discretize in an exact way. The correct symmetry should be translation of a vertex not in f\/lat space but following the curved solution. Hence, one should use curved simplices instead of the f\/lat ones, so as to get the constraint
\[
p_i = -\pi + \thet^{(\Lambda)}_{s s_i}(l_i, l_{ij}),
\]
where the dihedral angle $\thet^{(\Lambda)}_{s s_i}$ is evaluated within the curved tetrahedron.

The main result of this approach is a canonical framework which reproduces the discrete action. The presence of symmetries is therefore determined by the choice of the discrete action. That view was further expanded in \cite{hohn-canonical-regge} by studying the expansion of Regge calculus around a~background metric, and in~\cite{bahr-improved-regge}. In the latter, the key idea is to restore the symmetries which are broken upon discretization by going to the \emph{perfect action}. That perfect action is reached, at least formally, in the continuum limit. Hence, it can be approached by a coarse-graining process at the classical level. Let $\Delta_2$ be a triangulation whose edges are denoted $E$, and $\Delta_1$ a f\/iner triangulation with edges $e$. We will obtained from the Regge action on $\Delta_1$ with f\/lat simplices, an \emph{improved action} on $\Delta_2$. Let us f\/ix the lengths $L_E$, and solve the equations of motion for~$l_e$ with a homogeneously curved geometry determined by $\Lambda$. The action on the coarse-grained triangulation will be obtained thanks to the constraint $\sum\limits_{e\subset E}l_e =L_E$. The improved action is then
\[
S_{\Delta_2}(L_E) = S_{\Delta_1}(l_e^*)_{|\sum\limits_{e\subset E}l^*_e =L_E},
\]
where the lengths $l^*_e$ are solutions of the curved geometry. The action for $\Delta_2$ consequently takes into account the dynamics of the f\/iner triangulation. Repeating the process many times, the internal geometry of the simplices of the most coarse triangulation receives more and more contributions from the curvature of the f\/iner ones, and thus in the appropriate limit, each simplex becomes homogeneously curved. That leads to the Regge discretization which has the exact symmetries.

While the topological BF model supports an exact discrete version, gauge symmetries are generically broken. Restoring them (and hence getting the correct number of degrees of freedom, in particular of propagating degrees of freedom) necessitates to go to the continuous limit. It can be reached as above by coarse-graining (recursive integration of the equations of motion) to yield at least formally a discrete action with exact symmetries, or by taking the number of simplices to inf\/inity.

The work of Dittrich and collaborators was recently re-interpreted by Rovelli \cite{carlo-ditt-invariance}. The desired reparametrization invariance (dif\/feomorphisms) to be recovered was there called Ditt-invariance. One key feature is that the continuum limit is reached by sending the number of discrete steps of the parameters of the theory to inf\/inity, while the physical time step is sent to zero in a~dynamical way. In the case of the harmonic oscillator, portions where the trajectory can be approximated by a free particle need very little time steps. On the contrary portions where the potential is important need more points to describe the trajectory accurately. It turns out that the discretization in the arbitrary time parameter together with energy conservation leads to that picture in terms of the physical time parameter.

The discretization and coarse-graining process is discussed for reparametrization invariant systems at the quantum level in \cite{bahr-improved-quantum}.

\subsubsection{The continuum limit through matrix and tensor models} \label{sec:tensors}

The notion of continuum limit at the quantum (or statistical) level has been given a sense in matrix models (for two-dimensional surfaces) \cite{revue-mm} and tensor models (in higher dimensions) \cite{ambjorn3dqg, gross-tensor, sasa1}. These are f\/ield theories which generate sums over triangulations, and as such the latter can be controlled. The free energy of a typical model reads
\[
F = \sum_{n\geq 0}\quad e^{-\mu n} \sum_{\substack{\text{triangulations }T_n\\ \text{with $n$ simplices}}} s(T_n) A(T_n).
\]
Here $T_n$ are triangulations of the sphere\footnote{They are isolated from the full sums over triangulations of all topologies using the large $N$ limit~\cite{david-mm, razvan-largeN, kazakov-mm}, where $N$ is the typical size of the matrices or tensors. The triangulations to be summed over may moreover be a~specif\/ic subset of all triangulations of the sphere~\cite{critical-tensor}.} with $n$ simplices, $s(T_n)$ is a combinatorial factor, and~$A(T_n)$ the amplitude assigned by the model to a given triangulation. The feature we want to stress here comes from the chemical potential~$\mu$. When it is very large, only small~$n$ contribute. However, when it decreases, the sum loses its summability. This is the regime where~$F$ is dominated by an inf\/inite number of simplices, and it is obtained for a (non-zero) critical value of $\mu\rightarrow\mu_c$. This continuum limit is approached like
$
F\simeq \big(e^{-\mu_c} - e^{-\mu}\big)^{2-\gamma}.
$
In the simplest models, where $A(T_n)=1$, the sum at f\/ixed $n$ just counts the number of triangulations with a given number of vertices, so that $\gamma$ is known as the entropy exponent. In the matrix model, it is $\gamma=-1/2$ and in higher dimensions, $\gamma=1/2$.

Such models have a natural interpretation as \emph{dynamical triangulations}, in the regime of small Newton constant and large cosmological constant. It is also possible to modify the amplitudes~$A(G)$ to accommodate loop quantum gravity, or rather spin foam, amplitudes. That is the purpose of group f\/ield theories, which re-sums spin foam amplitudes, and much of the recent progress in tensor models have been motivated by the group f\/ield theory program.

\section{Conclusion} \label{sec:con}

Since the original def\/inition of Hamiltonian constraint by Thiemann, several new approaches to tackling quantum dynamics in LQG have emerged, and the quantum constraint, still the most popular of all its rivals in mini-superspace models, has taken a backseat in the full theory. The inf\/inite number of ambiguities, lack of spacetime covariance at quantum level, and ultra-locality are some prime reasons why it is even now considered to be a highly unsatisfactory def\/inition.

We have argued that the quantum spacetime covariance, i.e.\ the of\/f-shell closure of the constraint algebra, could be achieved if one was ready to broaden one's perspective on all possible choices that are available in def\/ining composite operators, thanks to the discontinuity of representations and its inherent non-local nature. Already in loop quantum cosmology and in parametrized f\/ield theory, one sees the fruits of such endeavours and ends up with regularization schemes which remarkably enough have never been considered in the full theory.

Applying these lessons to the dif\/feomorphism constraint bears some real fruits. A precise def\/inition of the quantum constraint which generates inf\/initesimal dif\/feomorphisms on states in the LM habitat, and whose kernel matches with the well-known dif\/feomorphism invariant space~$\mathcal{H}_{\rm dif\/f}$. Once again the choices one has to make in order to get such an operator are very dif\/ferent then the ones which have been looked at in the literature so far.

We have then presented a dif\/ferent road based on the loop quantization of the BF topological model. It includes pure 2+1 gravity, where a recent Hamiltonian, \cite{semiclass-paper, 3d-wdw}, was found to mimic the scalar constraint while reproducing the standard results. That shows that the framework supports non-trivial operators, with a nice geometric content: the result is a quantization of f\/lat (Euclidean) geometry. More generally, that model shows that classical properties of f\/lat geometry take the form of dif\/ference equations from group representation theory, in the spin network quantization scheme. It also shed some light on the quantization ambiguities, such as curvature regularization and its spin, and is close to Smolin's proposal~\cite{smolin-ultralocality} to avoid ultra-locality. However, to go further, the continuum limit should be investigated, beyond the topological model. We have presented some directions in the context of discrete gravity, where symmetries are broken, but can be restored upon coarse-graining, leading to the perfect action.

The remaining issues in 2+1 gravity are mainly to write explicitly the quantum algebra of the new Hamiltonian. It should be interesting since it does not implement the splitting between spatial dif\/feomorphisms and the scalar component in the usually expected way. That was a~surprise and it should certainly be considered seriously also in 3+1 general relativity.

This unexpected splitting of the algebra observed in 2+1 gravity speaks in favor of the spin foam formalism, because it is a space-time formulation on cell decompositions, and hence naturally compatible with the geometry arising from the new Hamiltonian. The dif\/ference equations coming from group representation to describe quantum f\/lat geometry also point towards spin foam models, since they are built from group Fourier transforms. In particular, the asymptotic form of the Wheeler--DeWitt equation derived through the new Hamiltonian (for the topological model) is satisf\/ied on the quantum 4-simplex by some spin foam models \cite{semiclass-paper}. However, in 3+1 dimensions, exhausting the gauge symmetries is subtler (due to reducibility). Hence, in spite of those quantum dif\/ference equations which match with the spin foam model for 3+1 BF theory~\cite{ooguri-4d}, the latter is not well-def\/ined yet: it is not known how to gauge-f\/ix it properly. We think it is a serious issue which should be resolved if one wants to understand what the dif\/feomorphism symmetry becomes in the f\/lat or almost f\/lat sector of the full theory, and how the internal gauge symmetry of BF theory is broken on non-f\/lat solutions.

\subsection*{Acknowledgements}

Research at Perimeter Institute is supported by the Government of Canada through Industry
Canada and by the Province of Ontario through the Ministry of Research and Innovation.
A.L.\ is grateful to the Penn-State gravity group, especially Abhay Ashtekar, Miguel Campiglia, Adam Handerson and Casey Tomlin for many discussions related to the dynamical issues in LQG over past year. He is indebted to Madhavan Varadarajan for countably inf\/inite discussions on the Hamiltonian constraint in LQG which crystallized his understanding about many of the issues discussed in this review. Work of A.L.\ is supported by the NSF grant, PHY-0854743.

\pdfbookmark[1]{References}{ref}
\LastPageEnding


\begin{thebibliography}{99}
\footnotesize\itemsep=0pt

\bibitem{alesci-noui}
Alesci E., Noui K., Sardelli F., Spin-foam models and the physical scalar
  product,
\href{http://dx.doi.org/10.1103/PhysRevD.78.104009}{\textit{Phys. Rev.~D}} \textbf{78} (2008), 104009, 16~pages,
  \href{http://arxiv.org/abs/0807.3561}{arXiv:0807.3561}.

\bibitem{alesci-rovelli-hamiltonian}
Alesci E., Rovelli C., Regularization of the Hamiltonian constraint compatible
  with the spinfoam dynamics, \href{http://dx.doi.org/10.1103/PhysRevD.82.044007}{\textit{Phys. Rev~D}} \textbf{82} (2010), 044007,
  17~pages, \href{http://arxiv.org/abs/1005.0817}{arXiv:1005.0817}.

\bibitem{alesci-thiemann}
Alesci E., Thiemann T., Zipfel A., Linking covariant and canonical LQG: new
  solutions to the Euclidean scalar constraint, \href{http://arxiv.org/abs/1109.1290}{arXiv:1109.1290}.

\bibitem{ambjorn3dqg}
Ambj{\o}rn J., Durhuus B., J{\'o}nsson T., Three-dimensional simplicial quantum
  gravity and generalized matrix models, \href{http://dx.doi.org/10.1142/S0217732391001184}{\textit{Modern Phys. Lett.~A}}
  \textbf{6} (1991), 1133--1146.

\bibitem{3nj-marzuoli}
Anderson R.W., Aquilanti V., Marzuoli A., $3nj$ morphogenesis and semiclassical
  disentangling, \href{http://dx.doi.org/10.1021/jp905212a}{\textit{J.~Phys. Chem.~A}} \textbf{113} (2009), 15106--15117,
  \href{http://arxiv.org/abs/1001.4386}{arXiv:1001.4386}.

\bibitem{spinnets-marzuoli}
Aquilanti V., Bitencourt A.C.P., da~S.~Ferreira C., Marzuoli A., Ragni M.,
  Quantum and semiclassical spin networks: from atomic and molecular physics to
  quantum computing and gravity, \href{http://dx.doi.org/10.1088/0031-8949/78/05/058103}{\textit{Phys. Scr.}} \textbf{78} (2008),
  058103, 7~pages, \href{http://arxiv.org/abs/0901.1074}{arXiv:0901.1074}.

\bibitem{qm6j}
Aquilanti V., Haggard H.M., Hedeman A., Jeevanjee N., Littlejohn R., Yu L.,
  Semiclassical mechanics of the Wigner $6j$-symbol, \href{http://arxiv.org/abs/1009.2811}{arXiv:1009.2811}.

\bibitem{alreview}
Ashtekar A., Lewandowski J., Background independent quantum gravity: a status
  report, \href{http://dx.doi.org/10.1088/0264-9381/21/15/R01}{\textit{Classical Quantum Gravity}} \textbf{21} (2004), R53--R152,
  \href{http://arxiv.org/abs/gr-qc/0404018}{gr-qc/0404018}.

\bibitem{aplimproved}
Ashtekar A., Pawlowski T., Singh P., Quantum nature of the big bang: improved
  dynamics, \href{http://dx.doi.org/10.1103/PhysRevD.74.084003}{\textit{Phys. Rev.~D}} \textbf{74} (2006), 084003, 23~pages,
  \href{http://arxiv.org/abs/gr-qc/0607039}{gr-qc/0607039}.

\bibitem{baez-sf-bf}
Baez J.C., An introduction to spin foam models of BF theory and quantum
  gravity, in Geometry and Quantum Physics ({S}chladming, 1999),
  \href{http://dx.doi.org/10.1007/3-540-46552-9_2}{\textit{Lecture Notes in Phys.}}, Vol.~543, Springer, Berlin, 25--93,
  \href{http://arxiv.org/abs/gr-qc/9905087}{gr-qc/9905087}.

\bibitem{baez-perez-strings-bf}
Baez J.C., Perez A., Quantization of strings and branes coupled to BF theory,
  \textit{Adv. Theor. Math. Phys.} \textbf{11} (2007), 451--469,
  \href{http://arxiv.org/abs/gr-qc/0605087}{gr-qc/0605087}.

\bibitem{baez-fractional-bf}
Baez J.C., Wise D.K., Crans A.S., Exotic statistics for strings in 4D BF
  theory, \textit{Adv. Theor. Math. Phys.} \textbf{11} (2007), 707--749,
  \href{http://arxiv.org/abs/gr-qc/0603085}{gr-qc/0603085}.

\bibitem{bahr-broken-sym}
Bahr B., Dittrich B., ({B}roken) gauge symmetries and constraints in {R}egge
  calculus, \href{http://dx.doi.org/10.1088/0264-9381/26/22/225011}{\textit{Classical Quantum Gravity}} \textbf{26} (2009), 225011,
  34~pages, \href{http://arxiv.org/abs/0905.1670}{arXiv:0905.1670}.

\bibitem{bahr-improved-regge}
Bahr B., Dittrich B., Improved and perfect actions in discrete gravity,
  \href{http://dx.doi.org/10.1103/PhysRevD.80.124030}{\textit{Phys. Rev.~D}} \textbf{80} (2009), 124030, 15~pages,
  \href{http://arxiv.org/abs/0907.4323}{arXiv:0907.4323}.

\bibitem{bahr-linearized-gauge}
Bahr B., Dittrich B., He S., Coarse graining free theories with gauge
  symmetries: the linearized case, \href{http://dx.doi.org/10.1088/1367-2630/13/4/045009}{\textit{New~J. Phys.}} \textbf{13} (2011),
  045009, 34~pages, \href{http://arxiv.org/abs/1011.3667}{arXiv:1011.3667}.

\bibitem{bianca-finite}
Bahr B., Dittrich B., Ryan J.P., Spin foam models with f\/inite groups,
  \href{http://arxiv.org/abs/1103.6264}{arXiv:1103.6264}.

\bibitem{bahr-improved-quantum}
Bahr B., Dittrich B., Steinhaus S.,
Perfect discretization of reparametrization invariant path integrals,
\href{http://dx.doi.org/10.1103/PhysRevD.83.105026}{\textit{Phys. Rev.~D}} \textbf{83} (2011), 19~pages,
\href{http://arxiv.org/abs/1101.4775}{arXiv:1101.4775}.

\bibitem{diffeo3dgft}
Baratin A., Girelli F., Oriti D., Dif\/feomorphisms in group f\/ield theories,
  \href{http://dx.doi.org/10.1103/PhysRevD.83.104051}{\textit{Phys. Rev.~D}} \textbf{83} (2011), 104051, 22~pages,
  \href{http://arxiv.org/abs/1101.0590}{arXiv:1101.0590}.

\bibitem{barrett-crane-wdw}
Barrett J.W., Crane L., An algebraic interpretation of the
  {W}heeler--{D}e{W}itt equation, \href{http://dx.doi.org/10.1088/0264-9381/14/8/011}{\textit{Classical Quantum Gravity}}
  \textbf{14} (1997), 2113--2121, \href{http://arxiv.org/abs/gr-qc/9609030}{gr-qc/9609030}.

\bibitem{barrett-asym-summary}
Barrett J.W., Dowdall R.J., Fairbairn W.J., Gomes H., Hellmann F., Pereira R.,
  Asymptotics of 4d spin foam models, \href{http://dx.doi.org/10.1007/s10714-010-0983-7}{\textit{Gen. Relativity Gravitation}}
  \textbf{43} (2011), 2421--2436, \href{http://arxiv.org/abs/1003.1886}{arXiv:1003.1886}.

\bibitem{barrett-asym15j}
Barrett J.W., Fairbairn W.J., Hellmann F., Quantum gravity asymptotics from the
  ${\rm SU}(2)$ $15j$-symbol, \href{http://dx.doi.org/10.1142/S0217751X10049281}{\textit{Internat.~J. Modern Phys.~A}} \textbf{25}
  (2010), 2897--2916, \href{http://arxiv.org/abs/0912.4907}{arXiv:0912.4907}.

\bibitem{barrett-pr}
Barrett J.W., Naish-Guzman I., The Ponzano--Regge model, \href{http://dx.doi.org/10.1088/0264-9381/26/15/155014}{\textit{Classical
  Quantum Gravity}} \textbf{26} (2011), 155014, 48~pages, \href{http://arxiv.org/abs/0803.3319}{arXiv:0803.3319}.

\bibitem{bergeron-fractional-bf}
Bergeron M., Semenof\/f G.W., Szabo R.J., Canonical BF-type topological f\/ield
  theory and fractional statistics of strings, \href{http://dx.doi.org/10.1016/0550-3213(94)00503-7}{\textit{Nuclear Phys.~B}}
  \textbf{437} (1995), 695--721, \href{http://arxiv.org/abs/hep-th/9407020}{hep-th/9407020}.

\bibitem{blau-thompson-torsion}
Blau M., Thompson G., A new class of topological f\/ield theories and the
  {R}ay--{S}inger torsion, \href{http://dx.doi.org/10.1016/0370-2693(89)90526-1}{\textit{Phys. Lett.~B}} \textbf{228} (1989), 64--68.

\bibitem{blau-thompson-bf}
Blau M., Thompson G., Topological gauge theories of antisymmetric tensor
  f\/ields, \href{http://dx.doi.org/10.1016/0003-4916(91)90240-9}{\textit{Ann. Physics}} \textbf{205} (1991), 130--172.

\bibitem{weinstein}
Blohmann C., Fernandes M.C.B., Weinstein A., Groupoid symmetry and constraints
  in general relativity, \href{http://arxiv.org/abs/1003.2857}{arXiv:1003.2857}.

\bibitem{semiclass-paper}
Bonzom V., Spin foam models and the Wheeler--DeWitt equation for the quantum
  4-simplex, \href{http://dx.doi.org/10.1103/PhysRevD.84.024009}{\textit{Phys. Rev.~D}} \textbf{84} (2011), 024009, 13~pages,
  \href{http://arxiv.org/abs/1101.1615}{arXiv:1101.1615}.

\bibitem{3njsmall}
Bonzom V., Fleury P., Asymptotics of Wigner $3nj$-symbols with small and large
  angular momenta: an elementary method, \href{http://arxiv.org/abs/1108.1569}{arXiv:1108.1569}.

\bibitem{3d-wdw}
Bonzom V., Freidel L., The Hamiltonian constraint in 3d Riemannian loop quantum
  gravity, \href{http://dx.doi.org/10.1088/0264-9381/28/19/195006}{\textit{Classical Quantum Gravity}} \textbf{28} (2011), 195006,
  24~pages, \href{http://arxiv.org/abs/1101.3524}{arXiv:1101.3524}.

\bibitem{critical-tensor}
Bonzom V., Gurau R., Riello A., Rivasseau V., Critical behavior of colored
  tensor models in the large $N$ limit, \href{http://dx.doi.org/10.1016/j.nuclphysb.2011.07.022}{\textit{Nuclear Phys.~B}} \textbf{853}
  (2011), 174--195, \href{http://arxiv.org/abs/1105.3122}{arXiv:1105.3122}.

\bibitem{yet-another}
Bonzom V., Livine E.R., Yet another recursion relation for the $6j$-symbol,
  \href{http://arxiv.org/abs/1103.3415}{arXiv:1103.3415}.

\bibitem{6jnlo}
Bonzom V., Livine E.R., Smerlak M., Speziale S., Towards the graviton from
  spinfoams: the complete perturbative expansion of the 3d toy model,
  \href{http://dx.doi.org/10.1016/j.nuclphysb.2008.05.012}{\textit{Nuclear Phys.~B}} \textbf{804} (2008), 507--526, \href{http://arxiv.org/abs/0802.3983}{arXiv:0802.3983}.

\bibitem{recurrence-paper}
Bonzom V., Livine E.R., Speziale S., Recurrence relations for spin foam
  vertices, \href{http://dx.doi.org/10.1088/0264-9381/27/12/125002}{\textit{Classical Quantum Gravity}} \textbf{27} (2010), 125002,
  32~pages, \href{http://arxiv.org/abs/0911.2204}{arXiv:0911.2204}.

\bibitem{sorting-out}
Bonzom V., Smerlak M., Bubble divergences: sorting out topology from cell
  structure, \href{http://arxiv.org/abs/1103.3961}{arXiv:1103.3961}.

\bibitem{cell-homology}
Bonzom V., Smerlak M., Bubble divergences from cellular cohomology,
  \href{http://dx.doi.org/10.1007/s11005-010-0414-4}{\textit{Lett. Math. Phys.}} \textbf{93} (2010), 295--305, \href{http://arxiv.org/abs/1004.5196}{arXiv:1004.5196}.

\bibitem{twisted}
Bonzom V., Smerlak M., Bubble divergences from twisted cohomology,
  \href{http://arxiv.org/abs/1008.1476}{arXiv:1008.1476}.

\bibitem{bf-anydim}
Bonzom V., Smerlak M.,
Gauge symmetries in spinfoam gravity: the case for `cellular quantization',
  \href{http://arxiv.org/abs/1201.4996}{arXiv:1201.4996}.

\bibitem{quantum-tet-marzuoli}
Carfora M., Marzuoli A., Rasetti M., Quantum tetrahedra, \href{http://dx.doi.org/10.1021/jp909824h}{\textit{J.~Phys.
  Chem.~A}} \textbf{113} (2009), 15376--15383, \href{http://arxiv.org/abs/1001.4402}{arXiv:1001.4402}.

\bibitem{cattaneo-3d-4dBF}
Cattaneo A.S., Cotta-Ramusino P., Fr{\"o}hlich J., Martellini M., Topological
  BF theories in 3 and 4 dimensions, \href{http://dx.doi.org/10.1063/1.531238}{\textit{J.~Math. Phys.}} \textbf{36}
  (1995), 6137--6160, \href{http://arxiv.org/abs/hep-th/9505027}{hep-th/9505027}.

\bibitem{bfym}
Cattaneo A.S., Cotta-Ramusino P., Fucito F., Martellini M., Rinaldi M., Tanzini
  A., Zeni M., Four-dimensional {Y}ang--{M}ills theory as a deformation of
  topological BF theory, \href{http://dx.doi.org/10.1007/s002200050465}{\textit{Comm. Math. Phys.}} \textbf{197} (1998),
  571--621, \href{http://arxiv.org/abs/hep-th/9705123}{hep-th/9705123}.

\bibitem{symmetries-bf-gravity}
Constantinidis C.P., Piguet O., Gieres F., Sarandy M.S., On the symmetries of
  BF models and their relation with gravity, \href{http://dx.doi.org/10.1088/1126-6708/2002/01/017}{\textit{J. High Energy Phys.}}
  \textbf{2002} (2002), no.~1, 017, 25~pages, \href{http://arxiv.org/abs/hep-th/0111273}{hep-th/0111273}.

\bibitem{david-mm}
David F., A model of random surfaces with nontrivial critical behaviour,
  \href{http://dx.doi.org/10.1016/0550-3213(85)90363-3}{\textit{Nuclear Phys.~B}} \textbf{257} (1985), 543--576.

\bibitem{bf-pleb}
De~Pietri R., Freidel L., {${\rm so}(4)$} {P}leba\'nski action and relativistic
  spin-foam model, \href{http://dx.doi.org/10.1088/0264-9381/16/7/303}{\textit{Classical Quantum Gravity}} \textbf{16} (1999),
  2187--2196, \href{http://arxiv.org/abs/gr-qc/9804071}{gr-qc/9804071}.

\bibitem{revue-mm}
Di~Francesco P., Ginsparg P., Zinn-Justin J., 2D gravity and random matrices,
  \href{http://dx.doi.org/10.1016/0370-1573(94)00084-G}{\textit{Phys. Rep.}} \textbf{254} (1995), no.~1--2, 133~pages,
  \href{http://arxiv.org/abs/hep-th/9306153}{hep-th/9306153}.

\bibitem{bianca-spinnets}
Dittrich B., Eckert F.C., Martin-Benito M., Coarse graining methods for spin
  net and spin foam models, \href{http://arxiv.org/abs/1109.4927}{arXiv:1109.4927}.

\bibitem{hohn-canonical-regge}
Dittrich B., H{\"o}hn P.A., From covariant to canonical formulations of
  discrete gravity, \href{http://dx.doi.org/10.1088/0264-9381/27/15/155001}{\textit{Classical Quantum Gravity}} \textbf{27} (2010),
  155001, 37~pages, \href{http://arxiv.org/abs/0912.1817}{arXiv:0912.1817}.

\bibitem{phase-space-dittrich}
Dittrich B., Ryan J.P., Phase space descriptions for simplicial 4D geometries,
  \href{http://dx.doi.org/10.1088/0264-9381/28/6/065006}{\textit{Classical Quantum Gravity}} \textbf{28} (2011), 065006, 34~pages,
  \href{http://arxiv.org/abs/0807.2806}{arXiv:0807.2806}.

\bibitem{simplicity-dittrich}
Dittrich B., Ryan J.P., Simplicity in simplicial phase space, \href{http://dx.doi.org/10.1103/PhysRevD.82.064026}{\textit{Phys.
  Rev.~D}} \textbf{82} (2010), 064026, 19~pages, \href{http://arxiv.org/abs/1006.4295}{arXiv:1006.4295}.

\bibitem{Dittrich:2004bn}
Dittrich B., Thiemann T., Testing the master constraint programme for loop
  quantum gravity. I.~General framework, \href{http://dx.doi.org/10.1088/0264-9381/23/4/001}{\textit{Classical Quantum Gravity}}
  \textbf{23} (2006), 1025--1065, \href{http://arxiv.org/abs/gr-qc/0411138}{gr-qc/0411138}.

\bibitem{Dittrich:2004bp}
Dittrich B., Thiemann T., Testing the master constraint programme for loop
  quantum gravity. II.~Finite-dimensional systems, \href{http://dx.doi.org/10.1088/0264-9381/23/4/002}{\textit{Classical Quantum
  Gravity}} \textbf{23} (2006), 1067--1088, \href{http://arxiv.org/abs/gr-qc/0411139}{gr-qc/0411139}.

\bibitem{Dittrich:2004bq}
Dittrich B., Thiemann T., Testing the master constraint programme for loop
  quantum gravity. III.~${\rm SL}(2,{\mathbb R})$ models, \href{http://dx.doi.org/10.1088/0264-9381/23/4/003}{\textit{Classical
  Quantum Gravity}} \textbf{23} (2006), 1089--1120, \href{http://arxiv.org/abs/gr-qc/0411140}{gr-qc/0411140}.

\bibitem{Dittrich:2004br}
Dittrich B., Thiemann T., Testing the master constraint programme for loop
  quantum gravity. IV.~Free f\/ield theories, \href{http://dx.doi.org/10.1088/0264-9381/23/4/004}{\textit{Classical Quantum Gravity}}
  \textbf{23} (2006), 1121--1142, \href{http://arxiv.org/abs/gr-qc/0411141}{gr-qc/0411141}.

\bibitem{Dittrich:2004bs}
Dittrich B., Thiemann T., Testing the master constraint programme for loop
  quantum gravity. V.~Interacting f\/ield theories, \href{http://dx.doi.org/10.1088/0264-9381/23/4/005}{\textit{Classical Quantum
  Gravity}} \textbf{23} (2006), 1143--1162, \href{http://arxiv.org/abs/gr-qc/0411142}{gr-qc/0411142}.

\bibitem{pushing6j}
Dupuis M., Livine E.R., Pushing the asymptotics of the $6j$-symbol further,
  \href{http://dx.doi.org/10.1103/PhysRevD.80.024035}{\textit{Phys. Rev.~D}} \textbf{80} (2009), 024035, 14~pages,
  \href{http://arxiv.org/abs/0905.4188}{arXiv:0905.4188}.

\bibitem{6jmaite}
Dupuis M., Livine E.R., The $6j$-symbol: recursion, correlations and
  asymptotics, \href{http://dx.doi.org/10.1088/0264-9381/27/13/135003}{\textit{Classical Quantum Gravity}} \textbf{27} (2010), 135003,
  15~pages, \href{http://arxiv.org/abs/0910.2425}{arXiv:0910.2425}.

\bibitem{fairbairn-perez-strings-bf}
Fairbairn W.J., Perez A., Extended matter coupled to BF theory, \href{http://dx.doi.org/10.1103/PhysRevD.78.024013}{\textit{Phys.
  Rev.~D}} \textbf{78} (2008), 024013, 21~pages, \href{http://arxiv.org/abs/0709.4235}{arXiv:0709.4235}.

\bibitem{freidel-gft}
Freidel L., Group f\/ield theory: an overview, \href{http://dx.doi.org/10.1007/s10773-005-8894-1}{\textit{Internat.~J. Theoret.
  Phys.}} \textbf{44} (2005), 1769--1783, \href{http://arxiv.org/abs/hep-th/0505016}{hep-th/0505016}.

\bibitem{freidel-puzio}
Freidel L., Krasnov K., Puzio R., BF description of higher-dimensional gravity
  theories, \textit{Adv. Theor. Math. Phys.} \textbf{3} (1999), 1289--1324,
  \href{http://arxiv.org/abs/hep-th/9901069}{hep-th/9901069}.

\bibitem{freidel-louapre-pr1}
Freidel L., Louapre D., Ponzano--{R}egge model revisited. I.~Gauge f\/ixing,
  observables and interacting spinning particles, \href{http://dx.doi.org/10.1088/0264-9381/21/24/002}{\textit{Classical Quantum
  Gravity}} \textbf{21} (2004), 5685--5726, \href{http://arxiv.org/abs/hep-th/0401076}{arXiv:hep-th/0401076}.

\bibitem{freidel-speziale-bf-gravity}
Freidel L., Speziale S., On the relations between gravity and BF theories,
  \href{http://arxiv.org/abs/1201.4247}{arXiv:1201.4247}.

\bibitem{twisted-freidel}
Freidel L., Speziale S., Twisted geometries: a geometric parametrisation of
  SU(2) phase space, \href{http://dx.doi.org/10.1103/PhysRevD.82.084040}{\textit{Phys. Rev.~D}} \textbf{82} (2010), 084040,
  16~pages, \href{http://arxiv.org/abs/1001.2748}{arXiv:1001.2748}.

\bibitem{bf-mdm}
Freidel L., Starodubtsev A., Quantum gravity in terms of topological
  observables, \href{http://arxiv.org/abs/hep-th/0501191}{hep-th/0501191}.

\bibitem{frohman-torsion}
Frohman C., Kania-Bartoszynska J., Dubois' torsion, A-polynomial and quantum
  invariants, \href{http://arxiv.org/abs/1101.2695}{arXiv:1101.2695}.

\bibitem{lm2}
Gambini R., Lewandowski J., Marolf D., Pullin J., On the consistency of the
  constraint algebra in spin network quantum gravity, \href{http://dx.doi.org/10.1142/S0218271898000103}{\textit{Internat.~J.
  Modern Phys.~D}} \textbf{7} (1998), 97--109, \href{http://arxiv.org/abs/gr-qc/9710018}{arXiv:gr-qc/9710018}.

\bibitem{symmetries-bf-bv}
Gieres F., Grimstrup J.M., Nieder H., Pisar T., Schweda M., Topological f\/ield
  theories and their symmetries within the {B}atalin--{V}ilkovisky framework,
  \href{http://dx.doi.org/10.1103/PhysRevD.66.025027}{\textit{Phys. Rev.~D}} \textbf{66} (2002), 025027, 14~pages,
  \href{http://arxiv.org/abs/hep-th/0111258}{hep-th/0111258}.

\bibitem{marolf}
Giulini D., Marolf D., On the generality of ref\/ined algebraic quantization,
  \href{http://dx.doi.org/10.1088/0264-9381/16/7/321}{\textit{Classical Quantum Gravity}} \textbf{16} (1999), 2479--2488,
  \href{http://arxiv.org/abs/gr-qc/9812024}{gr-qc/9812024}.

\bibitem{gross-tensor}
Gross M., Tensor models and simplicial quantum gravity in $>2$-D,
  \href{http://dx.doi.org/10.1016/S0920-5632(05)80015-5}{\textit{Nuclear Phys.~B Proc. Suppl.}} \textbf{25A} (1992), 144--149.

\bibitem{razvan-largeN}
Gurau R., The complete $1/N$ expansion of colored tensor models in arbitrary
  dimension, \href{http://arxiv.org/abs/1102.5759}{arXiv:1102.5759}.

\bibitem{gurau-review}
Gurau R., Ryan J.P., Colored tensor models -- a review, \href{http://arxiv.org/abs/1109.4812}{arXiv:1109.4812}.

\bibitem{hajicekpft}
H{\'a}j{\'{\i}}cek P., Isham C.J., The symplectic geometry of a parametrized
  scalar f\/ield on a curved background, \href{http://dx.doi.org/10.1063/1.531578}{\textit{J.~Math. Phys.}} \textbf{37}
  (1996), 3505--3521, \href{http://arxiv.org/abs/gr-qc/9510028}{gr-qc/9510028}.

\bibitem{Han:2009aw}
Han M., Thiemann T., On the relation between operator constraint, master
  constraint, reduced phase space and path integral quantization,
  \href{http://dx.doi.org/10.1088/0264-9381/27/22/225019}{\textit{Classical Quantum Gravity}} \textbf{27} (2010), 225019, 46~pages,
  \href{http://arxiv.org/abs/0911.3428}{arXiv:0911.3428}.

\bibitem{horowitz-bf}
Horowitz G.T., Exactly soluble dif\/feomorphism invariant theories, \href{http://dx.doi.org/10.1007/BF01218410}{\textit{Comm.
  Math. Phys.}} \textbf{125} (1989), 417--437.

\bibitem{chern-simons-jeffrey}
Jef\/frey L.C., Chern--{S}imons--{W}itten invariants of lens spaces and torus
  bundles, and the semiclassical approximation, \href{http://dx.doi.org/10.1007/BF02097243}{\textit{Comm. Math. Phys.}}
  \textbf{147} (1992), 563--604.

\bibitem{kazakov-mm}
Kazakov V.A., Bilocal regularization of models of random surfaces,
  \href{http://dx.doi.org/10.1016/0370-2693(85)91011-1}{\textit{Phys. Lett.~B}} \textbf{150} (1985), 282--284.

\bibitem{kitaev-code}
Kitaev A.Y., Fault-tolerant quantum computation by anyons, \href{http://dx.doi.org/10.1016/S0003-4916(02)00018-0}{\textit{Ann.
  Physics}} \textbf{303} (2003), 2--30, \mbox{\href{http://arxiv.org/abs/quant-ph/9707021}{quant-ph/9707021}}.

\bibitem{karelpft}
Kuchar K., Parametrized scalar f\/ield on ${\bf R}\times S^1$: dynamical
  pictures, spacetime dif\/feomorphisms, and conformal isometries, \href{http://dx.doi.org/10.1103/PhysRevD.39.1579}{\textit{Phys.
  Rev.~D}} \textbf{39} (1989), 1579--1593.

\bibitem{amhamconstraint}
Laddha A., Varadarajan M., Hamiltonian constraint in polymer parametrized f\/ield
  theory, \href{http://dx.doi.org/10.1103/PhysRevD.83.025019}{\textit{Phys. Rev~D}} \textbf{83} (2011), 025019, 27~pages,
  \href{http://arxiv.org/abs/1011.2463}{arXiv:1011.2463}.

\bibitem{alokme2}
Laddha A., Varadarajan M., Polymer quantization of the free scalar f\/ield and
  its classical limit, \href{http://dx.doi.org/10.1088/0264-9381/27/17/175010}{\textit{Classical Quantum Gravity}} \textbf{27} (2010),
  175010, 45~pages, \href{http://arxiv.org/abs/1001.3505}{arXiv:1001.3505}.

\bibitem{amdiffconstraint}
Laddha A., Varadarajan M., The dif\/feomorphism constraint operator in loop
  quantum gravity, \href{http://dx.doi.org/10.1088/0264-9381/28/19/195010}{\textit{Classical Quantum Gravity}} \textbf{28} (2011),
  195010, 29~pages, \href{http://arxiv.org/abs/1105.0636}{arXiv:1105.0636}.

\bibitem{levin-wen-condensation}
Levin M.A., Wen X.G., String-net condensation: a physical mechanism for
  topological phases, \href{http://dx.doi.org/10.1103/PhysRevB.71.045110}{\textit{Phys. Rev.~B}} \textbf{71} (2005), 045110,
  21~pages, \href{http://arxiv.org/abs/cond-mat/0404617}{cond-mat/0404617}.

\bibitem{lm1}
Lewandowski J., Marolf D., Loop constraints: a habitat and their algebra,
  \href{http://dx.doi.org/10.1142/S0218271898000231}{\textit{Internat.~J. Modern Phys.~D}} \textbf{7} (1998), 299--330,
  \href{http://arxiv.org/abs/gr-qc/9710016}{gr-qc/9710016}.

\bibitem{lost}
Lewandowski J., Okol{\'o}w A., Sahlmann H., Thiemann T., Uniqueness of
  dif\/feomorphism invariant states on holonomy-f\/lux algebras, \href{http://dx.doi.org/10.1007/s00220-006-0100-7}{\textit{Comm.
  Math. Phys.}} \textbf{267} (2006), 703--733, \href{http://arxiv.org/abs/gr-qc/0504147}{gr-qc/0504147}.

\bibitem{livine-coherentBF}
Livine E.R., Speziale S., New spinfoam vertex for quantum gravity,
  \href{http://dx.doi.org/10.1103/PhysRevD.76.084028}{\textit{Phys. Rev.~D}} \textbf{76} (2007), 084028, 14~pages,
  \href{http://arxiv.org/abs/0705.0674}{arXiv:0705.0674}.

\bibitem{spinor-lqg}
Livine E.R., Tambornino J., Spinor representation for loop quantum gravity,
  \href{http://dx.doi.org/10.1063/1.3675465}{\textit{J.~Math. Phys.}} \textbf{53} (2012), 012503, 33~pages,
  \href{http://arxiv.org/abs/1105.3385}{arXiv:1105.3385}.

\bibitem{renormalisability-bf}
Lucchesi C., Piguet O., Sorella S.P., Renormalization and f\/initeness of
  topological BF theories, \href{http://dx.doi.org/10.1016/0550-3213(93)90219-F}{\textit{Nuclear Phys.~B}} \textbf{395} (1993),
  325--353, \href{http://arxiv.org/abs/hep-th/9208047}{hep-th/9208047}.

\bibitem{maggiore-sorella-perturbative-4dbf}
Maggiore N., Sorella S.P., Perturbation theory for antisymmetric tensor f\/ields
  in four dimensions, \href{http://dx.doi.org/10.1142/S0217751X93000369}{\textit{Internat.~J. Modern Phys.~A}} \textbf{8} (1993),
  929--945, \href{http://arxiv.org/abs/hep-th/9204044}{hep-th/9204044}.

\bibitem{nicolaireview}
Nicolai H., Peeters K., Zamaklar M., Loop quantum gravity: an outside view,
  \href{http://dx.doi.org/10.1088/0264-9381/22/19/R01}{\textit{Classical Quantum Gravity}} \textbf{22} (2005), R193--R247,
  \href{http://arxiv.org/abs/hep-th/0501114}{hep-th/0501114}.

\bibitem{noui-perez-ps3d}
Noui K., Perez A., Three-dimensional loop quantum gravity: physical scalar
  product and spin-foam models, \href{http://dx.doi.org/10.1088/0264-9381/22/9/017}{\textit{Classical Quantum Gravity}} \textbf{22}
  (2005), 1739--1761, \href{http://arxiv.org/abs/gr-qc/0402110}{gr-qc/0402110}.

\bibitem{perez-pranzetti1}
Noui K., Perez A., Pranzetti D., Canonical quantization of non-commutative
  holonomies in 2+1 loop quantum gravity, \href{http://arxiv.org/abs/1105.0439}{arXiv:1105.0439}.

\bibitem{ooguri-3d}
Ooguri H., Partition functions and topology-changing amplitudes in the
  three-dimensional lattice gravity of {P}onzano and {R}egge, \href{http://dx.doi.org/10.1016/0550-3213(92)90188-H}{\textit{Nuclear
  Phys.~B}} \textbf{382} (1992), 276--304, \href{http://arxiv.org/abs/hep-th/9112072}{hep-th/9112072}.

\bibitem{ooguri-4d}
Ooguri H., Topological lattice models in four dimensions, \href{http://dx.doi.org/10.1142/S0217732392004171}{\textit{Modern Phys.
  Lett.~A}} \textbf{7} (1992), 2799--2810, \mbox{\href{http://arxiv.org/abs/hep-th/9205090}{hep-th/9205090}}.

\bibitem{oriti-review}
Oriti D., The group f\/ield theory approach to quantum gravity: some recent
  results, \href{http://arxiv.org/abs/0912.2441}{arXiv:0912.2441}.

\bibitem{alejandroreview}
Perez A., Introduction to loop quantum gravity and spin foams,
  \href{http://arxiv.org/abs/gr-qc/0409061}{gr-qc/0409061}.

\bibitem{perezhamconstraint}
Perez A., Regularization ambiguities in loop quantum gravity, \href{http://dx.doi.org/10.1103/PhysRevD.73.044007}{\textit{Phys.
  Rev.~D}} \textbf{73} (2006), 044007, 18~pages, \href{http://arxiv.org/abs/gr-qc/0509118}{gr-qc/0509118}.

\bibitem{perez-pranzetti2}
Perez A., Pranzetti D., On the regularization of the constraint algebra of
  quantum gravity in 2+1 dimensions with a nonvanishing cosmological constant,
  \href{http://dx.doi.org/10.1088/0264-9381/27/14/145009}{\textit{Classical Quantum Gravity}} \textbf{27} (2010), 145009, 20~pages,
  \href{http://arxiv.org/abs/1001.3292}{arXiv:1001.3292}.

\bibitem{PR}
Ponzano G., Regge T., Semi-classical limit of Racah coef\/f\/icients, in
  Spectroscopic and Group Theoretical Methods in Physics, Editor F.~Bloch,
  North-Holland, Amsterdam, 1968, 1--58.

\bibitem{roberts}
Roberts J., Classical $6j$-symbols and the tetrahedron, \href{http://dx.doi.org/10.2140/gt.1999.3.21}{\textit{Geom. Topol.}}
  \textbf{3} (1999), 21--66, \href{http://arxiv.org/abs/math-ph/9812013}{math-ph/9812013}.

\bibitem{carlo-newlook}
Rovelli C., A new look at loop quantum gravity, \href{http://dx.doi.org/10.1088/0264-9381/28/11/114005}{\textit{Classical Quantum
  Gravity}} \textbf{28} (2011), 114005, 24~pages, \href{http://arxiv.org/abs/1004.1780}{arXiv:1004.1780}.

\bibitem{carlo-ditt-invariance}
Rovelli C., Discretizing parametrized systems: the magic of Ditt-invariance,
  \href{http://arxiv.org/abs/1107.2310}{arXiv:1107.2310}.

\bibitem{twisted-carlo-simone}
Rovelli C., Speziale S., On the geometry of loop quantum gravity on a graph,
  \href{http://dx.doi.org/10.1103/PhysRevD.82.044018}{\textit{Phys. Rev.~D}} \textbf{82} (2010), 044018, 6~pages,
  \href{http://arxiv.org/abs/1005.2927}{arXiv:1005.2927}.

\bibitem{chern-simons-rozanski}
Rozansky L., A large $k$ asymptotics of Witten's invariant of Seifert
  manifolds, \href{http://dx.doi.org/10.1007/BF02099272}{\textit{Comm. Math. Phys.}} \textbf{171} (1995), 279--322,
  \href{http://arxiv.org/abs/hep-th/9303099}{hep-th/9303099}.

\bibitem{sasa1}
Sasakura N., Tensor model for gravity and orientability of manifold,
  \href{http://dx.doi.org/10.1142/S0217732391003055}{\textit{Modern Phys. Lett.~A}} \textbf{6} (1991), 2613--2623.

\bibitem{schulten-gordon2}
Schulten K., Gordon R.G., Semiclassical approximations to $3j$- and
  $6j$-coef\/f\/icients for quantum-mechanical coupling of angular momenta,
  \href{http://dx.doi.org/10.1063/1.522427}{\textit{J.~Math. Phys.}} \textbf{16} (1975), 1971--1988.

\bibitem{smolin-ultralocality}
Smolin L., The classical limit and the form of the Hamiltonian constraint in
  nonperturbative quantum general relativity, \href{http://arxiv.org/abs/gr-qc/9609034}{gr-qc/9609034}.

\bibitem{ttbook}
Thiemann T., Modern canonical quantum general relativity, Cambridge Monographs
  on Mathematical Physics, \href{http://dx.doi.org/10.1017/CBO9780511755682}{Cambridge University Press}, Cambridge, 2007,
  \href{http://arxiv.org/abs/gr-qc/0110034}{gr-qc/0110034}.

\bibitem{qsd1}
Thiemann T., Quantum spin dynamics (QSD), \href{http://dx.doi.org/10.1088/0264-9381/15/4/011}{\textit{Classical Quantum Gravity}}
  \textbf{15} (1998), 839--873, \href{http://arxiv.org/abs/gr-qc/9606089}{gr-qc/9606089}.

\bibitem{qsd2}
Thiemann T., Quantum spin dynamics ({QSD}). {II}.~The kernel of the
  {W}heeler--{D}e{W}itt constraint operator, \href{http://dx.doi.org/10.1088/0264-9381/15/4/012}{\textit{Classical Quantum Gravity}}
  \textbf{15} (1998), 875--905, \href{http://arxiv.org/abs/gr-qc/9606090}{gr-qc/9606090}.

\bibitem{qsd3}
Thiemann T., Quantum spin dynamics ({QSD}). {III}.~Quantum constraint algebra
  and physical scalar product in quantum general relativity, \href{http://dx.doi.org/10.1088/0264-9381/15/5/010}{\textit{Classical
  Quantum Gravity}} \textbf{15} (1998), 1207--1247, \href{http://arxiv.org/abs/gr-qc/9705017}{gr-qc/9705017}.

\bibitem{thiemann-3d}
Thiemann T., Quantum spin dynamics (QSD). IV.~2+1 Euclidean quantum gravity as
  a model to test 3+1 Lorentzian quantum gravity, \href{http://dx.doi.org/10.1088/0264-9381/15/5/011}{\textit{Classical Quantum
  Gravity}} \textbf{15} (1998), 1249--1280, \href{http://arxiv.org/abs/gr-qc/9705018}{gr-qc/9705018}.

  \bibitem{Thiemann:2005zg}
Thiemann T., Quantum spin dynamics. VIII.~The master constraint,
  \href{http://dx.doi.org/10.1088/0264-9381/23/7/003}{\textit{Classical Quantum Gravity}} \textbf{23} (2006), 2249--2265,
  \href{http://arxiv.org/abs/gr-qc/0510011}{gr-qc/0510011}.

\bibitem{thiemann-master-constraint}
Thiemann T., The {P}hoenix {P}roject: master constraint programme for loop
  quantum gravity, \href{http://dx.doi.org/10.1088/0264-9381/23/7/002}{\textit{Classical Quantum Gravity}} \textbf{23} (2006),
  2211--2247, \href{http://arxiv.org/abs/gr-qc/0305080}{gr-qc/0305080}.


\bibitem{varshalovich}
Varshalovich D.A., Moskalev A.N., Khersonskii V.K., Quantum theory of angular
  momentum, World Scientif\/ic Publishing Co. Inc., Teaneck, NJ, 1988.

\bibitem{witten-3d-gravity}
Witten E., 2+1 dimensional gravity as an exactly soluble system,
  \href{http://dx.doi.org/10.1016/0550-3213(88)90143-5}{\textit{Nuclear Phys.~B}} \textbf{311} (1988), 46--78.

\bibitem{witten-2dym}
Witten E., On quantum gauge theories in two dimensions, \href{http://dx.doi.org/110.1007/BF02100009}{\textit{Comm. Math.
  Phys.}} \textbf{141} (1991), 153--209.

\bibitem{witten-amplitude-3d}
Witten E., Topology-changing amplitudes in (2+1)-dimensional gravity,
  \href{http://dx.doi.org/10.1016/0550-3213(89)90591-9}{\textit{Nuclear Phys.~B}} \textbf{323} (1989), 113--140.



\bibitem{yu15j}
Yu L., Asymptotic limits of the Wigner $15j$-symbol with small quantum numbers,
  \href{http://arxiv.org/abs/1104.3641}{arXiv:1104.3641}.

\bibitem{yu12j}
Yu L., Semiclassical analysis of the Wigner $12j$ symbol with one small angular
  momentum, \href{http://dx.doi.org/10.1103/PhysRevA.84.022101}{\textit{Phys. Rev.~A}} \textbf{84} (2011), 022101, 13~pages,
\href{http://arxiv.org/abs/1104.3275}{arXiv:1104.3275}.

\bibitem{yu9j}
Yu L., Littlejohn R.G., Semiclassical analysis of the Wigner $9j$ symbol with
  small and large angular momenta, \href{http://dx.doi.org/10.1103/PhysRevA.83.052114}{\textit{Phys. Rev.~A}} \textbf{83} (2011),
  052114, 16~pages, \href{http://arxiv.org/abs/1104.1499}{arXiv:1104.1499}.

\bibitem{jucys}
Yutsis A.P., Levinson I.B., Vanagas V.V., Mathematical apparatus of the theory
  of angular momentum, Israel Program for Scientif\/ic Translations, Jerusalem,
  1962.

\end{thebibliography}
\end{document}